\documentclass[journal]{IEEEtran}
\ifCLASSINFOpdf
  \usepackage[pdftex]{graphicx}
  \DeclareGraphicsExtensions{.pdf,.jpeg,.png}
\else
\fi
\usepackage{amsmath}
\ifCLASSOPTIONcompsoc
  \usepackage[caption=false,font=normalsize,labelfont=sf,textfont=sf]{subfig}
\else
  \usepackage[caption=false,font=footnotesize]{subfig}
\fi
\usepackage{url}
\usepackage{xcolor}
\usepackage{booktabs, makecell, multirow}
\usepackage{amsfonts} %mathbb
\usepackage{hyperref}
\usepackage{threeparttable}

\usepackage[nolist,nohyperlinks]{acronym}

\newacro{fom}[FoM]{Figure of Merit}
\newacroplural{fom}[FoMs]{Figures of Merit}
\newacro{mlem}[ML-EM]{Maximum-Likelihood Expectation-Maximization}
\newacro{ml}[ML]{Maximum Likelihood}
\newacro{lmmlem}[LM-MLEM]{List-Mode Maximum-Likelihood Expectation-Maximization}
\newacro{fov}[FOV]{Field of View}
\newacro{mc}[MC]{Monte-Carlo}
\newacro{pet}[PET]{Positron Emission Tomography}
\newacro{ics}[ICS]{inter-crystal scattering}
\newacro{lor}[LOR]{line of response}
\newacroplural{lor}[LORs]{lines of response}
\newacro{nist}[NIST]{National Institute of Standards and Technology}
\newacro{iq}[IQ]{image quality}
\newacro{roi}[ROI]{Region-Of-Interest}
\newacroplural{roi}[ROIs]{Regions-Of-Interest}
\newacro{rc}[RC]{recovery coefficient}
\newacroplural{rc}[RCs]{recovery coefficients}
\newacro{sor}[SOR]{Spill-Over-Ratio}

\begin{document}
\bstctlcite{IEEEexample:BSTcontrol} % to enforce limited number of authors in the bib

% ------------------------------------------------------------------
% First page with notice
% ------------------------------------------------------------------
\begin{titlepage}
\thispagestyle{empty}
    {\Large Notice}\\[1cm]

This article has been accepted for publication in IEEE Transactions on Radiation and Plasma Medical Sciences. This is the author's version which has not been fully edited and content may change prior to final publication. Citation information: DOI 10.1109/TRPMS.2026.3691192

\vspace*{\fill} 
This work is licensed under a Creative Commons Attribution 4.0 License. For more information, see https://creativecommons.org/licenses/by/4.0/

\end{titlepage}

%
% paper title
% Titles are generally capitalized except for words such as a, an, and, as,
% at, but, by, for, in, nor, of, on, or, the, to and up, which are usually
% not capitalized unless they are the first or last word of the title.
% Linebreaks \\ can be used within to get better formatting as desired.
% Do not put math or special symbols in the title.
%\title{An Analytical Model for Inter-crystal Scattering in PET }
\title{Inclusion of Inter-crystal Scattering in PET: Analytical Models and Dedicated Reconstruction}

%
%
% author names and IEEE memberships
% note positions of commas and nonbreaking spaces ( ~ ) LaTeX will not break
% a structure at a ~ so this keeps an author's name from being broken across
% two lines.
% use \thanks{} to gain access to the first footnote area
% a separate \thanks must be used for each paragraph as LaTeX2e's \thanks
% was not built to handle multiple paragraphs
%

\author{Jorge~Roser,~\IEEEmembership{Member,~IEEE,}
        Hong~Phuc~Vo,~\IEEEmembership{Graduate Student Member,~IEEE,}
        Rebecca~Kantorek,
        Steven~Seeger,~\IEEEmembership{Member,~IEEE,}and~Magdalena~Rafecas,~\IEEEmembership{Senior~Member,~IEEE}% <-this % stops a space
\thanks{This work did not involve human subjects or animals in its research.}
\thanks{J. Roser, H. P. Vo, R. Kantorek and M. Rafecas are with the Institute of Medical Engineering, Universit{\"a}t
zu L{\"u}beck, Germany (e-mail: jorge.rosermartinez@uni-luebeck.de).}% <-this % stops a space
\thanks{R. Kantorek and S. Seeger are with the Isotope Laboratory of the Natural Science Section, Universit{\"a}t zu L{\"u}beck, Germany.}% <-this % stops a space
%\thanks{Manuscript received xxxx xx, xxxx; revised August 26, 2015.}
}

% note the % following the last \IEEEmembership and also \thanks - 
% these prevent an unwanted space from occurring between the last author name
% and the end of the author line. i.e., if you had this:
% 
% \author{....lastname \thanks{...} \thanks{...} }
%                     ^------------^------------^----Do not want these spaces!
%
% a space would be appended to the last name and could cause every name on that
% line to be shifted left slightly. This is one of those "LaTeX things". For
% instance, "\textbf{A} \textbf{B}" will typeset as "A B" not "AB". To get
% "AB" then you have to do: "\textbf{A}\textbf{B}"
% \thanks is no different in this regard, so shield the last } of each \thanks
% that ends a line with a % and do not let a space in before the next \thanks.
% Spaces after \IEEEmembership other than the last one are OK (and needed) as
% you are supposed to have spaces between the names. For what it is worth,
% this is a minor point as most people would not even notice if the said evil
% space somehow managed to creep in.

% The paper headers
%\markboth{Journal of \LaTeX\ Class Files,~Vol.~14, No.~8, August~2015}%
%{Shell \MakeLowercase{\textit{et al.}}: Bare Demo of IEEEtran.cls for IEEE Journals}
%\markboth{IEEE TRANSACTIONS ON RADIATION AND PLASMA MEDICAL SCIENCES,~Vol.~XX, No.~XX, January~2026}%
\markboth{THIS WORK HAS BEEN SUBMITTED TO THE IEEE FOR POSSIBLE PUBLICATION}%
{Roser \MakeLowercase{\textit{et al.}}}
% The only time the second header will appear is for the odd numbered pages
% after the title page when using the twoside option.
% 
% *** Note that you probably will NOT want to include the author's ***
% *** name in the headers of peer review papers.                   ***
% You can use \ifCLASSOPTIONpeerreview for conditional compilation here if
% you desire.

% If you want to put a publisher's ID mark on the page you can do it like
% this:
%\IEEEpubid{0000--0000/00\$00.00~\copyright~2015 IEEE}
% Remember, if you use this you must call \IEEEpubidadjcol in the second
% column for its text to clear the IEEEpubid mark.

% use for special paper notices
%\IEEEspecialpapernotice{(Invited Paper)}

% make the title area
\maketitle

% As a general rule, do not put math, special symbols or citations
% in the abstract or keywords.
\begin{abstract}
Inter-crystal scattering (ICS) in Positron Emission Tomography (PET) is commonly regarded as a degradation effect that might compromise the image spatial resolution. In parallel, the inclusion of ICS events has also been recognized as a potential approach to increase PET sensitivity, which could be especially beneficial in scenarios where the latter is a limiting factor, such as very small animal imaging. Several methods for the recovery of ICS events have been proposed, many of which aim to locate the first interaction, i.e., the Compton scattering site, usually limited by  their success rate, computational burden or data and training dependency. Conversely, this work proposes a physics-based model for ICS events, leading to analytical expressions of the sensitivity image and the system matrix (required by statistical reconstruction algorithms), without the need to identify the original line of response. After validating the model, the work shows how ICS events can be integrated into a joint image reconstruction algorithm (based on list-mode MLEM) together with conventional PET events, for which dedicated analytical models are also developed. To assess the performance of the proposed approach, Monte-Carlo simulated and experimental data of an image quality phantom were obtained with the MERMAID small-fish PET scanner prototype. Both simulation and experimental results indicate that, while slightly decreasing the recovery coefficient values, the inclusion of ICS clearly reduces statistical noise and
improves uniformity.
\end{abstract}

% Note that keywords are not normally used for peerreview papers.
\begin{IEEEkeywords}
Inter-crystal scattering, detector scatter, ICS,  PET,  image reconstruction, sensitivity
\end{IEEEkeywords}

% For peer review papers, you can put extra information on the cover
% page as needed:
% \ifCLASSOPTIONpeerreview
% \begin{center} \bfseries EDICS Category: 3-BBND \end{center}
% \fi
%
% For peerreview papers, this IEEEtran command inserts a page break and
% creates the second title. It will be ignored for other modes.
\IEEEpeerreviewmaketitle

\section{Introduction}

\IEEEPARstart{T}{he} quest for increased sensitivity has motivated advances in \ac{pet} since the very beginning of this nuclear imaging technique. Improving the sensitivity of \ac{pet} systems is essential to allow for dose and/or scan time reduction \cite{ENLOW2023}; in addition, sensitivity can be a limiting factor in low-count \ac{pet} scenarios. These scenarios include imaging certain segments of the population (e.g. pediatric \cite{MINGELS2024} or pregnant \cite{BURTON2023}), dynamic \ac{pet},  \ac{pet}-based ${}^{90}$Y dosimetry \cite{CARLIER2015}, as well as some specific applications like range verification in particle therapy by means of \ac{pet} \cite{OZOEMELAM2020,PARODI2023}, and preclinical \ac{pet} with very small animals such as zebrafish \cite{ZVOLSKY2022}, where the amount of radiotracer that can be administered to research animals is very small.  

To increase the sensitivity of \ac{pet} scanners, several approaches have been proposed, some of them already implemented in commercial systems, like the extension of the axial field of view (total-body \ac{pet} \cite{GODINEZ2025}), the use of organ-specific designs \cite{SANAAT2024,VO2025} and larger scintillation crystals \cite{SANAAT2020} or the application of wider energy windows \cite{KUANG2022}. A different, albeit complementary approach is the inclusion of \ac{ics} events. In an \ac{ics} event, at least one of the annihilation photons undergoes Compton scattering in a given \ac{pet} detector, prior to the absorption of the scattered photon in a different detector. It follows that an \ac{ics} event is characterized by at least three hits in different \ac{pet} detectors, as opposed to those conventional events in which both annihilation photons are directly absorbed, yielding two hits and hereafter referred to as \textit{golden} events. The term \textit{hit} refers here to the physical photon interaction, regardless of whether it is eventually detected or not\footnote{For instance, depending on the energy window, some Compton scattering interactions might remain undetected with the subsequent second interaction being assigned to the coincidence event, thus \textit{disguising} an \ac{ics} event as a golden one; this effect becomes more pronounced the worse the energy resolution.}; furthermore, \textit{\ac{pet} detector} refers here to the smallest detection unit capable of discriminating hits. Therefore, the concept of \ac{ics} also excludes that of \textit{intra-crystal scattering}, in which the Compton scattering and the eventual absorption of the scattered photon happen in the same PET detector and thus cannot be identified separately.

In the last years, \ac{ics} in \ac{pet} has received renewed attention, as some modern high-resolution detector designs - with two or more layers of small, tightly packed crystals - promote Compton scattering of the photons \cite{ZOU2025,STORTZ2018,GU2020,KANG2025}. Indeed, if properly handled, the detected \ac{ics} events fraction can significantly contribute to increase the system efficiency \cite{PETERSEN2024,RAFECAS2003,LEE2018b}. A common problem encountered with such events is the need to identify the Compton scattering site. For this reason, most of the \ac{ics}-recovery strategies focus on determining the position of the Compton scattering in order to obtain the correct \ac{lor} \cite{LEE2024}, either exploiting the energy information \cite{COMANOR1996}, Compton kinematics \cite{RAFECAS2003, PRATX2009}, optimization algorithms \cite{LAGE2015, LEE2018} or neural networks \cite{MICHAUD2015, WU2020, LEE2021,PETERSEN2024}. These strategies, however, entail the risk of incorrectly estimating the primary photon trajectory, thus introducing a number of \textit{wrong} \acp{lor} into the reconstruction, i.e., events carrying incorrect information about the spatial emission distribution. Such wrong \acp{lor} can compromise the attainable image quality and alter the activity quantification, especially for \ac{ics}-recovery methods with low accuracy. Even though some of the aforementioned event-selection strategies have reported notably high \acs{ics} identification accuracy (e.g. $98~\%$ \cite{LEE2018}), they require either significant computational resources, prior information, a particular PET geometry, or extensive data and training \cite{LEE2024}.

Conversely, an alternative approach is to model the system response to \ac{ics} events, which leads naturally to V-shaped \acp{lor} \cite{GILLAM2014}. Response models are a key element of statistical iterative reconstruction, such as \ac{lmmlem}, through the so-called \textit{system matrix} $\mathbf{H}$ and \textit{sensitivity image} $\mathbf{s}$:
\begin{IEEEeqnarray}{lCl} 
\label{eq:plainLMMLEM}
\boldsymbol{\lambda}^{(n+1)} 
&=& \boldsymbol{\lambda}^{(n)} 
\oslash \mathbf{s} \odot 
\left( \mathbf{H}^{\top}  \Big( \mathbf{1} \oslash (\mathbf{H} \boldsymbol{\lambda}^{(n)}) \right)
\end{IEEEeqnarray}
where $\boldsymbol{\lambda}^{(n)}$ is the image vector after the $n$-th iteration, $\mathbf{1}$ is a column vector of ones of length equal to the number of detected events;  $\odot$ and  $\oslash$  represent element-wise  multiplication and element-wise division, respectively. 
An element of the system matrix, $h_{iv}$, represents the probability that an emission from image element $v$ (usually a voxel) is detected as the measurement element $i$; whereas an element of the sensitivity image, $s_{v}$, denotes the overall detection probability of an emission from $v$. The ability of the models behind $\mathbf{H}$ and $\mathbf{s}$ to capture the underlying physics significantly determines the image quality.

Measurement elements described by V-\acp{lor} are less informative than regular \acp{lor}, or equivalently, the rows of the corresponding \ac{ics} system matrix $\mathbf{H}^\text{I}$  have higher intricacy\footnote{Another interpretation would be that the \ac{ics} system matrix $\mathbf{H}^\text{I}$ is less sparse than the golden one, $\mathbf{H}^\text{G}$.} than those of the golden system matrix $\mathbf{H}^\text{G}$; nonetheless, V-\acp{lor} carry useful information about the spatial emission distribution. In this way, golden and \ac{ics} events can be treated as two separated channels of information from which independent, accurate estimations of the object spatial emission distribution can be obtained; for instance, by applying (\ref{eq:plainLMMLEM}) separately to golden events modeled as \acp{lor} through $\mathbf{H}^\text{G}$, and $\mathbf{s}^\text{G}$, and to \ac{ics} events modeled as V-\acp{lor} through $\mathbf{H}^\text{I}$, and $\mathbf{s}^\text{I}$. However, and importantly, both channels \textit{together} can provide an estimate of the activity distribution via a joint image reconstruction algorithm \cite{GILLAM2014,GRKOVSKI2015,ROSER2022}, the resulting image quality being potentially superior to that obtained using only golden or \ac{ics} events, especially in the low-count regime.

 The present work explores and extends this V-LOR approach by developing a dedicated analytical model for \ac{ics} events based on the physics behind the photon emission and detector interactions, which is integrated into the \ac{lmmlem} algorithm and implemented for a very small-animal PET scanner (although it is neither restricted to this algorithm nor to this application). The study shows how the new model provides reasonably accurate estimations of the \ac{ics} sensitivity image $\mathbf{s}^\text{I}$ and the system matrix $\mathbf{H}^\text{I}$. As the concept of V-\acp{lor} takes into account the two possible trajectories, while only one is the original one, some degradation of the spatial resolution and other metrics could be expected. Therefore, this study also assesses the image quality obtained with the inclusion of \ac{ics} events modeled in this way, compared to reconstructing only golden events. In particular, the developed model targets three-hit \ac{ics} events, i.e., those featuring a single Compton scattering, as such events are typically far more frequent than four-or-more hit \ac{ics} events. For a fair comparison and also to correctly implement the joint reconstruction approach, this work also introduces a dedicated analytical model for the golden events, also guided by a probabilistic formulation of physical phenomena analogous to that of the \ac{ics} model. The methods are applied to both simulated and experimental data, with a focus on very low-count acquisitions.

The remainder of the work is structured as follows. In section \ref{sec:materials}, the analytical model for \ac{ics} events is formulated, mathematical expressions for the sensitivity image and the system matrix are derived, and the conducted experiments and Monte Carlo simulations are described. Section \ref{sec:results} shows the obtained results, which are subsequently discussed in section \ref{sec:discussion}. Finally, section \ref{sec:conclusion} summarizes the conclusions. In addition, four appendices have been included: Appendix \ref{sec:nomenclature} describes the employed nomenclature and Appendix \ref{sec:golden} provides details of the analytical model used for golden events; Appendix \ref{sec:sor} expand some of the results. Lastly, Appendix \ref{sec:conventionalICS} shows a comparison  between the proposed approach (dedicated V-LOR and golden modeling with joint reconstruction) and a LOR-based reconstruction of golden events plus ICS events, after estimation of the original ICS trajectory using conventional methods.

\section{Methods and Materials}\label{sec:materials}
In the following, the analytical models and their integration into a dedicated joint image reconstruction formalism are presented. The nomenclature employed can be found in table \ref{tab:nomenclature} (Appendix \ref{sec:nomenclature}).

\subsection{Analytical model for \ac{ics} events}

\begin{figure}[!t]
\centering
\includegraphics[width=2.8in]{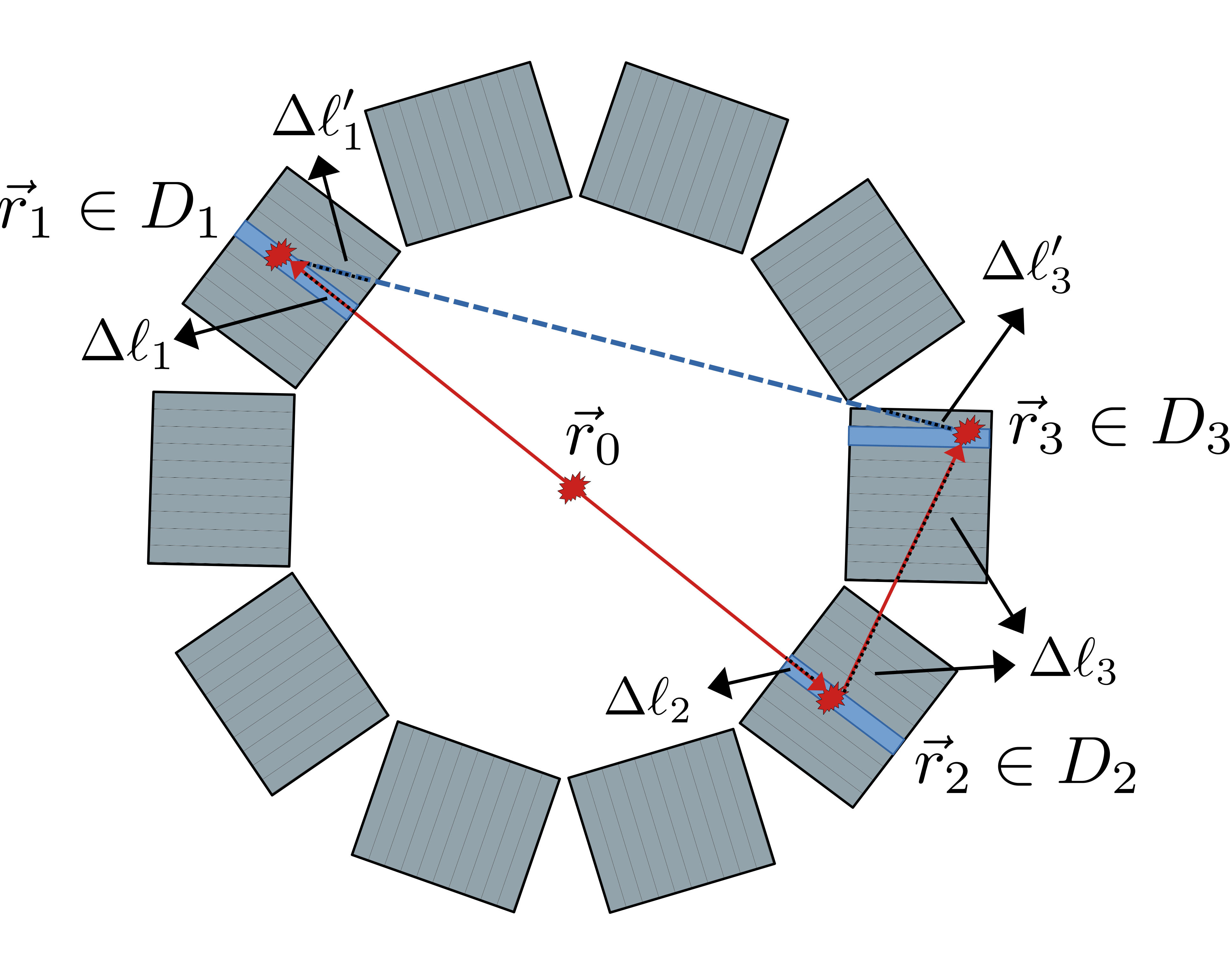}
\caption{Scheme of an \ac{ics} event with relevant quantities of the proposed analytical model. The red lines indicate the photon trajectories. The PET detector elements where the physical interactions occur are highlighted in light blue. In particular, the Compton scattering takes place in $\vec{r}_2$. }
\label{fig:ICSevent}
\end{figure}

Similar to \cite{ROSER2020}, the expressions for the system matrix and the sensitivity image are derived from the differential probability of an emission to become a true event in the detector. In this work, an emission refers to a positron-electron annihilation yielding two photons, and the true event is a three-hit \ac{ics} in which one annihilation photon is photoabsorbed in detector $D_1$ whereas the other annihilation photon undergoes a Compton scatter in detector $D_2$ and is photoabsorbed in detector $D_3$; a scheme of such event is provided in figure \ref{fig:ICSevent}. The corresponding differential probability, denoted as $dP^{\text{I}}_{123}$, can be formulated as 
\begin{IEEEeqnarray}{lCl}
dP^{\text{I}}_{123} &= &
2\cdot dP(\vec{r}_0 | v ) dP(\Omega_0 \Omega_0' | v \vec{r}_0 \Omega_0 = - \Omega_0' ) dP(\ell_1 | v \vec{r}_0 \Omega_0' ) 
\nonumber \\ 
&\cdot  &  
 dP(\ell_2 d\Omega_2 d\ell_3 | v \vec{r}_0 \Omega_0 ).
\label{eq:modelICS}
\end{IEEEeqnarray}
In (\ref{eq:modelICS}), the leading factor $2$ accounts for the fact that one annihilation photon undergoes direct photoabsorption whereas the other photon does not, and it is irrelevant which one undergoes which of the interactions.
The term 
\begin{equation}
dP(\vec{r}_0 | v )
=
\frac{d^3r_0}{V} \Theta_v(\vec{r}_0)
\end{equation}
corresponds to the differential probability of both annihilation photons to be emitted at position $\vec{r_0}$ inside image element $v$ (with volume V), whose limits are given by the indicator function $\Theta_v$. Furthermore,
\begin{equation}
\label{eq:omegas}
dP(\Omega_0 \Omega_0' | v \vec{r}_0 \Omega_0 = - \Omega_0' )
=
\frac{d\Omega_0}{4\pi} \frac{d\Omega_0'}{4\pi} 4\pi \delta(\Omega_0 + \Omega_0')
\end{equation}
corresponds to the differential probability for one annihilation photon to be emitted in a direction and the other in the exact opposite direction, being  $\Omega_0$ and $\Omega_0'$ the solid angles representing the two directions, respectively. In other words, acollinearity is not considered in the model.

Next, the differential probability of one annihilation photon to survive a distance $\Delta \ell_1$ in any detector material prior to photolectric absorption in $D_1$ is accounted for by
\begin{equation}
dP(\ell_1 | v \vec{r}_0 \Omega_0 ) = e^{-\mu_{0} \Delta \ell_1}  \mu_{0}^{e}d\ell_1 
,
\label{eq:exp1}
\end{equation}
whereas the \ac{ics} part is represented by
\begin{equation}
dP(\ell_2 \Omega_2 d\ell_3 | v \vec{r}_0 \Omega_0' ) = 
e^{-\mu_{0}  \Delta \ell_2}  n_e \frac{d\sigma^C}{d\Omega_2}d\Omega_2 d\ell_2 
e^{-\mu_{2} \Delta \ell_3} \mu_{2}^{e} d\ell_3, 
\label{eq:exp2}
\end{equation}
with $\Delta \ell_2$ being the distance over which the annihilation photon survives in any detector material before undergoing Compton scattering in $D_2$, and $\Delta \ell_3$ being the distance over which the Compton-scattered photon survives in any detector material before undergoing photoelectric effect in $D_3$. For the sake of simplicity, this formulation of the model assumes a single-material PET surrounded by vacuum; however, further materials could be included by replacing the exponents in (\ref{eq:exp1}) and (\ref{eq:exp2}) by the corresponding sum over the involved distances and attenuation coefficients.

By integrating $dP_{123}^{\text{I}}$ over the entire voxel volume V, all possible emission and Compton scattering directions, and all possible lengths $\ell_1, \ell_2, \ell_3$ over the whole PET extent, an expression for the probability of an annihilation originated in voxel $v$ to be detected anywhere in the PET scanner as an \ac{ics} event is obtained. This is precisely the definition of the \ac{ics} sensitivity image element,
\begin{IEEEeqnarray}{lCl}
s^{\text{I}}_v  &=&  \frac{2}{4\pi V}
\int_{\vec{r}_0 \in v} d^3r_0
\int_{\Omega_0 \in S^2} d\Omega_0 
\underset{\substack{ \vec{r}_0 - \ell_1\hat{\Omega}_0 \in \text{PET}}}{\int}
d\ell_1
e^{-\mu_0 \Delta \ell_1}\mu_0^e \nonumber \\ 
& & \cdot 
\underset{\substack{ \vec{r}_0 + \ell_2\hat{\Omega}_0 \in \text{PET} \\ \vec{r}_2 + \ell_3\hat{\Omega}_2 \in \text{PET}  \\  \Omega_2 \in S^2}}{\int}
d\ell_2 d\ell_3 d\Omega_2
e^{-\mu_0 \Delta \ell_2}
n_e \frac{d\sigma_0^C}{d\Omega_2} 
e^{-\mu_2 \Delta \ell_3} \mu_2^e, \nonumber \\*
\label{eq:sens1}
\end{IEEEeqnarray}
which can be readily computed e.g. via Monte Carlo numerical integration methods.

Now, the change of variables $\left\{ 
\vec{r}_0, \Omega_0, \ell_1, \ell_2, \Omega_2, \ell_3 
\right\}
\longrightarrow 
\left\{ 
\vec{r}_1, \vec{r}_2, \vec{r}_3, \xi  
\right\}$ allows $dP_{123}^{\text{I}}$ to be expressed as

\begin{IEEEeqnarray}{lCl}
dP_{123}^{\text{I}} &=& \frac{2}{4\pi V}
 d^3r_1
 d^3r_2
 d\xi \Theta_v(\xi)  
\frac{e^{-\mu_0 (\Delta\ell_1 + \Delta\ell_2) } \mu_0^e}{|\vec{r}_1 - \vec{r}_2 |^2} d^3r_3
\nonumber \\ 
&\cdot  &
n_e \frac{d\sigma_0^C}{d\Omega_2} 
\frac{e^{-\mu_2 (\Delta \ell_3)} \mu_2^e }{|\vec{r}_3 - \vec{r}_2|^2}
, \nonumber \\*
\label{eq:dPchangevar}
\end{IEEEeqnarray}
which, upon integration over the whole detector extent, leads to the following alternative expression for the sensitivity image:

\begin{IEEEeqnarray}{lCl}
s^{\text{I}}_v &=& \frac{2}{4\pi V}
\int_{\vec{r}_1 \in PET} d^3r_1
\int_{\vec{r}_2 \in PET} d^3r_2
\int_{\xi \in \mathbb{R}^+} d\xi \Theta_v(\xi) 
\nonumber \\ 
&\cdot  & 
\frac{e^{-\mu_0 (\Delta\ell_1 + \Delta\ell_2) } \mu_0^e}{|\vec{r}_1 - \vec{r}_2 |^2}
\int_{\vec{r}_3 \in PET} d^3r_3
n_e \frac{d\sigma_0^C}{d\Omega_2} 
\frac{e^{-\mu_2 (\Delta \ell_3)} \mu_2^e }{|\vec{r}_3 - \vec{r}_2|^2}. \nonumber \\*
\label{eq:sens2}
\end{IEEEeqnarray}

Equation (\ref{eq:sens2}) is particularly well-suited for numerical computation via ray-tracing methods \cite{SIDDON1985}. Furthermore, the aforementioned change of variables allows obtaining the probability density:

\begin{IEEEeqnarray}{lCl}
\frac{dP_{123}^{\text{I}}}{d^3r_1 d^3r_2 d^3r_3} &= & \nonumber 
\frac{2}{4\pi V}
\int_{\xi \in \mathbb{R}^+}  d\xi \frac{e^{-\mu_0 (\Delta\ell_1 + \Delta\ell_2)}\mu_0^e}{|\vec{r}_2 - \vec{r}_1|^2 |\vec{r}_3 - \vec{r}_2|^2 }\Theta_v(\xi)
\\
&&n_e \frac{d\sigma_0^C}{d\Omega_2} 
e^{-\mu_2 \Delta \ell_3} \mu_2^{e}.
\label{eq:sm1}
\end{IEEEeqnarray}

An element of the \ac{ics} system matrix is defined as the probability of an annihilation originated in voxel $v$ to be detected as an ICS event in the volume associated to the measurement element $\Delta \eta_i = \{\Delta\vec{r}_1, \Delta\vec{r}_2, \Delta \vec{r}_3 \}$. In absence of further knowledge over the true Compton scattering site, the system matrix must be recognized as the sum of two terms, each one related to a possible detection sequence: $dP_{123}^{I}$, and  $dP_{132}^{I}$, where the latter describes the differential probability of an \ac{ics} events in which the Compton scattering occurs in detector $D_3$ and the scattered photon is absorbed in $D_2$. This second probability density is expressed as

\begin{IEEEeqnarray}{lCl}
\frac{dP_{132}^{\text{I}}}{d^3r_1 d^3r_2 d^3r_3} &= & \nonumber 
\frac{2}{4\pi V}
\int_{\xi \in \mathbb{R}^+}  d\xi \frac{e^{-\mu_0 (\Delta\ell_1' + \Delta\ell_3')}\mu_0^e}{|\vec{r}_3 - \vec{r}_1|^2 |\vec{r}_3 - \vec{r}_2|^2 }\Theta_v(\xi)
\\
&&n_e \frac{d\sigma_0^C}{d\Omega_3} 
e^{-\mu_3 \Delta \ell_3} \mu_3^{e}.
\label{eq:sm2}
\end{IEEEeqnarray}

Finally, an element of the system matrix can be formulated as 
\begin{IEEEeqnarray}{lCl}
h^{\text{I}}_{iv} &=&  \nonumber
w_1\int_{\vec{r}_1, \vec{r}_2, \vec{r}_3  \in \Delta \eta_i} d^3r_1 d^3r_2 d^3r_3 
\frac{dP_{123}^{\text{I}}}{d^3r_1 d^3r_2 d^3r_3} \\
&+&  
w_2\int_{\vec{r}_1, \vec{r}_2, \vec{r}_3 \in \Delta \eta_i} d^3r_1 d^3r_2 d^3r_3 
\frac{dP_{132}^{\text{I}}}{d^3r_1 d^3r_2 d^3r_3} \nonumber \\*
\label{eq:sm3}
\end{IEEEeqnarray}
%
%\underset{\substack{ \vec{r}_0 - \ell_1\hat{\Omega}_0 \in D_1}}{\int}
%
which leads to V-shaped LORs with different weights arising from the physical factors in (\ref{eq:sm1}) and (\ref{eq:sm2}). The weights $w_1$ and $w_2$ allow conveying any additional knowledge over the Compton scattering site; if no such additional knowledge exists, then $w_1=w_2=1/2$. The integration over the interaction positions is to be performed over the volume associated to the measurement element $\Delta \eta_i = \{\Delta\vec{r}_1, \Delta\vec{r}_2, \Delta \vec{r}_3 \}$ which, in the absence of further discretization, given e.g. by depth-of-interaction information, might be equal to the whole extent of the involved PET detectors, i.e. $\Delta \eta_i = \{V_{D_1,} V_{D_2}, V_{D_3} \}$. In this case, and under the assumption that (\ref{eq:sm1}) and  (\ref{eq:sm2}) are slowly varying functions inside the integration limits, the system matrix element can be approximated as:
\begin{IEEEeqnarray}{lCl}
h_{iv}^{\text{I}} &\approx&  \nonumber
V_{D_1} V_{D_2} V_{D_3}
\left(
w_1
\frac{dP_{123}^{\text{I}}}{d^3r_1 d^3r_2 d^3r_3} +
w_2 
\frac{dP_{132}^{\text{I}}}{d^3r_1 d^3r_2 d^3r_3} 
\right). \nonumber \\*
\label{eq:sm4}
\end{IEEEeqnarray}

Remarkably, the expressions derived here are not constrained to any particular PET geometry.

It must be noted that the estimation of the \ac{ics} sensitivity image expressions (\ref{eq:sens1}) and (\ref{eq:sens2}) involves the computation of angles, length or interaction positions that allow determining the Compton-scattered photon energy, $E_{2}$, by using the corresponding Compton equation \cite{COMPTON1923}. This, in turn, allows implementing strategies to account for possible energy cuts applied over the \ac{ics} events. For instance, if (\ref{eq:sens2}) is estimated via Monte Carlo integration, each sampled point $\{\vec{r}_1, \vec{r}_2, \vec{r}_3, \xi\}_n$ yields an energy $E_{2,n}$, which determines the deposited energy in $D_2$ (equal to $511~\textrm{keV}$ - $E_{2,n}$) and in $D_3$ (equal to $E_{2,n}$); if any of these deposited energies lies outside of the applied energy window used to acquire the \ac{ics} data, the contribution of the sampled point to the estimated integral is zero.

\subsection{Image reconstruction}\label{sec:mm:ir}

The \ac{lmmlem} algorithm was implemented as in (\ref{eq:plainLMMLEM}) to obtain images with only \ac{ics} events (using the analytical sensitivity image (\ref{eq:sens2}) and system matrix (\ref{eq:sm4}) with equal weights $w_1 = w_2 = 1/2$), and only golden events. For the latter, a dedicated model was also developed (see analytical expressions for the system matrix $\mathbf{H}^{\text{G}}$ and sensitivity image $\mathbf{s}^{\text{G}}$ in Appendix \ref{sec:golden}). Furthermore, and similar to \cite{GILLAM2014,GRKOVSKI2015,ROSER2022}, the \ac{lmmlem} was extended to allow for the joint reconstruction of \ac{ics} and golden events. The corresponding update equation is:

\begin{IEEEeqnarray}{lCl}
\label{eq:joint}
\lambda_v^{(n+1)} = 
\frac{\lambda_v^{(n)}}{s_v^{\text{G}} + s_v^{\text{I}}} 
\left(
\sum_{i \in \text{G}}
\frac{h_{iv}^{\text{G}}}{\sum_w h_{iw}^{\text{G}} \lambda_{w}^{(n)}}  
+
\sum_{i \in \text{I}}
\frac{h_{iv}^\text{I}}{\sum_w h_{iw}^{\text{I}} \lambda_{w}^{(n)}}  
\right), \nonumber \\*
\end{IEEEeqnarray}
where $\lambda_v^{(n)}$ refers to the image value at the image element $v$ and iteration $n$. Images were discretized using a grid of $150\times150\times180$ voxels, except when specified otherwise. In all cases, cubic voxels with side length $0.25~\textrm{mm}$ were employed. Images obtained at different iteration numbers were analyzed, and no post-reconstruction smoothing was applied. No corrections (attenuation, object-scattering, normalization) were included into the reconstruction algorithm.

\subsection{Experimental setup}\label{sec:materials:exp}

The derived analytical models were tested on MERMAID data. MERMAID stands for \textit{Multi-Emission Radioisotopes - Marine Animal Imaging Device}, a small aquatic animal PET scanner prototype being developed at the Universit\"at zu L\"ubeck \cite{SEEGER2022}. The current prototype consists of four modules mounted on a rotating gantry (diameter $66$~mm), with a linear stage allowing for bed (axial) motion. The axial coverage of a single axial step amounts to $6$~mm. Each module contains $16 \times 8$ pixelated LYSO crystals ($1.12 \times 1.12 \times 15~\textrm{mm}^3$ each) with a crystal pitch of $1.20$~mm, and  3M Enhanced Specular Reflector Film (ESR) as reflector material. The crystals are coupled one-to-one to Silicon Photomultipliers (Hamamatsu S13615-1050N-08), with a photosensitive area of $1.0 \times 1.0~\textrm{mm}^2$,  $369$ pixels (pixel pitch of $50~\mu$m), and readout by the PETsys Electronics TOFPET 2C ASIC. In this study, the SiPM breakdown voltage and overvoltage was set to $53.0 + 3.0$~V and the thresholds $Vth\_t1, Vth\_t2$ were set to $20$; the energy threshold $Vth\_e$ was individually adjusted for each SiPM channel. The average energy resolution at $511~\textrm{keV}$ is $21.6~\%$. The individual  readout allows retrieving the list of fired SiPM channels ordered by time. Coincidences were subsequently obtained offline by applying a temporal coincidence window of $2~\textrm{ns}$, given a coincidence timing resolution of $920~\textrm{ps}$ (FWHM), and classified according to the number of fired SiPM channels and the energy deposited. Coincidences involving two SiPM channels depositing energy within $[450, 550]$~\text{keV} were classified as golden, whereas coincidences involving three SiPM channels, with one energy deposition within $[450, 550]~\textrm{keV}$ and the other two within $[180, 380]~\textrm{keV}$ and $[180, 550]~\textrm{keV}$, were classified as \ac{ics} events. The lower threshold of $180~\textrm{keV}$ was chosen as the lowest energy value that all the SiPM channels of the current prototype can reliably measure without being compromised by electronic noise and was taken into account in the corresponding \ac{ics} sensitivity image model. The upper threshold of $380~\textrm{keV}$ encompasses the maximum energy that a $511~\textrm{keV}$ photon can deposit in a Compton scattering (approximately $341~\textrm{keV}$), with a margin to account for the energy resolution of the scanner. The remaining coincidences were discarded. Compensation for radioactive decay was considered in the acquisition time per rotation and axial step. 

The performance of the proposed models was assessed by means of an \ac{iq} phantom adapted to the very-small-animal scenario. To this end, the design of the NU4-2008 (NU4) \ac{iq} phantom proposed by the National Electrical Manufacturers Association (NEMA) \cite{NEMA2008} was downscaled to the $50~\%$ of its size; i.e. including a rods region with rod diameters of $2.50, 2.00, 1.50, 1.00~\textrm{and}~0.50~\textrm{mm}$ and equal height of $10~\textrm{mm}$, a uniform region of diameter $15.0~\textrm{mm}$ and height of $15.0~\textrm{mm}$, and an insert region with the same dimensions as the uniform region and with two inner, air-filled rods with diameter $4.0~\textrm{mm}$. This downscaled \ac{iq} phantom was 3D-printed with ${}^{18}$F-FDG-labeled resin \cite{ELMOUJARKACH2024} with a total activity of $0.27~\textrm{MBq}$ at the time of measurement. This very low amount aims to mirror the challenging conditions that may arise in small-fish PET imaging. The full scan involved three rotation steps ($60^\circ$ each) at ten different bed/axial positions separated $3~\textrm{mm}$ each; the total measurement time was approximately $50$~minutes. The selected phantom activity and measurement parameters resulted in a low-count PET scenario, where image noise clearly strongly degraded the images (when compared to higher-activity phantom measurements conducted in preliminary experiments).

The reconstructed images of the rod and uniform regions of the \ac{iq} phantom were used to obtain the \acp{rc} and the uniformity percentage standard deviation ($\%~\textrm{STD}$), respectively. For this purpose, the protocols described in \cite{NEMA2008} were adapted to the downscaled size of the phantom. Note that the metrics were aimed to compare the proposed approach with the conventional reconstruction of golden events, and not for comparison with other PET systems. 
To compute the uniformity metric, a cylindrical \ac{roi} centered in the uniform region of the phantom was defined with diameter $11.25~\textrm{mm}$ and height $5.00~\textrm{mm}$. For the estimation of the \acp{rc}, the NEMA protocol was further modified in order to compensate for the degrading effects of the low counts. Specifically, cylindrical \acp{roi} were defined for each rod with shared height of $5~\textrm{mm}$ and $75~\%$ of their physical radii, their means being used for the RC computation. This procedure departs from the NEMA protocol, where the axial profiles at the maximum value of the averaged transversal \ac{roi} slices are used. As pointed out elsewhere \cite{HALLEN2020}, such a protocol might be sub-optimal as it overestimates systematically the \acp{rc}, especially in high-variance cases such as the ones expected in a low-count PET scenario. By contrast, the procedure employed in this work, also suggested in \cite{HALLEN2020}, was found to alleviate the overestimation of the RC values.
Additionally, the inserts region of the \ac{iq} phantom was used to compute the \ac{sor}, again adapting the NEMA protocol to the downscaled size of the phantom. These results with the \ac{sor} metric, which is mainly indicative of the accuracy of attenuation of scatter corrections (not implemented in this work), can be found in Appendix \ref{sec:sor}.

\subsection{Monte-Carlo simulations}

Simulations were conducted in \textsc{GATE} version 9.2 \cite{SARRUT2021,SARRUT2022} reproducing MERMAID's geometry. Several back-to-back $511~\textrm{keV}$ photon sources were simulated, with different purposes.

\subsubsection*{One-voxel sources} Seven one-voxel sources (i.e. cubic-shaped, with $0.25~\textrm{mm}$ length size) were independently simulated at different positions of the $X$-axis (centered at $\pm 15.125, \pm 10.125, \pm 5.125~\textrm{and}~0.125~\textrm{mm}$), with source activity of $1~\textrm{MBq}$. Five realizations were acquired at each position, each realization comprising three rotation steps ($1~\textrm{s}$ and $60^{\circ}$ each) in a single axial step. These simulations were used to validate the analytical sensitivity image expression (\ref{eq:sens2}), by comparing the predictions of the model to the number of ICS events detected in the simulations at the voxels where the source was placed. 

\subsubsection*{Cylinder source} A cylinder with diameter $8~\textrm{mm}$ and length $4~\textrm{mm}$ was simulated at the center of the \ac{fov}, with uniform activity of $1~\textrm{MBq}$. The simulation comprised three rotation steps ($10~\textrm{s}$ and $60^{\circ}$ each) in a single axial step. As such, the axial extent of the images was reduced in this case to $50$ voxels. The acquired ICS and golden events, obtained using the energy windows applied experimentally, were reconstructed in order to assess the quantitative accuracy of the obtained images, thus indirectly supporting the validation of the models.  

\subsubsection*{\ac{iq} phantom} The aforementioned \ac{iq} phantom was simulated, reproducing its downscaled size and activity. GATE's material database \textit{Plastic} was chosen as phantom material. The purpose was not to provide validated simulations, i.e., simulations which are in fully agreement with experimental results, but to assess the best attainable improvement in the images under close-to-ideal conditions. 

Table \ref{tab:materials} summarizes the properties of the materials employed. In all simulation scenarios, the physics list emstandard\_opt3 was used, together with production cuts of $1~\textrm{mm}$. The golden and ICS datasets were extracted using the list of detected GATE \textit{Hits} and an external C++ code. In particular, events with both annihilation photons undergoing photoelectric effect were classified as golden events; events with one annihilation photon undergoing Compton scattering and with the subsequent scattered photon undergoing photoelectric effect in a different crystal were classified as ICS events. The index of the involved crystals was registered and used to identify the volume associated to the measurement elements (i.e. the integration limits $\Delta\eta_i$ in the system matrix model), thus reproducing MERMAID's detector spatial resolution, limited to the pixelated crystal size. However, and given that the purpose of the work was not to provide validated simulations, no further degradation effects (e.g. energy resolution or optical crosstalk) were modeled\footnote{Time-related degradation effects (pile-up, temporal resolution, randoms) were equally not considered, although they are not expected to be relevant in this work given the low source activities involved.}.

\setlength{\tabcolsep}{0.03mm}
\begin{table}[h]
\centering
\begin{threeparttable}
\caption{Material properties used in the GATE $9.2$ simulations.}\label{tab:materials}
\begin{tabular}{cccc} 
\toprule%
Material  & Chemical formula & Density (g cm$^{-3}$)  & State \\
\midrule
LYSO  &  $\textrm{Lu}_{2( 1 - 0.2)}\textrm{Y}_{0.4}\textrm{SiO}_5$   & $7.36$ &  Solid\\
Plastic  &  $\textrm{C}_{5}\textrm{H}_{8}\textrm{O}_2$   & $1.18$ &  Solid\\
Air  &  \tnote{a}$\;\;\textrm{N}_{(0.75)}\textrm{O}_{(0.23)}\textrm{Ar}_{(0.01)}\textrm{C}_{(0.0001)}$   & $0.00129$ &  Gas\\
\bottomrule
\end{tabular}
\begin{tablenotes}
    \item [a] Parentheses show the mass fractions
\end{tablenotes}
\end{threeparttable}
\end{table}

\subsection{Numerical estimation of the analytical expressions}

The analytical expression for the \ac{ics} sensitivity image (\ref{eq:sens2}) was estimated via numerical \ac{mc} integration. Sample points not fulfilling the event selection criteria described above (e.g. events outside the employed energy windows) did not contribute to the estimated integrals. All reconstructions were performed on a $2$x AMD EPYC~$7552$~@~$2.20$~GHz ($2\times 48$~cores) CPU. 

The attenuation coefficients in the models were obtained by using linear interpolation over the XCOM photon cross section database of the \ac{nist} \cite{BERGER2010}, assuming the same LYSO composition and density (see Table \ref{tab:materials}) as defined in the materials database of \textsc{GATE} version 9.2. 

\section{Results}\label{sec:results}

\subsection{ICS sensitivity image and system matrix}

\begin{figure}[!t]
\centering
\subfloat[]{\includegraphics[height=1.6in]{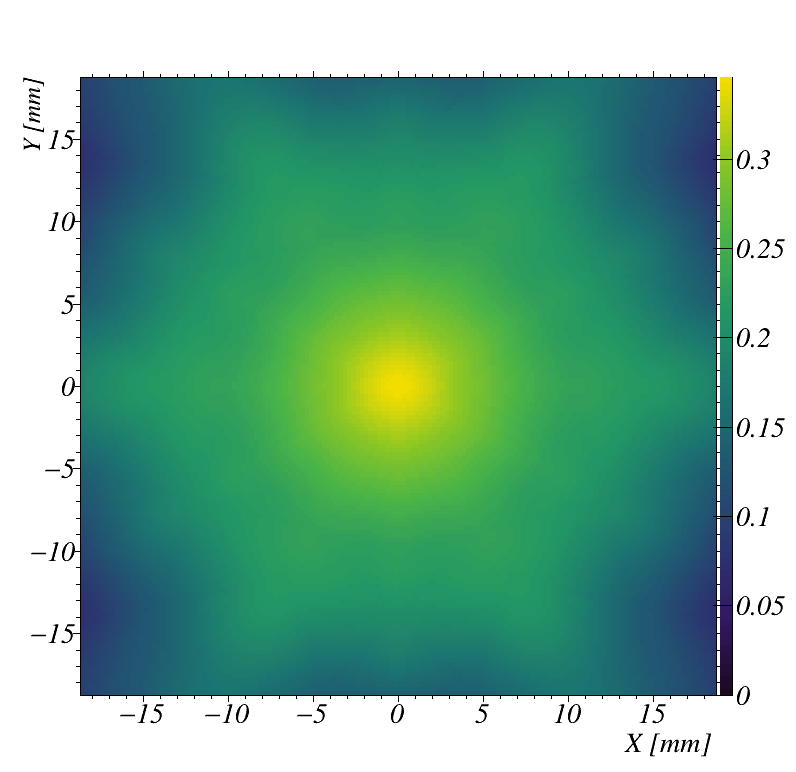}
\label{fig:sens:1}}
\hfil
\subfloat[]{\includegraphics[height=1.6in]{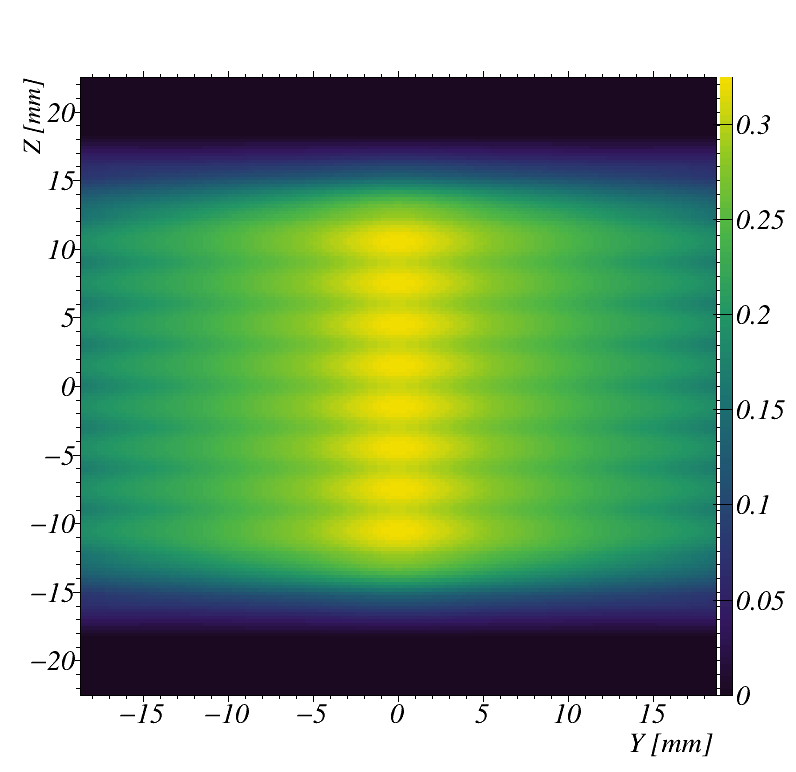}
\label{fig:sens:2}}
\caption{Transversal ($XY$) and sagittal ($YZ$) projections over the analytical \ac{ics} sensitivity image, computed for the MERMAID geometry with $3$ rotation steps and $10$ bed positions, using (\ref{eq:sens2}) and assuming an energy threshold of $180~\textrm{keV}$ over the \ac{ics} events. The scale represent the probability of an annihilation emitted in a given group of voxels (e.g. a $Z$-row of voxels for the transversal projection) to be detected as an \ac{ics} event in MERMAID. }
\label{fig:sens}
\end{figure}
\begin{figure}[!t]
\centering
\subfloat[]{\includegraphics[height=1.6in]{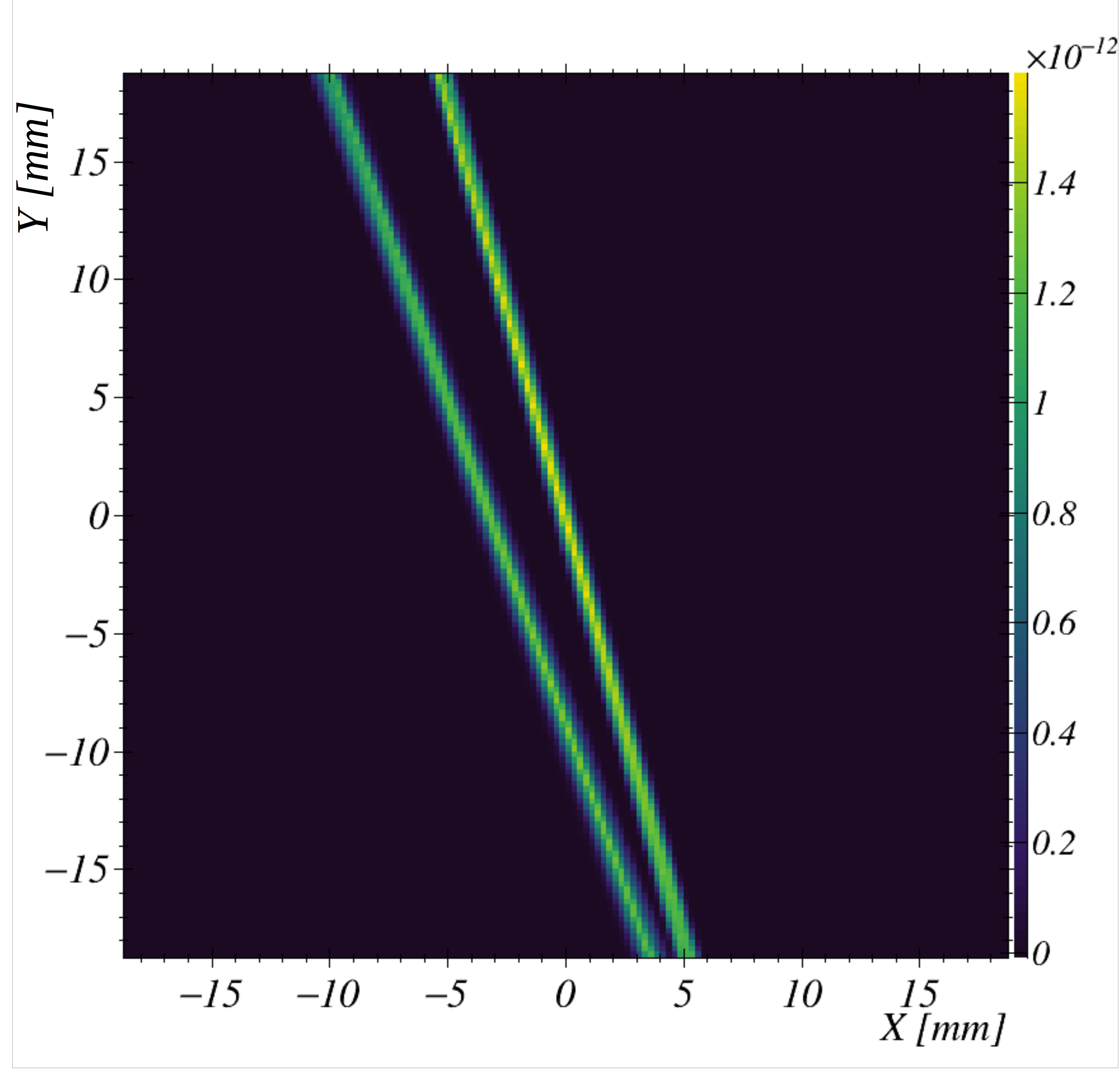}
\label{fig:sm:1}}
\hfil
\subfloat[]{\includegraphics[height=1.6in]{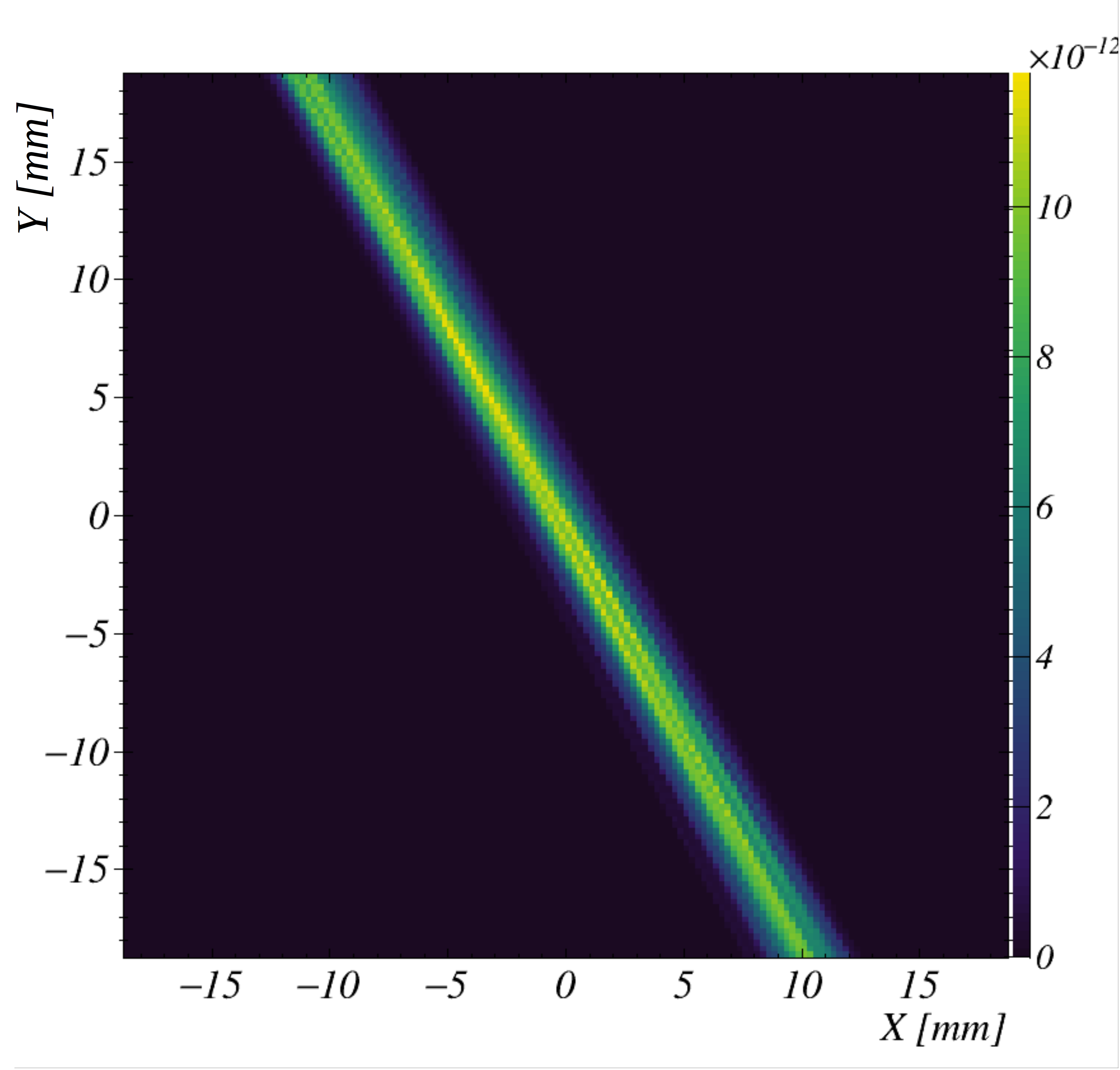}
\label{fig:sm:2}}
\caption{Transversal ($XY$) projections over two different analytical \ac{ics} system matrix rows, computed for the MERMAID geometry using (\ref{eq:sm4}). The scale represents the probability of annihilation emitted in a $Z$-row of voxels to be detected in terms of the particular measurement element considered.}
\label{fig:sm}
\end{figure}

As a representative example of the outcome of (\ref{eq:sens2}), figures \ref{fig:sens:1} and \ref{fig:sens:2} show the transversal (XY) and sagittal (YZ) projections over the analytical \ac{ics} sensitivity image, for three rotation steps ($60^{\circ}$), ten bed/axial positions with axial shift of $3~\textrm{mm}$ (i.e., with overlapping), and energy threshold of $180~\textrm{keV}$ over the ICS events. In particular, this \ac{ics} sensitivity image was subsequently used to obtain images of the downscaled \ac{iq} phantom, both with experimental and \ac{mc}-simulated data.

Likewise, figure \ref{fig:sm} shows two examples of the estimation of \ac{ics} system matrix rows via (\ref{eq:sm4}); particularly, figure \ref{fig:sm:1} illustrates the V-shape with which  \ac{ics} events are modeled with the proposed approach. Such V-shape is not always conspicuous, especially when the interaction positions $\vec{r}_2$ and $\vec{r}_3$ occurs in closer-neighboring crystals, which is the case illustrated by figure \ref{fig:sm:2}.

\subsection{Validation of the analytical model}

Table \ref{tab:validationICSneigh1} lists the results of the validation of the analytical sensitivity image (\ref{eq:sens2}) with and without a threshold in $180~\textrm{keV}$ as well as including \ac{ics} events occurring in neighboring crystals, whereas table \ref{tab:validationICSneigh2} lists the corresponding results excluding such neighboring-crystal \ac{ics} events. In all cases, the model predictions at seven different positions of the central $X$-axis are compared with the number of detected \ac{ics} events per back-to-back emission in the simulation. The five realizations simulated at each $X$-position allowed for an estimation of the mean and its standard error; as for the model, the Monte Carlo integration error was estimated.

\setlength{\tabcolsep}{0.60mm}
\begin{table}[h]
\centering
\caption{\ac{ics} sensitivity image values at seven different positions over the central X-axis, for the simulation and the analytical model, including \ac{ics} events occurring in neighboring crystals, with (without) energy threshold at $180~\textrm{keV}$}\label{tab:validationICSneigh1}
\begin{tabular}{ccc} 
\toprule%
X-position [mm]  & Simulation $(\times 10^{-3})$ & $\textrm{Model} \pm  0.002~(\times 10^{-3})$  \\
\midrule
$-15.125$  & \makecell{$1.109 \pm 0.014$ \\ ($1.78 \pm 0.02$)}    & \makecell{$1.143$\\ ($1.827$)}  \\
$-10.125$  & \makecell{$1.249 \pm 0.018$ \\ ($2.006 \pm 0.018$)}  & \makecell{$1.324$\\ ($2.100$)} \\
$-5.125$   & \makecell{$1.652 \pm 0.014$ \\ ($2.645 \pm 0.02$)}   & \makecell{$1.731$\\ ($2.725$)} \\
$0.125$    & \makecell{$2.115 \pm 0.011$ \\ ($3.38 \pm 0.03$)}    & \makecell{$2.219$\\ ($3.502$)} \\
$5.125$    & \makecell{$1.653 \pm 0.013$ \\ ($2.64 \pm 0.02$)}    & \makecell{$1.733$\\ ($2.730$)} \\
$10.125$   & \makecell{$1.249 \pm 0.005$ \\ ($2.006 \pm 0.009$)}  & \makecell{$1.337$\\ ($2.108$)} \\
$15.125$   & \makecell{$1.077 \pm 0.014$ \\ ($1.727 \pm 0.012$)}  & \makecell{$1.146$\\ ($1.837$)} \\
\bottomrule
\end{tabular}
\end{table}
\setlength{\tabcolsep}{0.60mm}
\begin{table}[h]
\centering
\caption{\ac{ics} sensitivity image values at seven different positions over the central X-axis, for the simulation and the analytical model, excluding \ac{ics} events occurring in neighboring crystals, with (without) energy threshold at $180~\textrm{keV}$}\label{tab:validationICSneigh2}
\begin{tabular}{ccc} 
\toprule%
X-position [mm]  & Simulation $(\times 10^{-3})$ & $\textrm{Model} \pm  0.002~(\times 10^{-3})$  \\
\midrule
$-15.125$  & \makecell{$0.575 \pm 0.007$ \\ ($0.904 \pm 0.014$)}    & \makecell{$0.576$\\ ($0.920$)}  \\
$-10.125$  & \makecell{$0.639 \pm 0.007$ \\ ($1.011 \pm 0.007$)}  & \makecell{$0.676$\\ ($1.059$)} \\
$-5.125$   & \makecell{$0.865 \pm 0.009$ \\ ($1.349 \pm 0.013$)}   & \makecell{$0.878$\\ ($1.364$)} \\
$0.125$    & \makecell{$1.084 \pm 0.010$ \\ ($1.70 \pm 0.02$)}    & \makecell{$1.116$\\ ($1.730$)} \\
$5.125$    & \makecell{$0.877 \pm 0.008$ \\ ($1.352 \pm 0.017$)}    & \makecell{$0.877$\\ ($1.364$)} \\
$10.125$   & \makecell{$0.642 \pm 0.006$ \\ ($1.010 \pm 0.012$)}  & \makecell{$0.678$\\ ($1.058$)} \\
$15.125$   & \makecell{$0.549 \pm 0.009$ \\ ($0.873 \pm 0.010$)}  & \makecell{$0.575$\\ ($0.924$)} \\
\bottomrule
\end{tabular}
\end{table}
\begin{figure}[!t]
\centering
\subfloat[]{\includegraphics[height=1.64in]{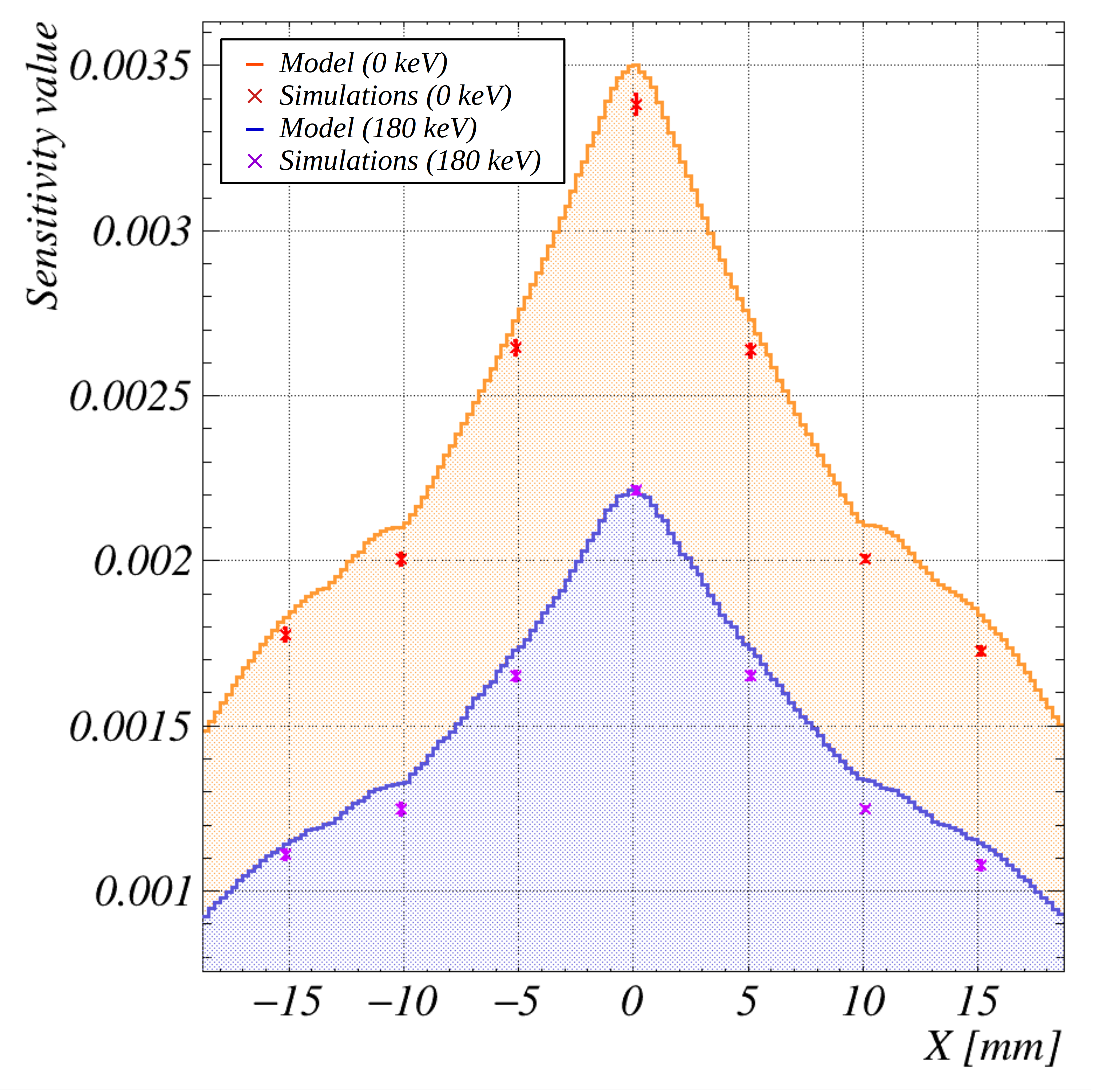}
\label{fig:validationICS:1}}
\hfil
\subfloat[]{\includegraphics[height=1.64in]{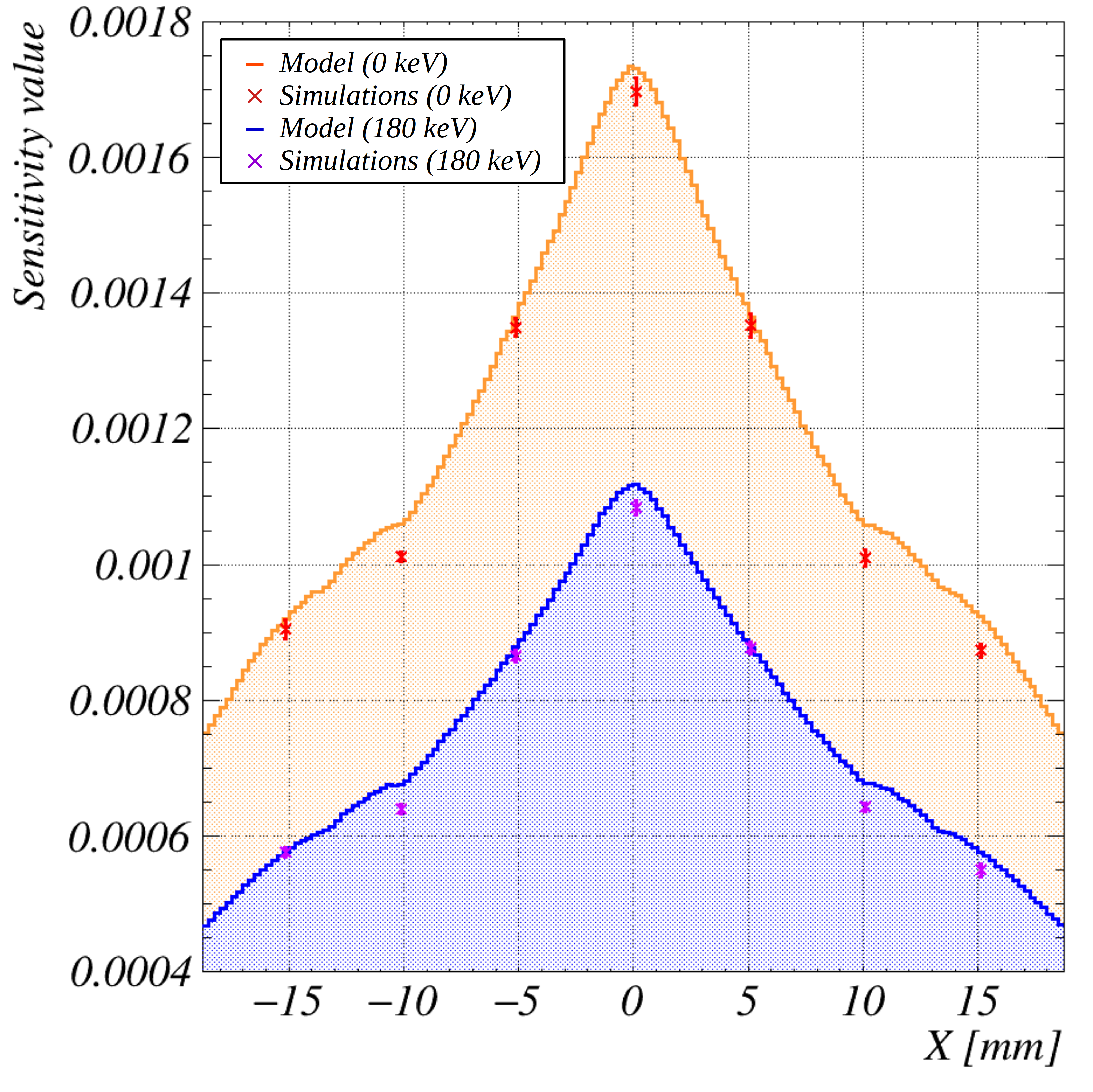}
\label{fig:validationICS:2}}
\caption{Validation of the \ac{ics} analytical sensitivity image model with no energy threshold over the ICS events (orange line) and with a threshold of $180~\textrm{keV}$ (blue line) against Monte Carlo simulations (red and violet crosses, respectively), (a) including \ac{ics} ocurring in neighboring crystals, and (b) discarding them. Each simulated $X$-position represents the mean of five independent simulations; the displayed error bars correspond to the standard error of the mean.}
\label{fig:validationICS}
\end{figure}

As expected, higher sensitivity values are obtained when including neighboring-crystal \ac{ics} events; indeed, when doing so, the sensitivity values almost increase twofold, independently of the energy threshold.

These results, which are also graphically displayed in figure \ref{fig:validationICS}, show a good agreement between the model predictions and the simulated data, although they also suggest that the \ac{ics} model slightly overestimates the probability of ICS detection when compared to the \ac{mc} simulations (see Discussion). When including the neighboring-crystal \ac{ics} events (figure \ref{fig:validationICS:1}), a mean discrepancy of $5.0~\%$ (maximum: $6.6~\%$) is found with the energy cut at $180~\textrm{keV}$; slightly lower discrepancy values are found without energy cuts (mean $4.0~\%$, maximum $5.9~\%$). The discrepancy is even lower when excluding the neighboring-crystal \ac{ics} events (figure \ref{fig:validationICS:2}), with a mean value of $2.8~\%$ (maximum: $5.4~\%$) found with the energy cut at $180~\textrm{keV}$; very similar discrepancy values are found without energy cuts (mean $2.9~\%$, maximum $5.5~\%$).

\begin{figure}[!t]
\centering
\subfloat[]{\includegraphics[height=1.72in]{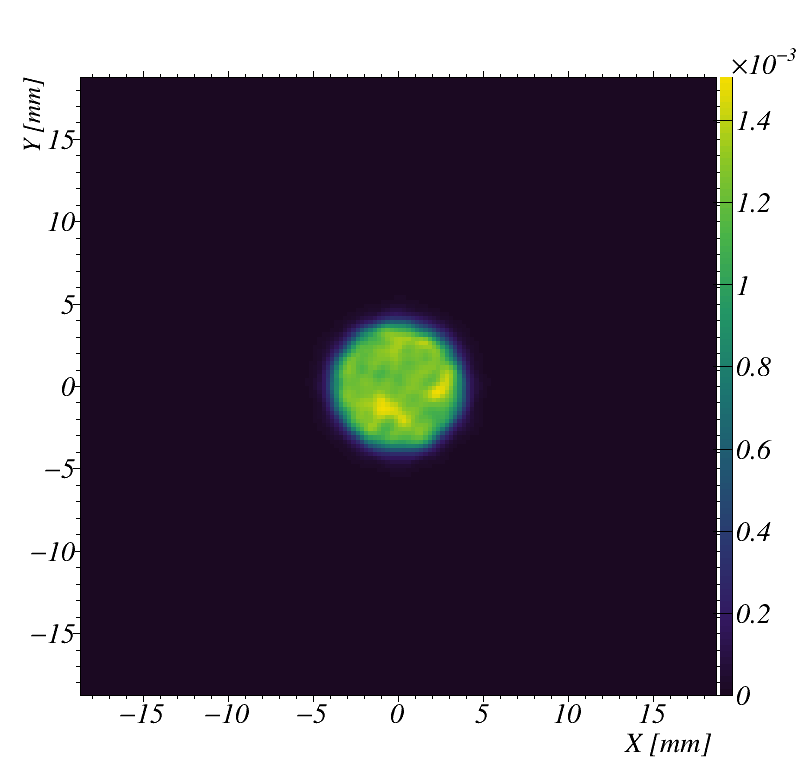}
\label{fig:cylinder:1}}
\hfil
\subfloat[]{\includegraphics[height=1.59in]{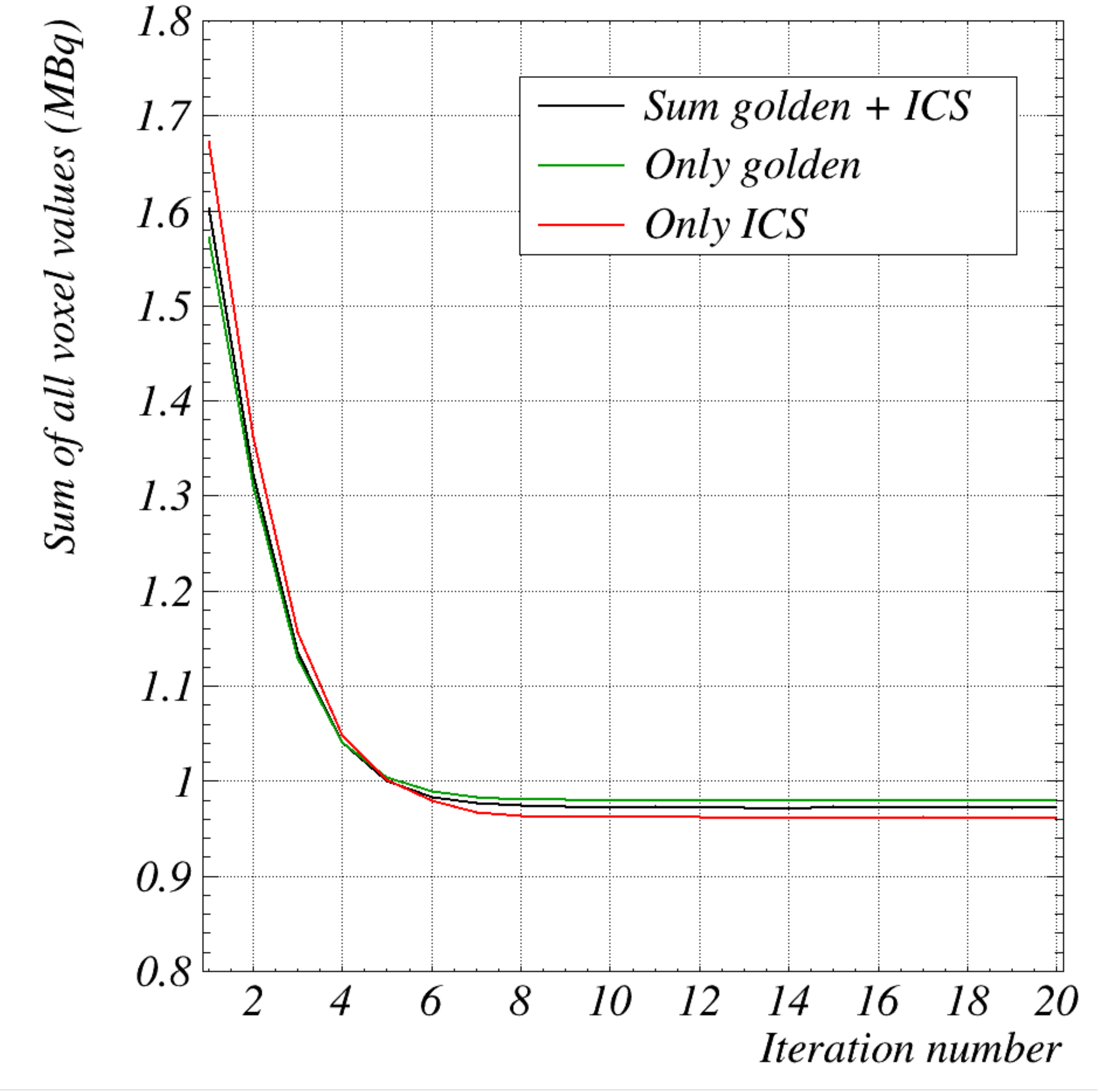}
\label{fig:cylinder:2}}
\caption{Results with the $1~\textrm{MBq}$ \ac{mc}-simulated cylinder source: (a) Transversal ($XY$) projection at iteration 10 for golden and \ac{ics} events, with the color scale representing the activity in MBq recovered in the sum of voxels involved in the projection; (b) evolution of the total activity recovered in the golden and \ac{ics}, only golden and only \ac{ics} images by summing all the voxel values in the three cases.}
\label{fig:cylinder}
\end{figure}

Figure \ref{fig:cylinder} shows some results regarding the $1~\textrm{MBq}$ \ac{mc}-simulated cylinder source, which yielded $29\thinspace205$ golden events and $16\thinspace058$ \ac{ics} events (i.e. $65~\%$ and $35~\%$ of total, respectively). In particular, figure \ref{fig:cylinder:1} shows the $XY$-projection at iteration $10$ when using both \ac{ics} and golden events through the joint algorithm (\ref{eq:joint}). Figure \ref{fig:cylinder:2} shows the evolution of the sum of all voxel values against the \ac{lmmlem} iteration number in the images obtained with the joint algorithm, and also when reconstructing independently with golden events and with \ac{ics} events. As the proposed model is quantitative (i.e., the system matrix and sensitivity elements represent absolute probabilities), the sum of intensity values represents the reconstructed activity of the simulated source (in MBq); the ground truth value was $1~\textrm{MBq}$ in this case. While very close to this reference value, figure \ref{fig:cylinder:2} shows that the algorithm converges to a slightly smaller value for all three sets of data (e.g. $0.96$ MBq at iteration $20$ using only \ac{ics} events). This observation is consistent with the aforementioned small overestimation of the model sensitivity image when compared with \ac{mc}-simulations.

\subsection{\ac{iq} phantom simulation results}\label{sec:results:simiq}

The \ac{iq} phantom simulation yielded a total of $126\thinspace071$ golden events and $68\thinspace338$ \ac{ics} events (i.e. $65~\%$ and $35~\%$ of total, respectively). The sensitivity (\ref{eq:sens2}) was computed using $2^{20}$~rays and took~$6.2$~hours. The subsequent reconstructions spanned approximately $21$ minutes in total using $2^{10}$ rays for the system matrix, which was recomputed on-the-fly at each iteration.

\begin{figure}[!t]
\centering
\subfloat[]{\includegraphics[height=1.08in]{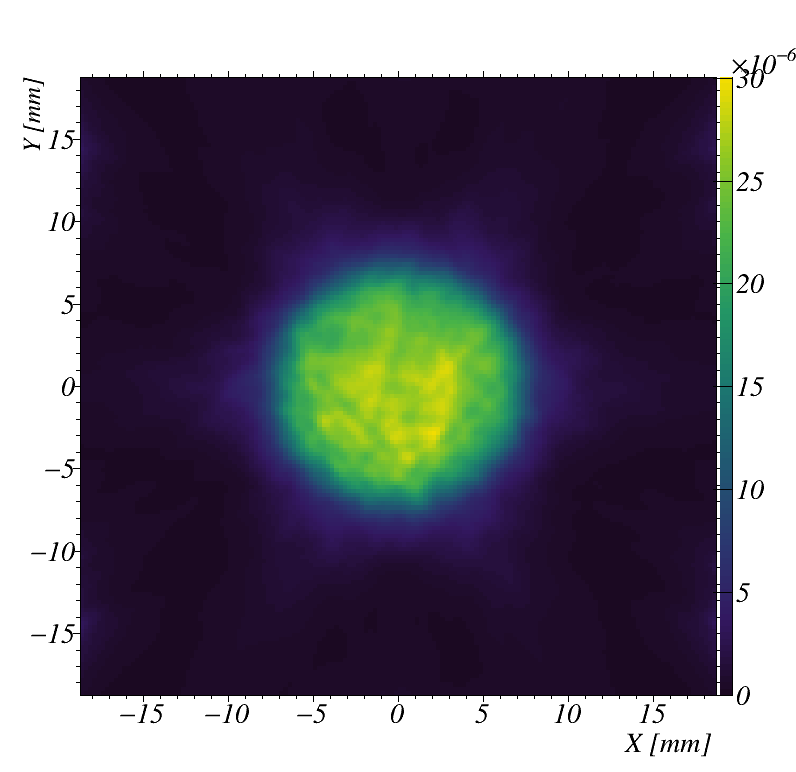}
\label{fig:simunif:1}}
\hfil
\subfloat[]{\includegraphics[height=1.08in]{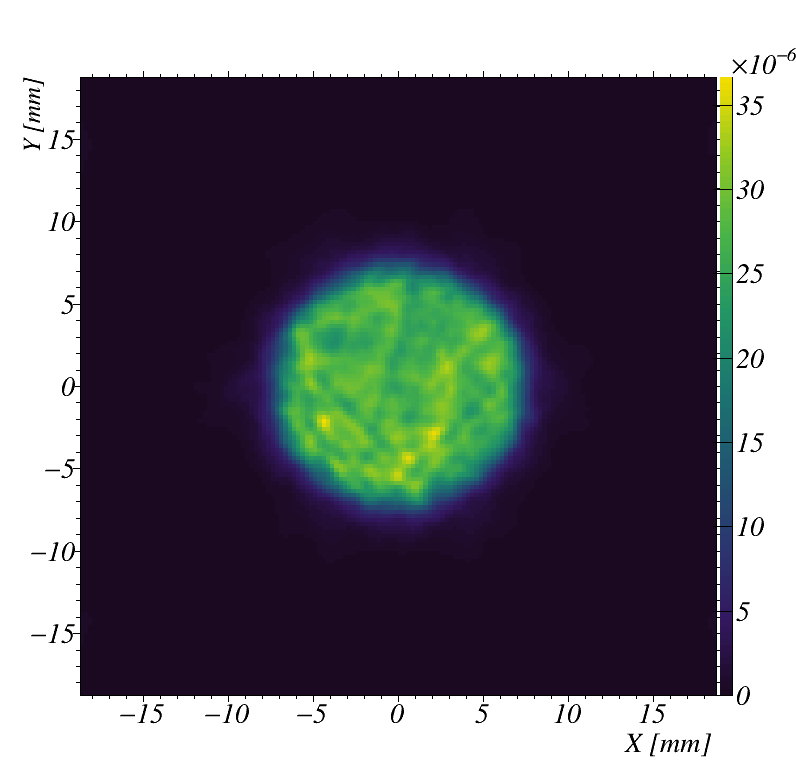}
\label{fig:simunif:2}}
\hfil
\subfloat[]{\includegraphics[height=1.08in]{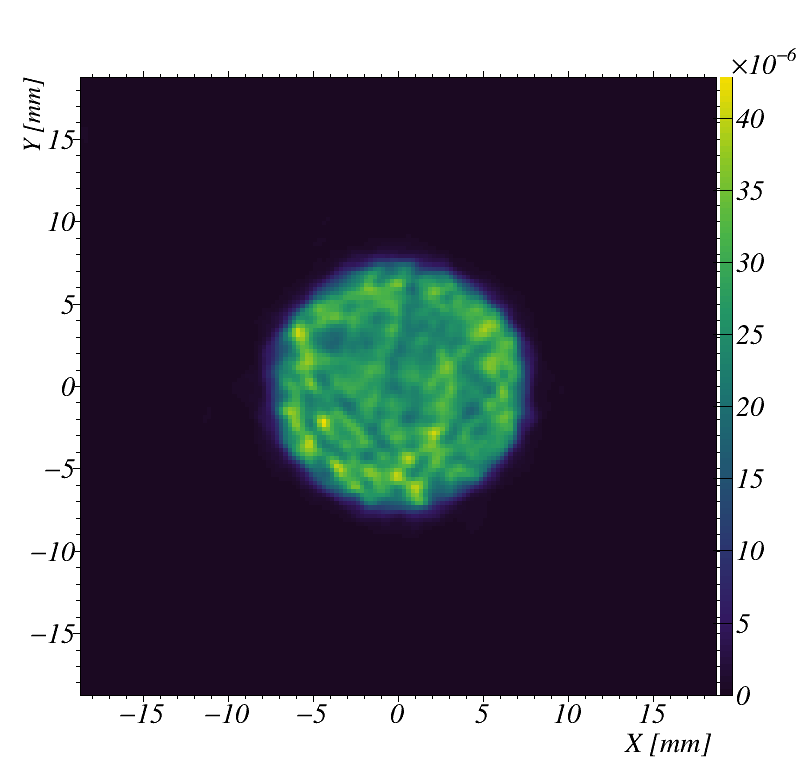}
\label{fig:simunif:3}}
\\ %forces next subfloat to the next line
\subfloat[]{\includegraphics[height=1.08in]{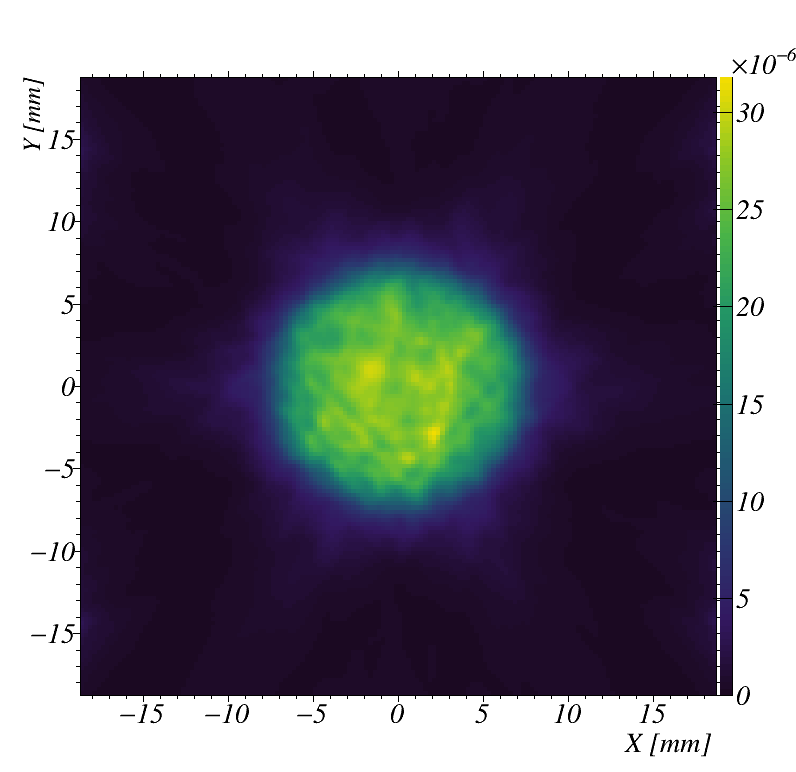}
\label{fig:simunif:4}}
\hfil
\subfloat[]{\includegraphics[height=1.08in]{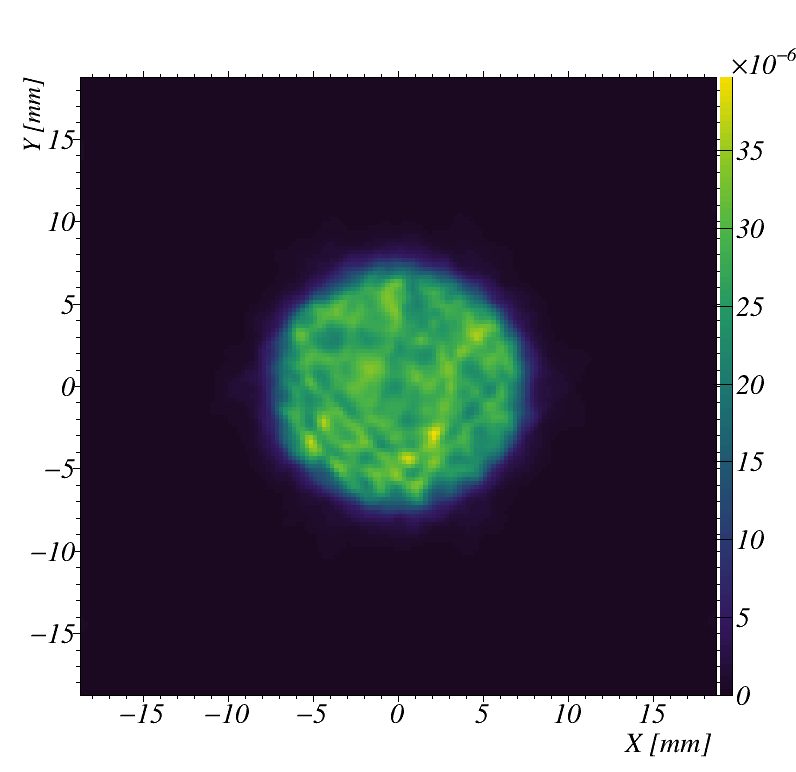}
\label{fig:simunif:5}}
\hfil
\subfloat[]{\includegraphics[height=1.08in]{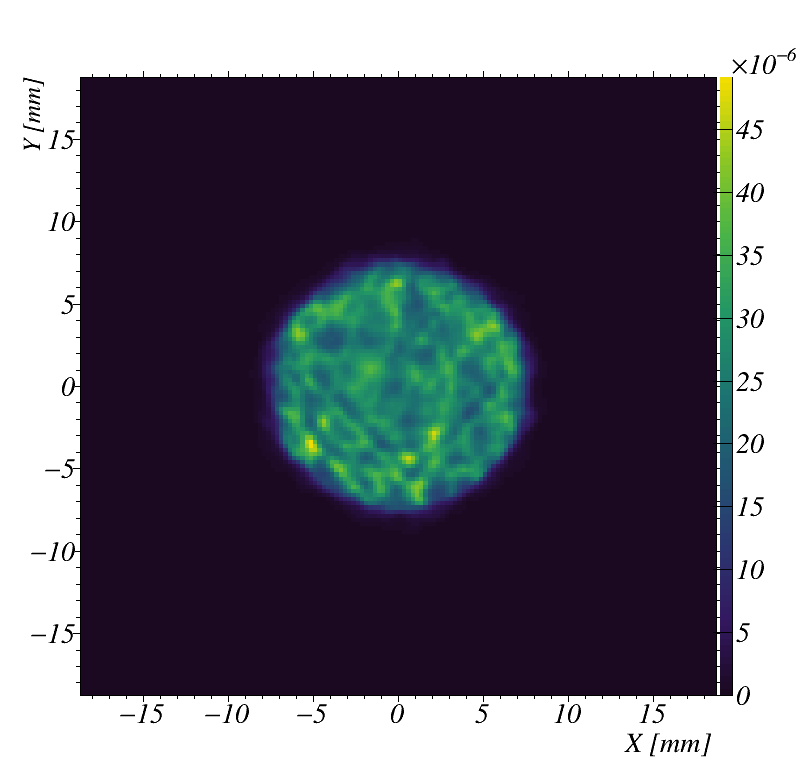}
\label{fig:simunif:6}}
\\ %forces next subfloat to the next line
\subfloat[]{\includegraphics[height=1.08in]{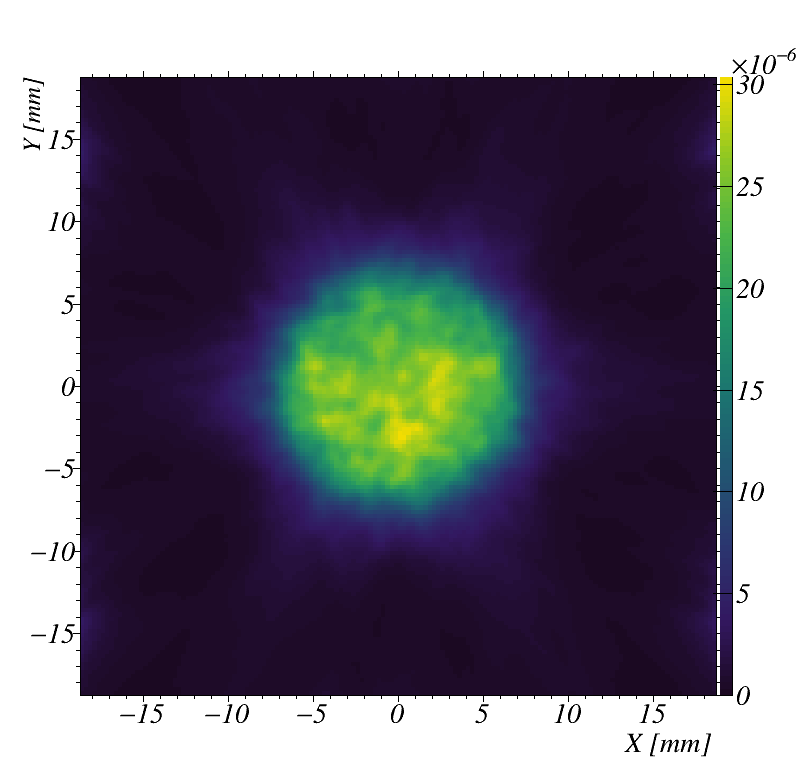}
\label{fig:simunif:7}}
\hfil
\subfloat[]{\includegraphics[height=1.08in]{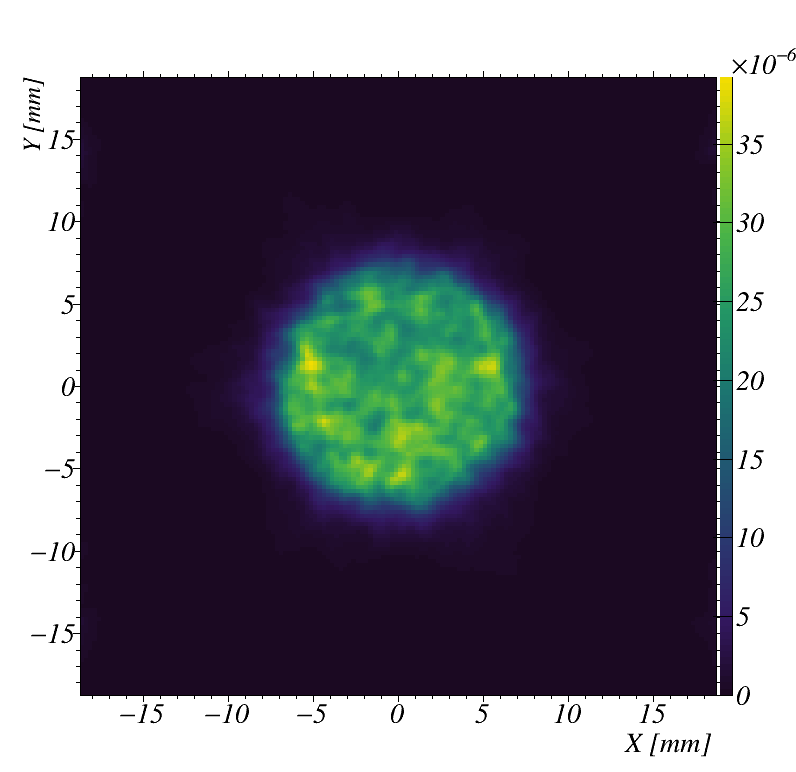}
\label{fig:simunif:8}}
\hfil
\subfloat[]{\includegraphics[height=1.08in]{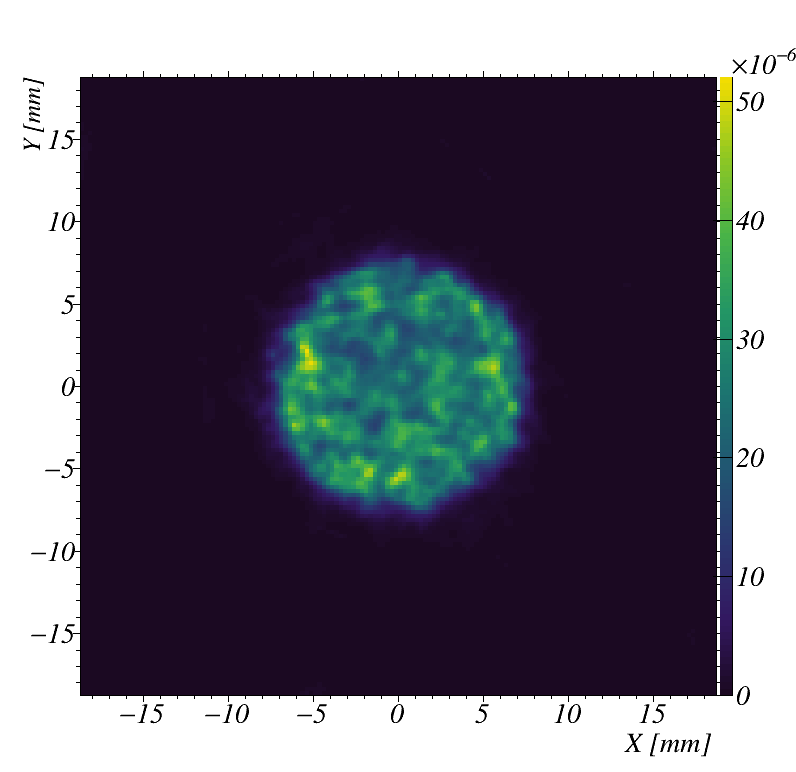}
\label{fig:simunif:9}}
\caption{Transversal ($XY$) projections over the \ac{iq}-phantom uniformity region at iteration $5$ (first column), $10$ (second column) and $20$ (third column) for golden and \ac{ics} (first row), only golden (second row) and only \ac{ics} (third row) events, obtained with simulations. The scale represents the activity in MBq recovered in the sum of voxels involved in each projection (e.g. a Z-row of voxels for the transversal projection).}
\label{fig:simunif}
\end{figure}

\begin{figure}[!t]
\centering
\subfloat[]{\includegraphics[height=1.08in]{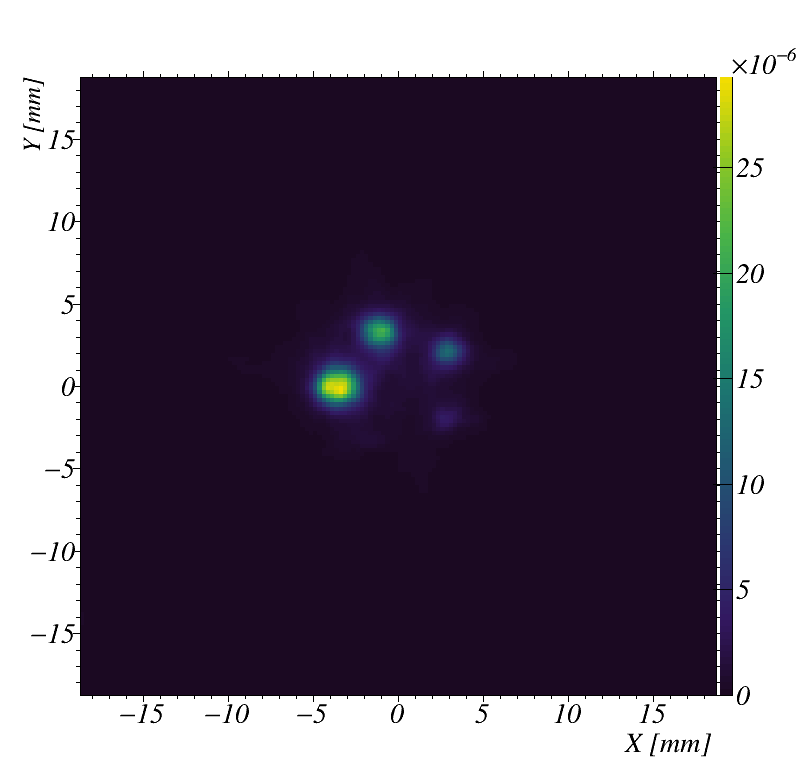}
\label{fig:simrods:1}}
\hfil
\subfloat[]{\includegraphics[height=1.08in]{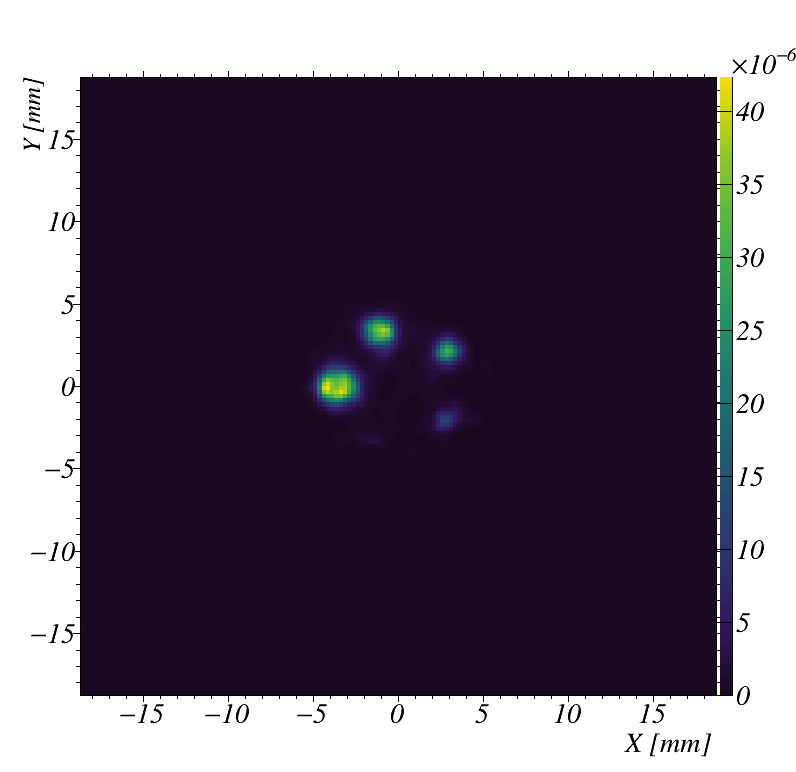}
\label{fig:simrods:2}}
\hfil
\subfloat[]{\includegraphics[height=1.08in]{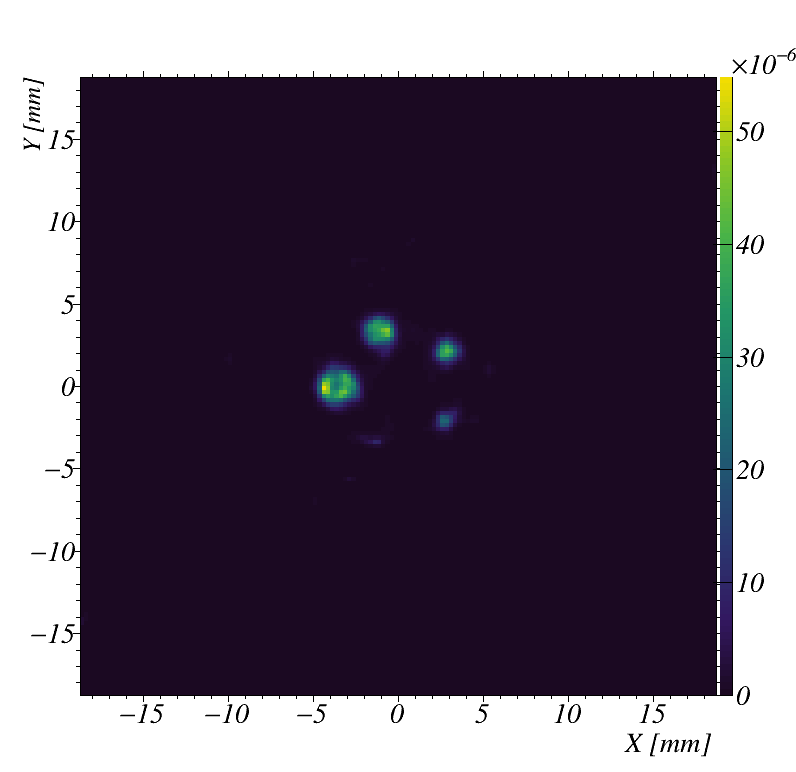}
\label{fig:simrods:3}}
\\ %forces next subfloat to the next line
\subfloat[]{\includegraphics[height=1.08in]{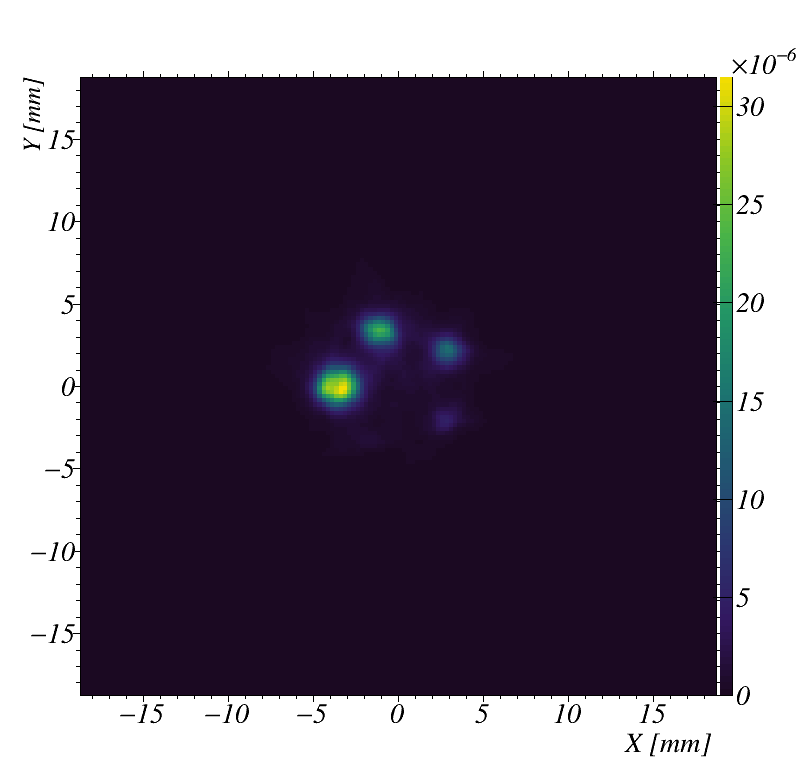}
\label{fig:simrods:4}}
\hfil
\subfloat[]{\includegraphics[height=1.08in]{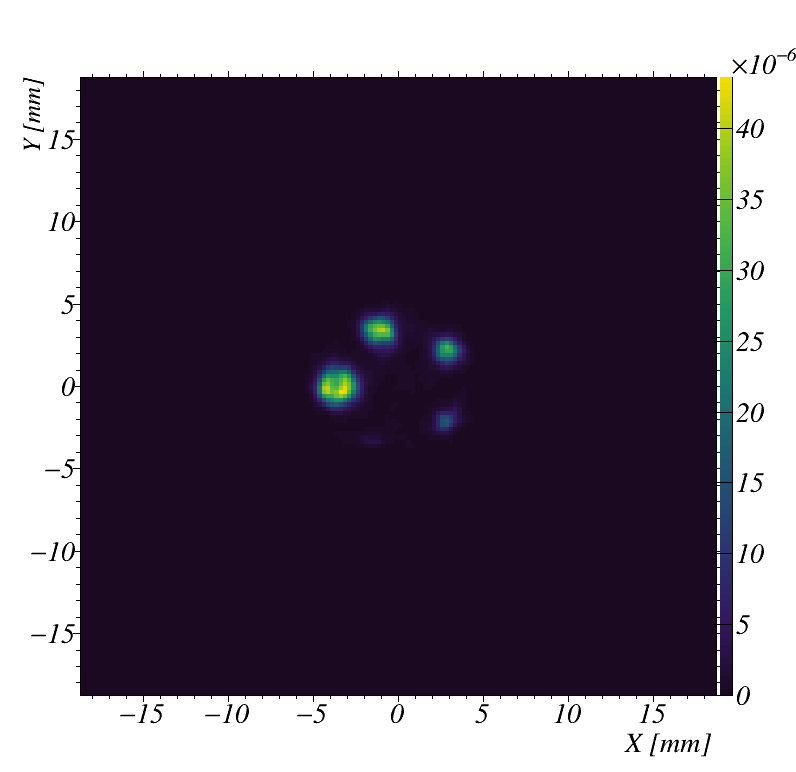}
\label{fig:simrods:5}}
\hfil
\subfloat[]{\includegraphics[height=1.08in]{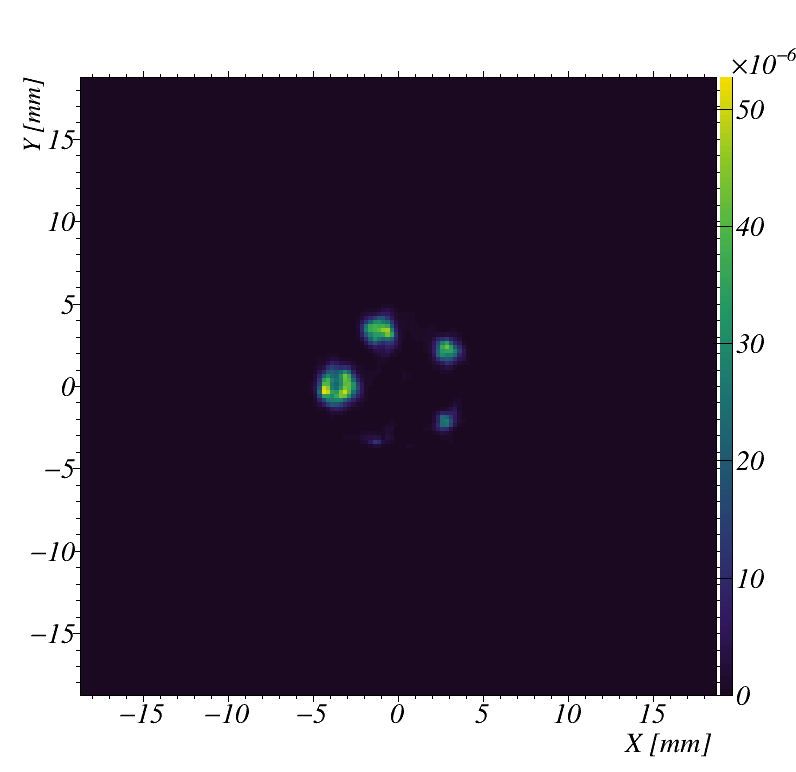}
\label{fig:simrods:6}}
\\ %forces next subfloat to the next line
\subfloat[]{\includegraphics[height=1.08in]{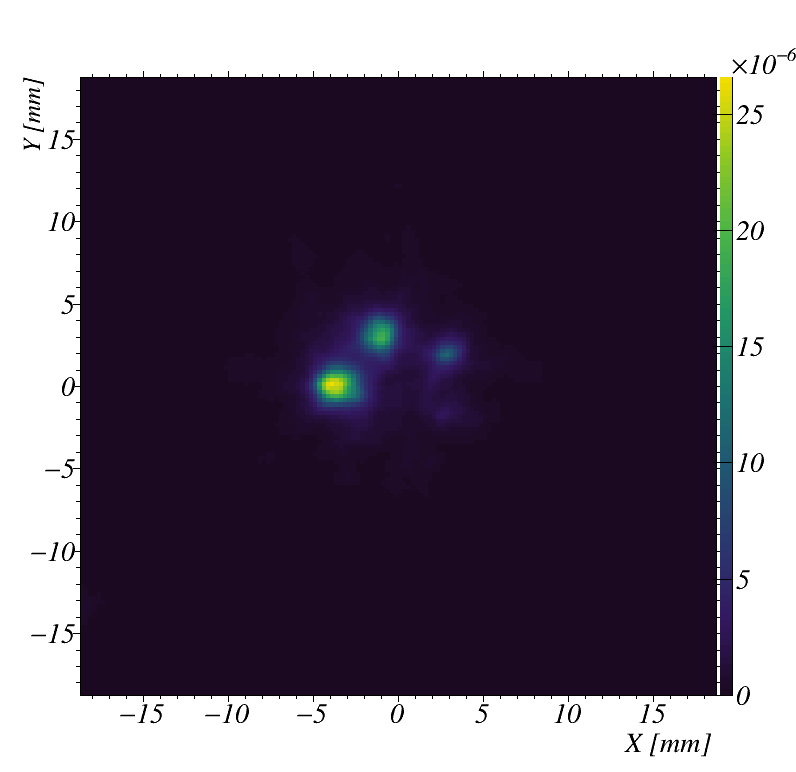}
\label{fig:simrods:7}}
\hfil
\subfloat[]{\includegraphics[height=1.08in]{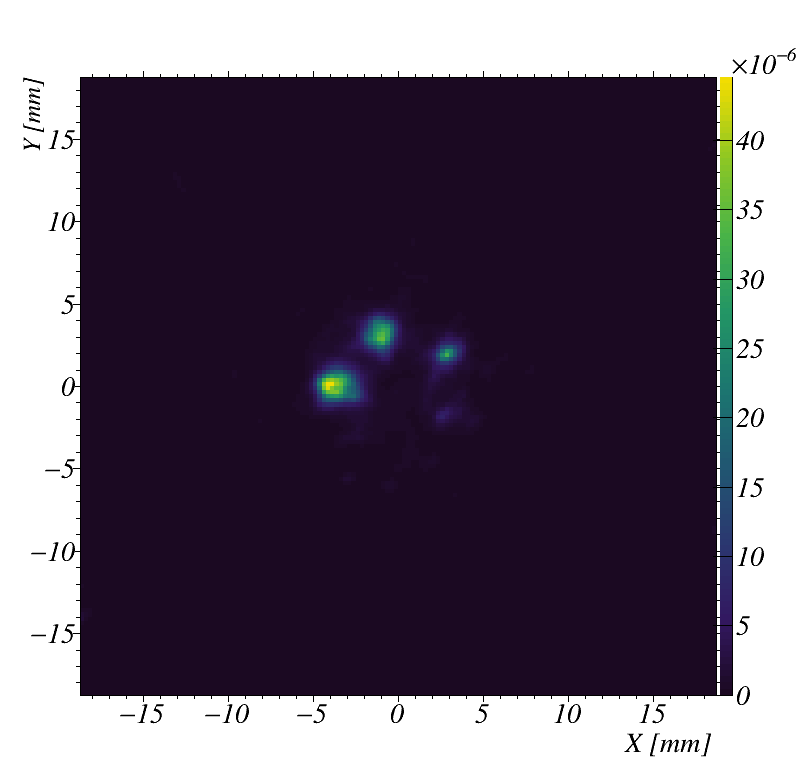}
\label{fig:simrods:8}}
\hfil
\subfloat[]{\includegraphics[height=1.08in]{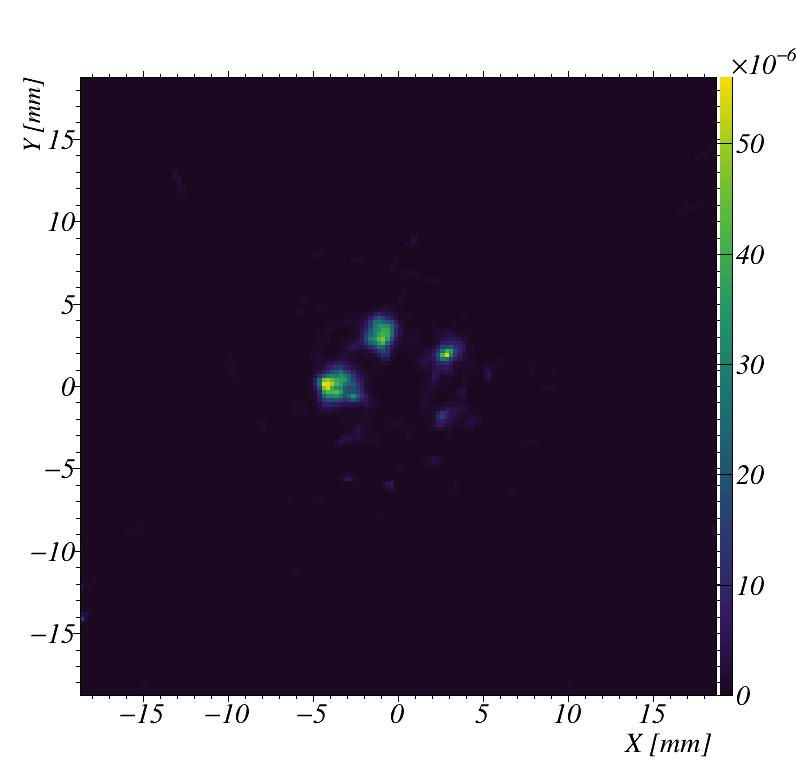}
\label{fig:simrods:9}}
\caption{Transversal ($XY$) projections over the NEMA rods region at iteration $5$ (first column), $10$ (second column) and $20$ (third column) for golden and \ac{ics} (first row), only golden (second row) and only \ac{ics} (third row) events, obtained with \ac{mc} simulations. The scale represents the activity in MBq recovered in the sum of voxels involved in each projection (e.g. a Z-row of voxels for the transversal projection).}
\label{fig:simrods}
\end{figure}

\begin{figure}[!t]
\centering
\subfloat[]{\includegraphics[height=1.45in]{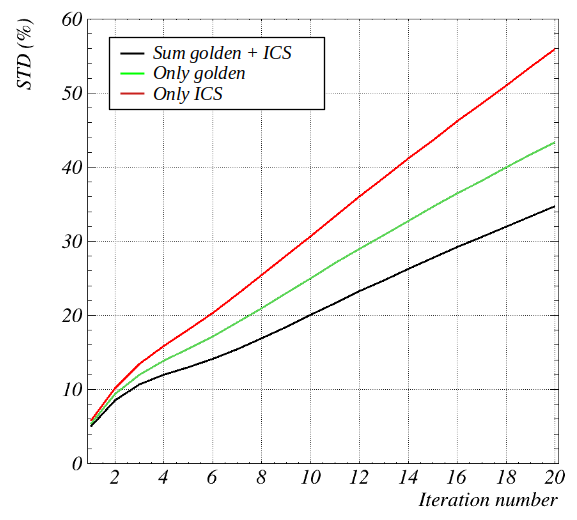}
\label{fig:simMetrics:1}}
\hfil
\subfloat[]{\includegraphics[height=1.45in]{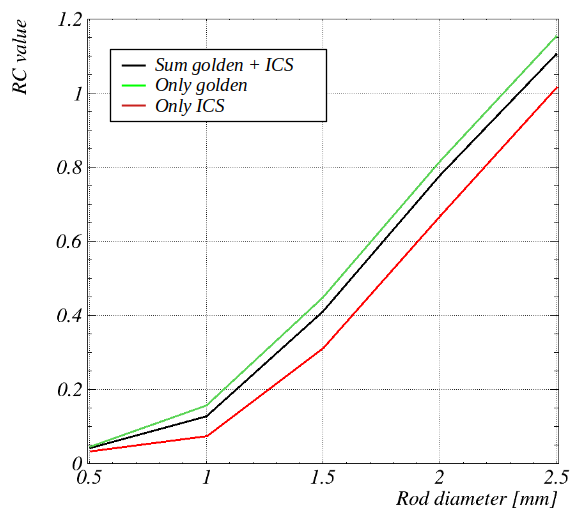}
\label{fig:simMetrics:2}}
\caption{\ac{iq} metrics obtained with \ac{mc} simulated data. (a) Evolution of the percentage of standard deviation in the uniformity \ac{roi} in a range of \ac{lmmlem} iterations. (b) Recovery coefficients for the five \acp{roi} in the rods region at iteration $5$.}
\label{fig:simMetrics}
\end{figure}

Figure \ref{fig:simunif} shows the $XY$-projections over the uniform region of the \ac{iq} phantom at three different iterations ($5$, $10$ and $20$) and for different sets of \ac{mc}-simulated data: only \ac{ics} events (bottom row), only golden events (central row) and both \ac{ics} and golden reconstructed through the joint algorithm (\ref{eq:joint}).  
Figure \ref{fig:simrods} shows the $XY$-projections over the rods region of the phantom, for the same aforementioned iterations and sets of data.
Visual inspection of the images reveals, as expected, an increased degradation at higher iterations of the iterative algorithm; but, importantly, it also reveals that the joint usage of \ac{ics} and golden events improves the uniformity in the corresponding phantom region, especially at higher iterations. 
The only-\ac{ics} images, even if they are not conceived for use, prove that these events contain valuable information about the activity distribution. Conversely, there is no appreciable degradation of the rods region when using \ac{ics} and golden events together, despite the use of V-LORs. Figure \ref{fig:simMetrics} provides quantitative grounds for these observations with the \ac{iq} metrics computed in the uniform and rods regions, following the procedure described in section \ref{sec:materials}; table \ref{tab:rcsim} shows the recovery coefficients and their corresponding $\%~\textrm{STD}$.

\setlength{\tabcolsep}{0.01mm}
\begin{table}[h]
\centering
\caption{Recovery coefficients obtained with simulated data for sum golden + ICS (S), only golden (G) and only ICS (I) events}\label{tab:rcsim}
\begin{tabular}{ccccccccccc} 
\toprule%
& \multicolumn{2}{c}{0.5 mm}   & \multicolumn{2}{c}{1.0 mm} & \multicolumn{2}{c}{1.5 mm}  & \multicolumn{2}{c}{2.0 mm}  & \multicolumn{2}{c}{2.5 mm}  \\
\midrule
 & RC & $\%~\textrm{STD}$ & RC & $\%~\textrm{STD}$  & RC & $\%~\textrm{STD}$  & RC & $\%~\textrm{STD}$ & RC & $\%~\textrm{STD}$  \\
\midrule
(S) & $0.04$  & $54.63$    & $0.13$  & $40.00$ & $0.41$  & $33.70$ & $0.78$  & $29.09$ & $1.11$  & $25.70$\\
(G) & $0.05$  & $71.71$    & $0.16$  & $55.06$ & $0.45$  & $41.24$ & $0.81$  & $34.60$ & $1.16$  & $31.12$\\
(I) & $0.03$  & $47.56$    & $0.07$  & $55.47$ & $0.31$  & $63.57$ & $0.67$  & $35.55$ & $1.02$  & $50.32$\\
\bottomrule
\end{tabular}
\end{table}

\subsection{\ac{iq} phantom experimental results}

The \ac{iq} phantom experimental measurement with MERMAID yielded a total of $46\thinspace413$ golden events and $21\thinspace368$ \ac{ics} events (i.e. $68~\%$ and $32~\%$ of total, respectively). The  sensitivity employed with simulated data was re-used and the subsequent reconstructions spanned approximately $10$ minutes in total; again, using $2^{10}$ rays for the system matrix.

\begin{figure}[!t]
\centering
\subfloat[]{\includegraphics[height=1.08in]{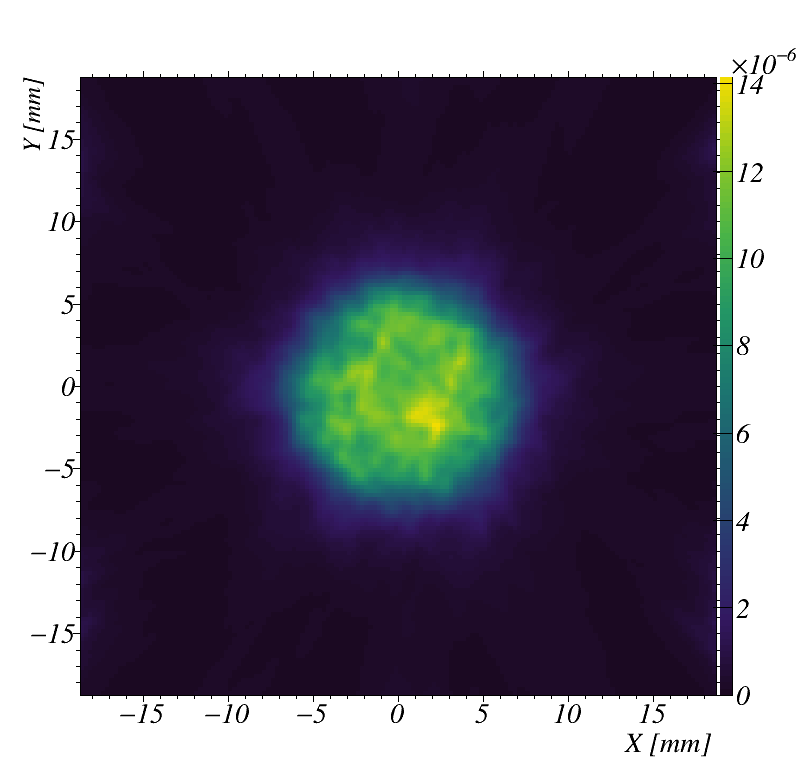}
\label{fig:expunif:1}}
\hfil
\subfloat[]{\includegraphics[height=1.08in]{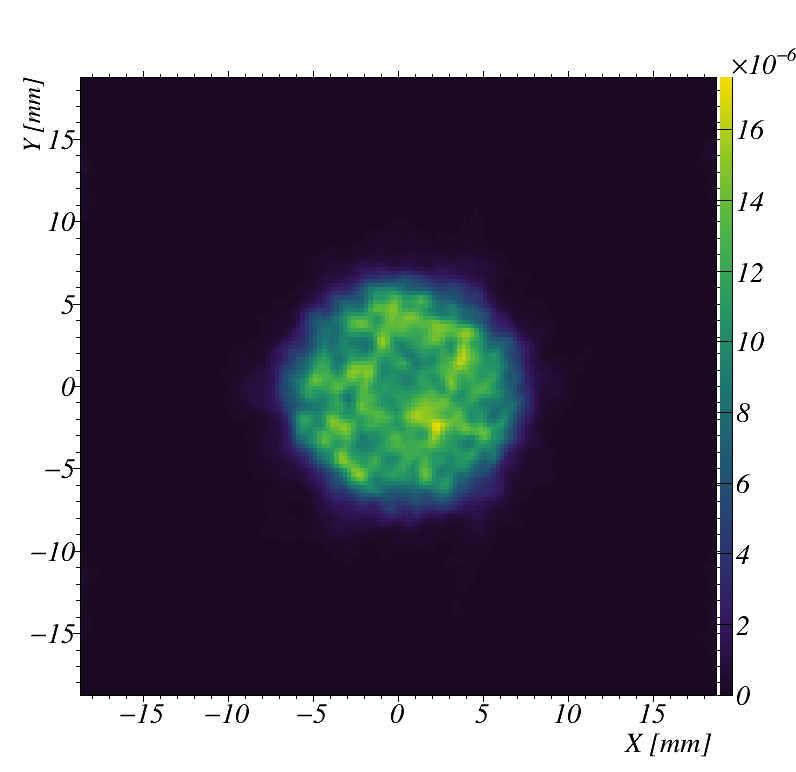}
\label{fig:expunif:2}}
\hfil
\subfloat[]{\includegraphics[height=1.08in]{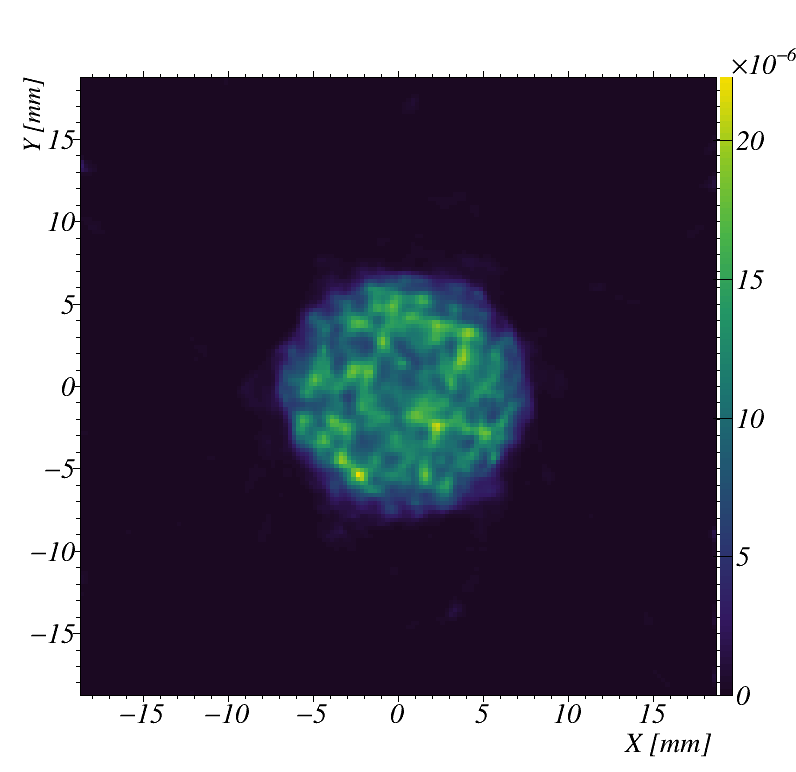}
\label{fig:expunif:3}}
\\ %forces next subfloat to the next line
\subfloat[]{\includegraphics[height=1.08in]{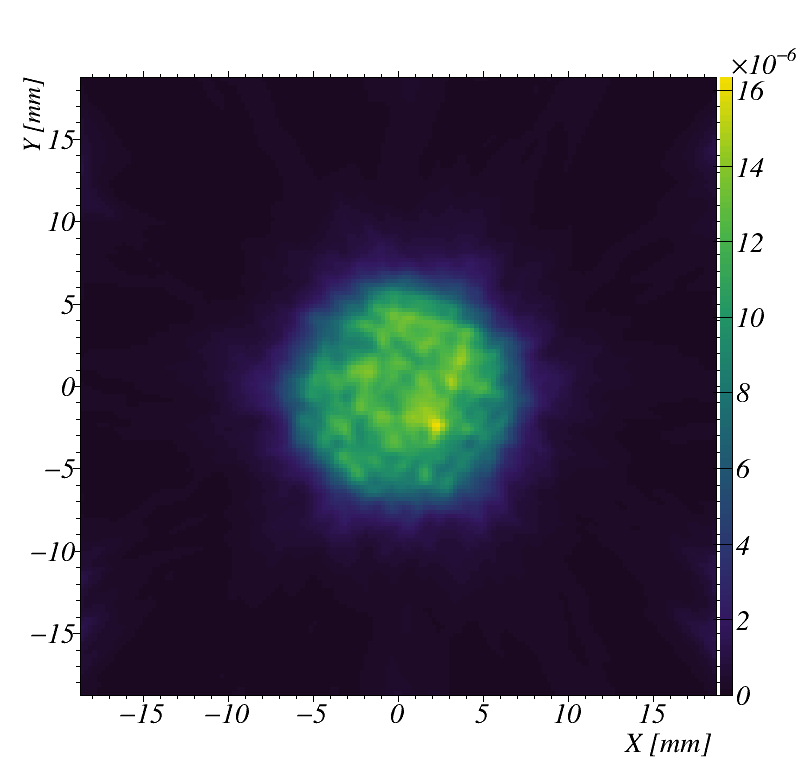}
\label{fig:expunif:4}}
\hfil
\subfloat[]{\includegraphics[height=1.08in]{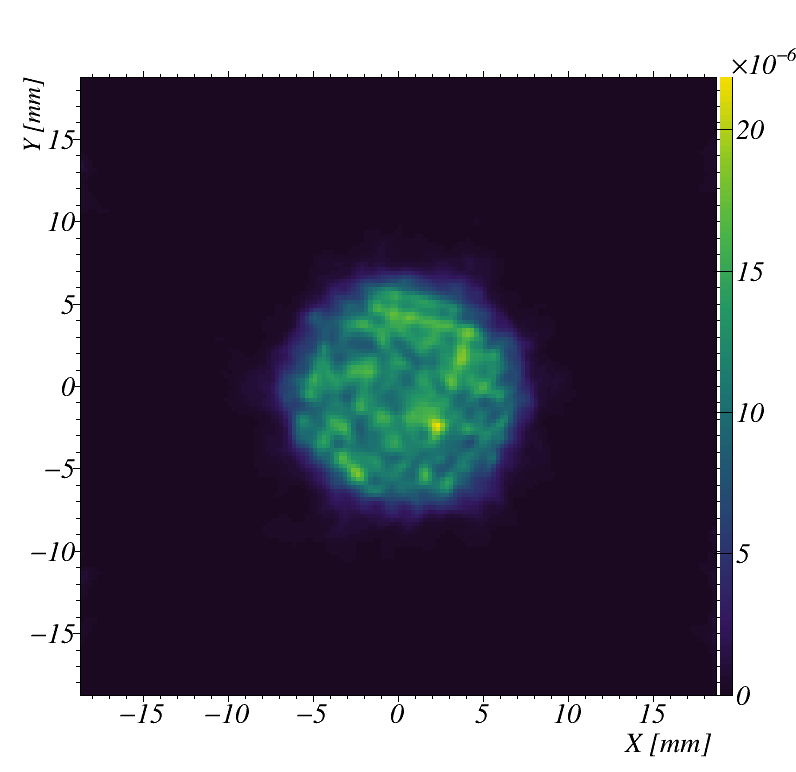}
\label{fig:expunif:5}}
\hfil
\subfloat[]{\includegraphics[height=1.08in]{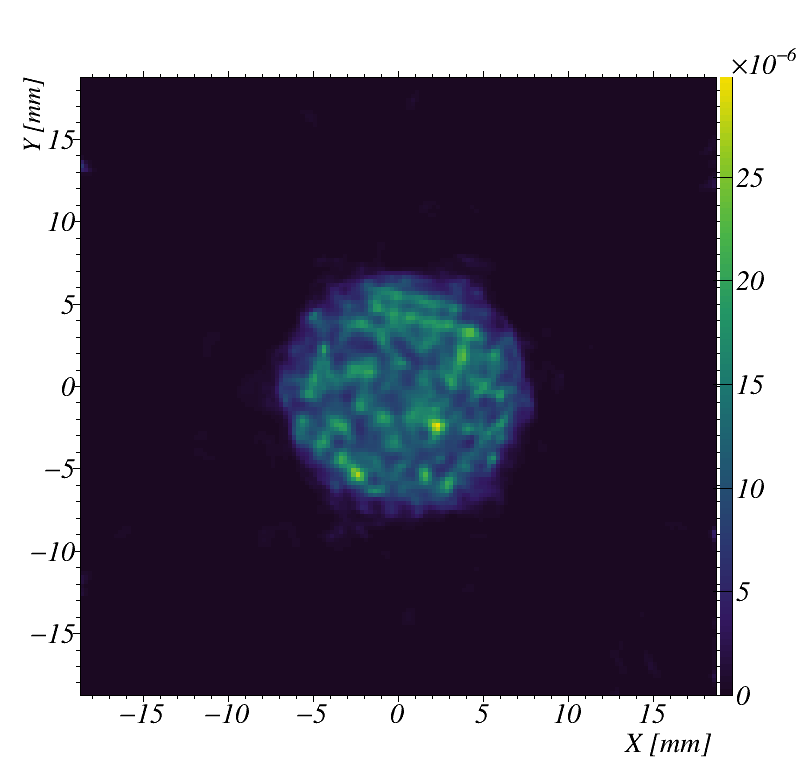}
\label{fig:expunif:6}}
\\ %forces next subfloat to the next line
\subfloat[]{\includegraphics[height=1.08in]{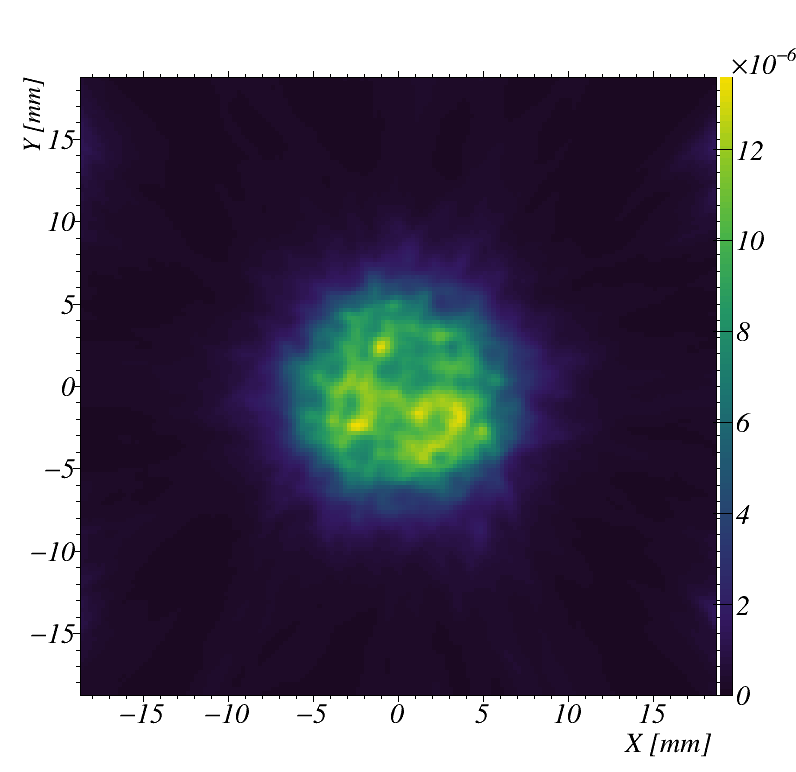}
\label{fig:expunif:7}}
\hfil
\subfloat[]{\includegraphics[height=1.08in]{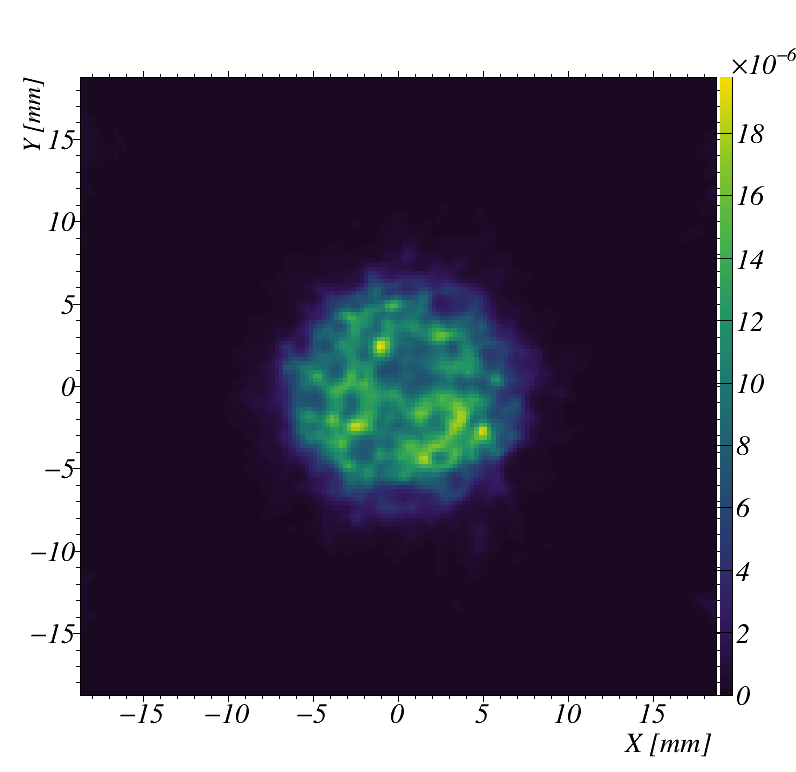}
\label{fig:expunif:8}}
\hfil
\subfloat[]{\includegraphics[height=1.08in]{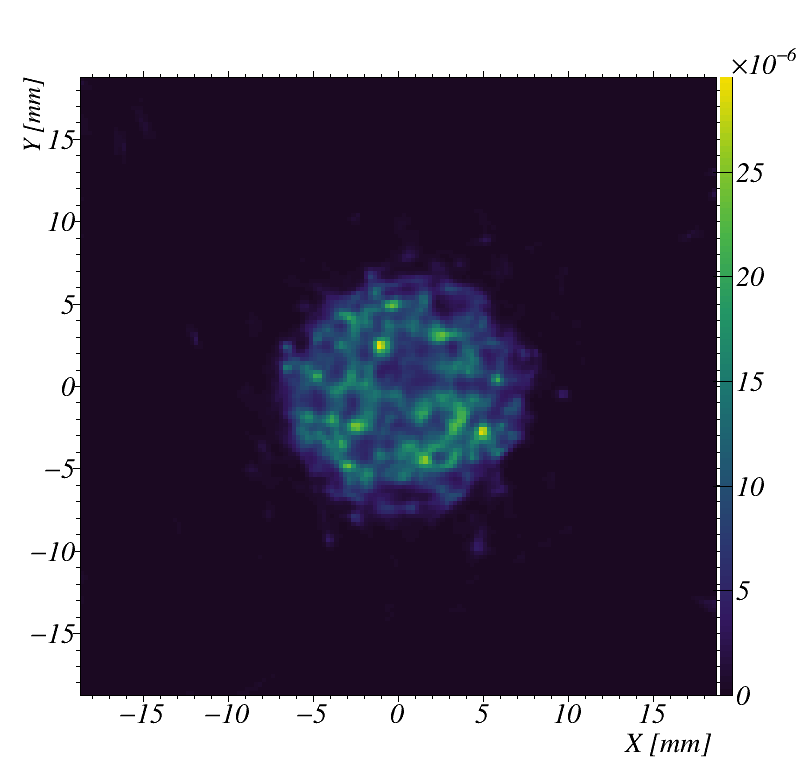}
\label{fig:expunif:9}}
\caption{Transversal ($XY$) projections over the NEMA uniformity region at iteration $5$ (first column), $10$ (second column) and $20$ (third column) for golden and \ac{ics} (first row), only golden (second row) and only \ac{ics} (third row) events, obtained with MERMAID. The scale represents the activity in MBq recovered in the sum of voxels involved in each projection (e.g. a Z-row of voxels for the transversal projection).}
\label{fig:expunif}
\end{figure}

\begin{figure}[!t]
\centering
\subfloat[]{\includegraphics[height=1.08in]{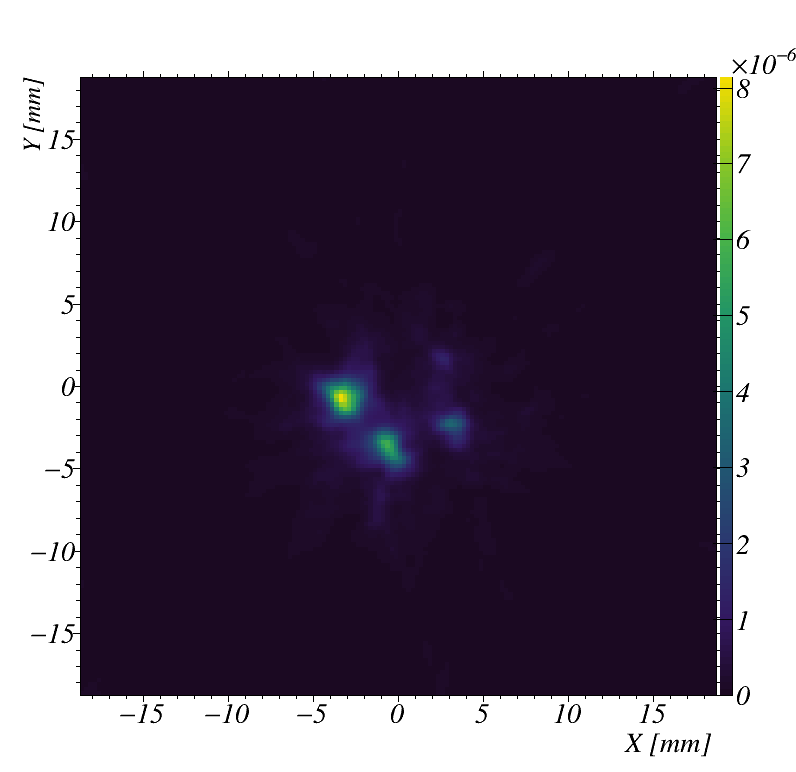}
\label{fig:exprods:1}}
\hfil
\subfloat[]{\includegraphics[height=1.08in]{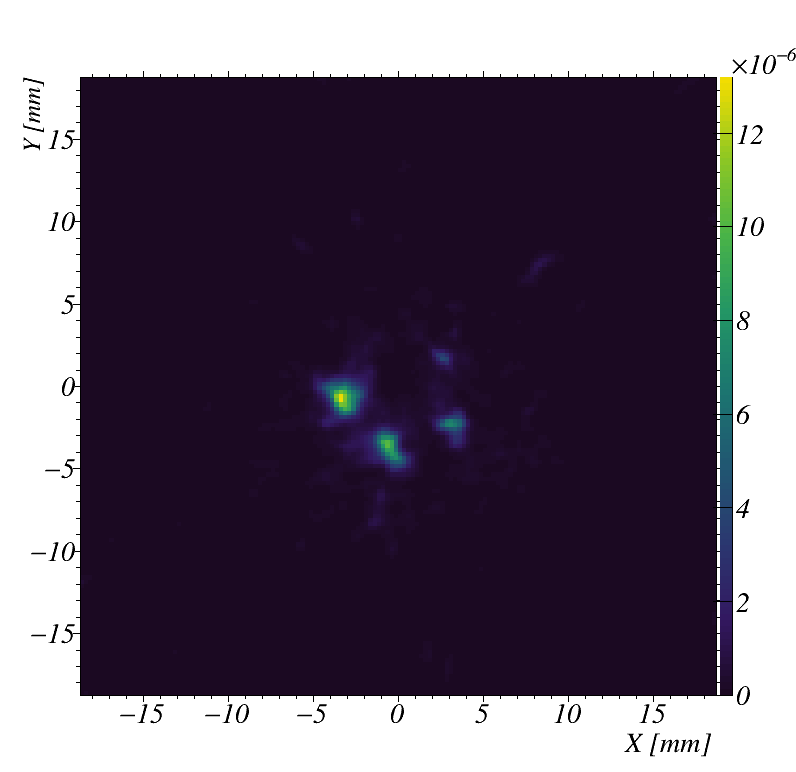}
\label{fig:exprods:2}}
\hfil
\subfloat[]{\includegraphics[height=1.08in]{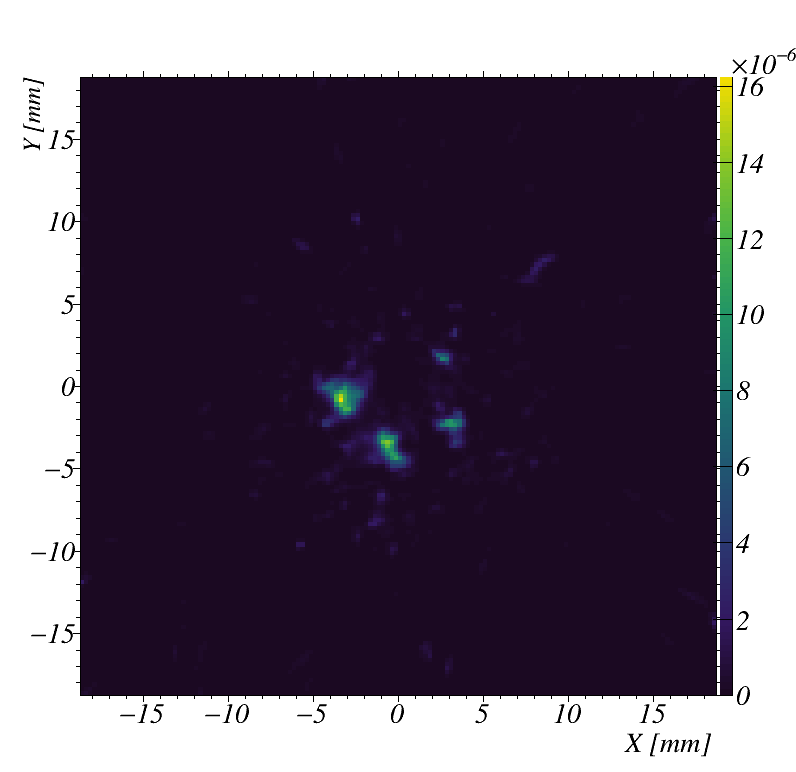}
\label{fig:exprods:3}}
\\ %forces next subfloat to the next line
\subfloat[]{\includegraphics[height=1.08in]{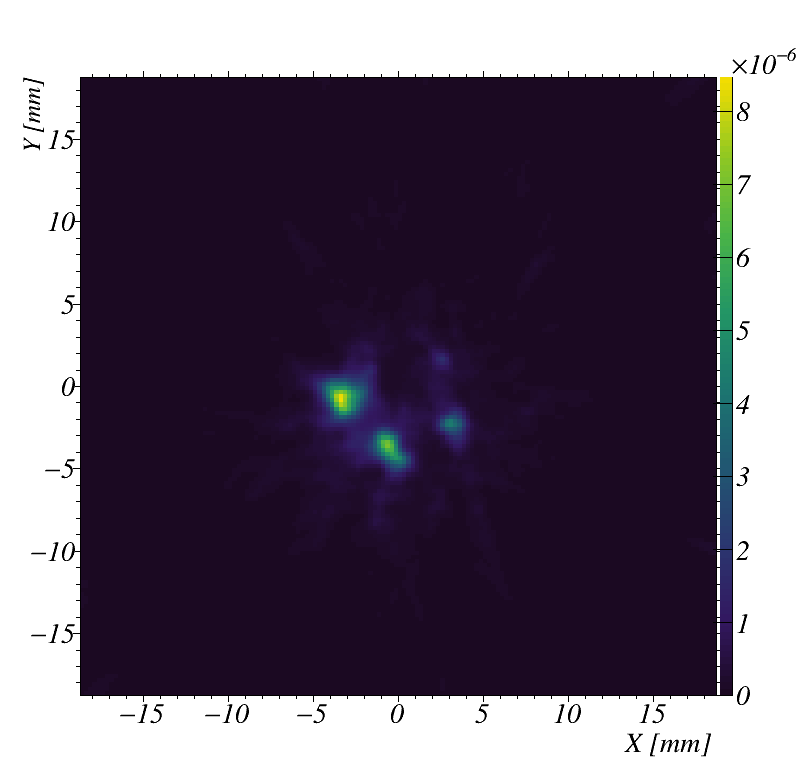}
\label{fig:exprods:4}}
\hfil
\subfloat[]{\includegraphics[height=1.08in]{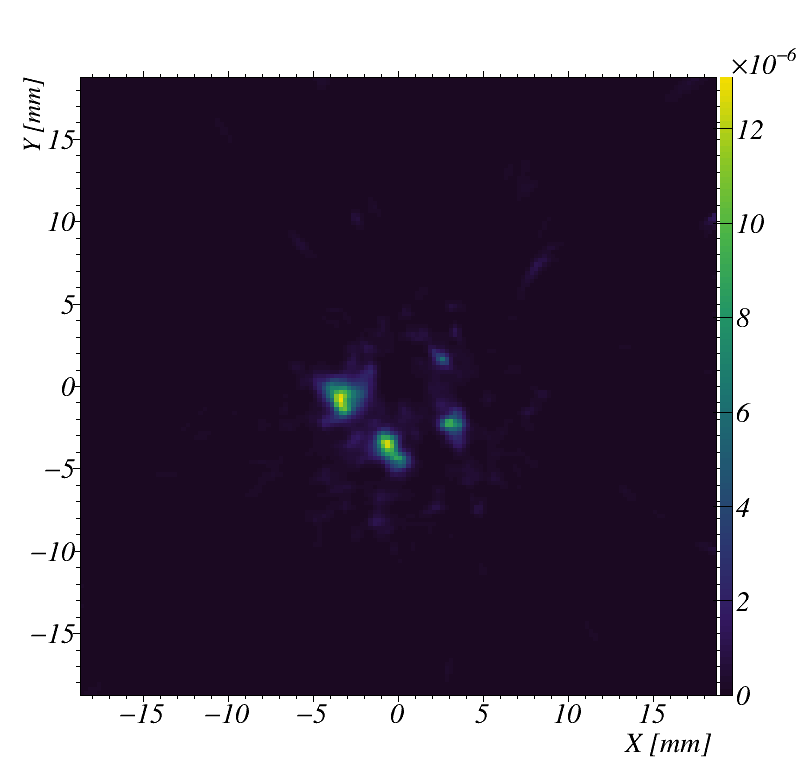}
\label{fig:exprods:5}}
\hfil
\subfloat[]{\includegraphics[height=1.08in]{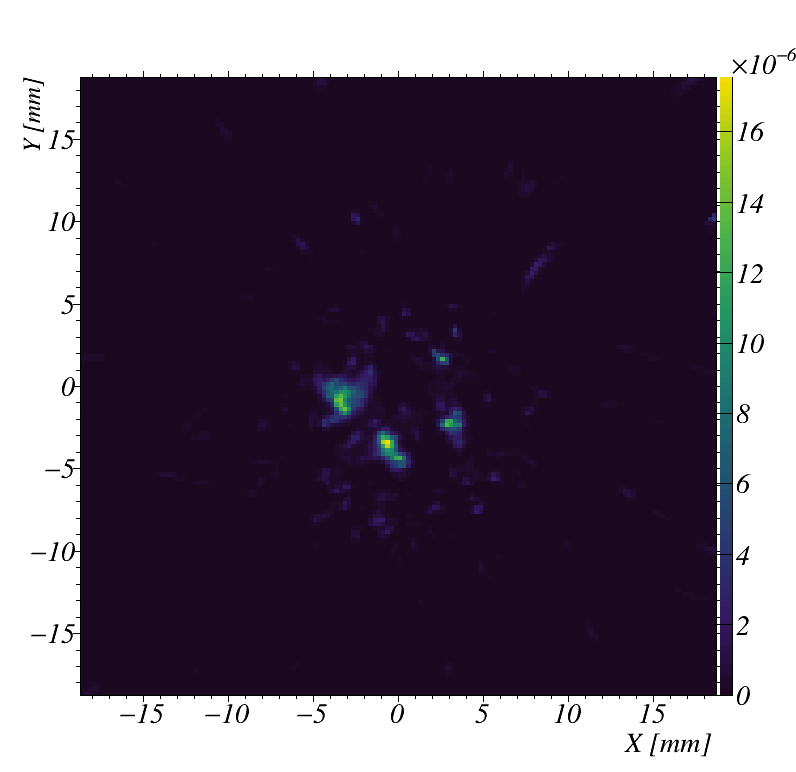}
\label{fig:exprods:6}}
\\ %forces next subfloat to the next line
\subfloat[]{\includegraphics[height=1.08in]{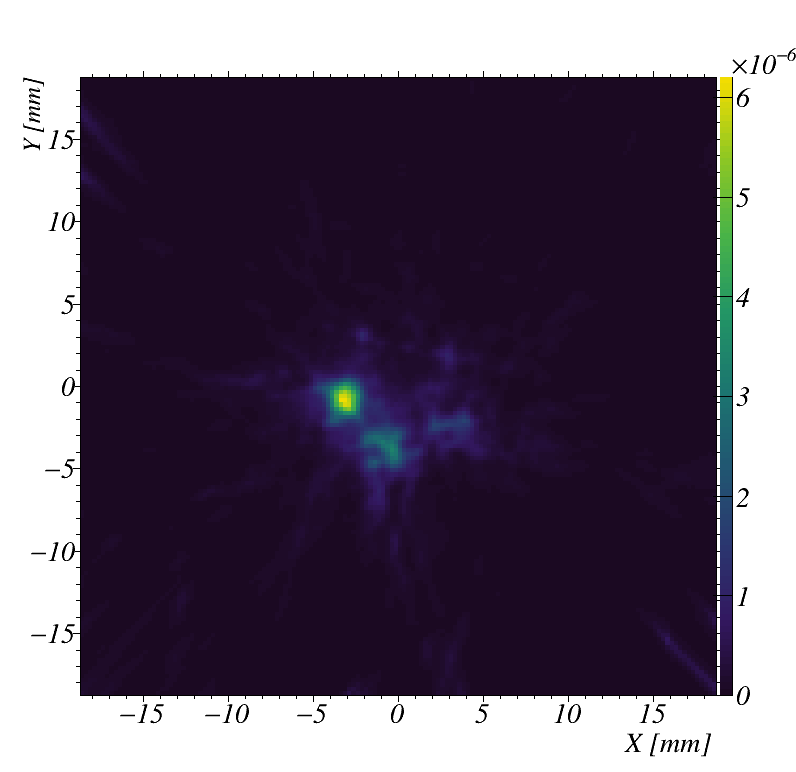}
\label{fig:exprods:7}}
\hfil
\subfloat[]{\includegraphics[height=1.08in]{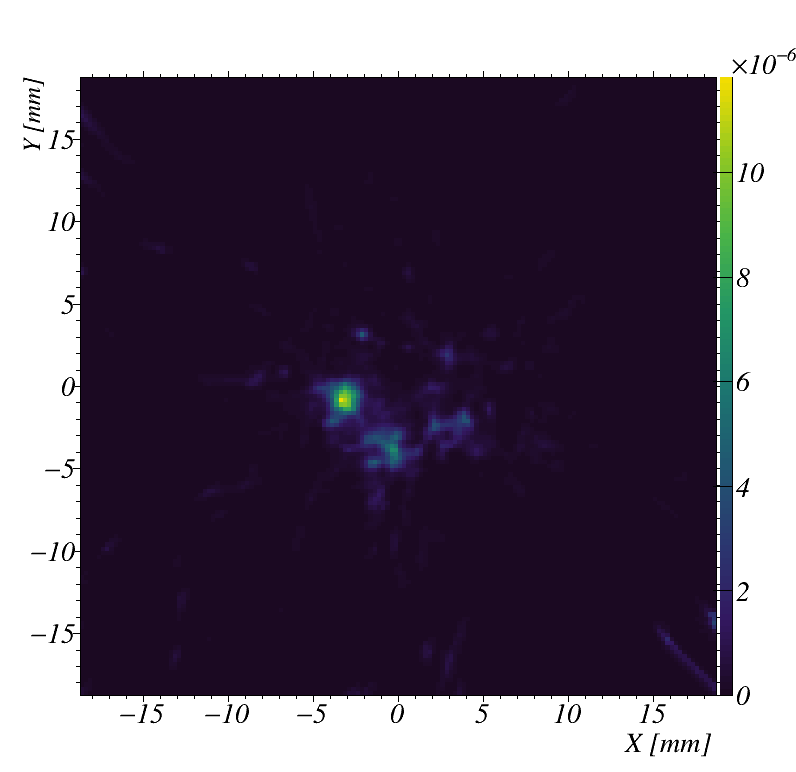}
\label{fig:exprods:8}}
\hfil
\subfloat[]{\includegraphics[height=1.08in]{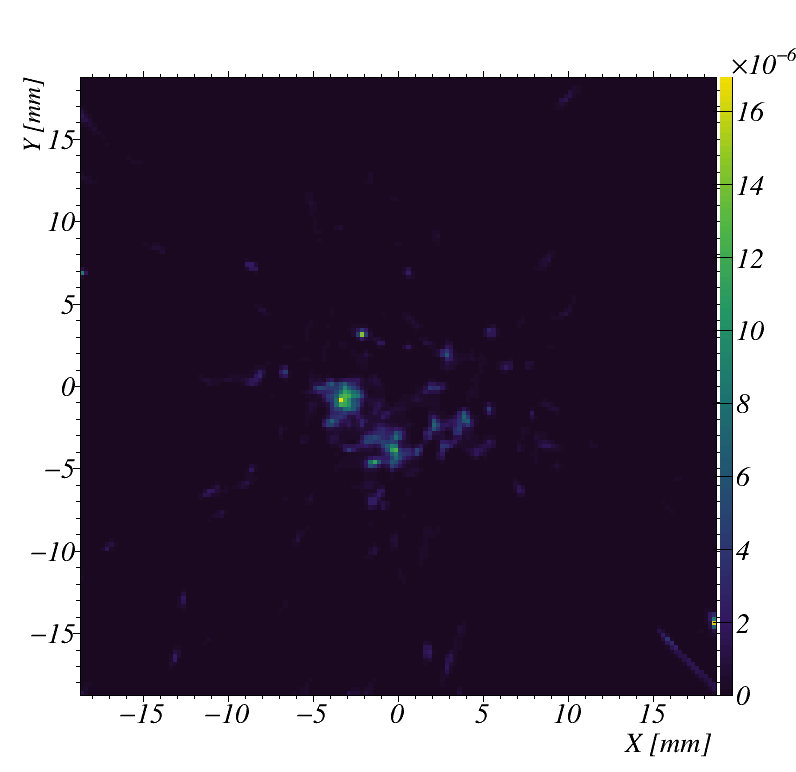}
\label{fig:exprods:9}}
\caption{Transversal ($XY$) projections over the NEMA rods region at iteration $5$ (first column), $10$ (second column) and $20$ (third column) for golden and \ac{ics} (first row), only golden (second row) and only \ac{ics} (third row) events, obtained with MERMAID. The scale represents the activity in MBq recovered in the sum of voxels involved in each projection (e.g. a Z-row of voxels for the transversal projection).}
\label{fig:exprods}
\end{figure}

Figure \ref{fig:expunif} shows the $XY$-projections over the uniform region of the \ac{iq} phantom at three different iterations ($5$, $10$ and $20$) and for different sets of experimental data acquired with MERMAID: only \ac{ics} events (bottom row), only golden events (central row) and both \ac{ics} and golden reconstructed through the joint algorithm (\ref{eq:joint}).  
%
%Figure \ref{fig:exprods} shows 
The $XY$-projections over the rods region of the phantom are shown in Figure \ref{fig:exprods}, for the same aforementioned number of iterations and data sets. 

\begin{figure}[!t]
\centering
\subfloat[]{\includegraphics[height=1.45in]{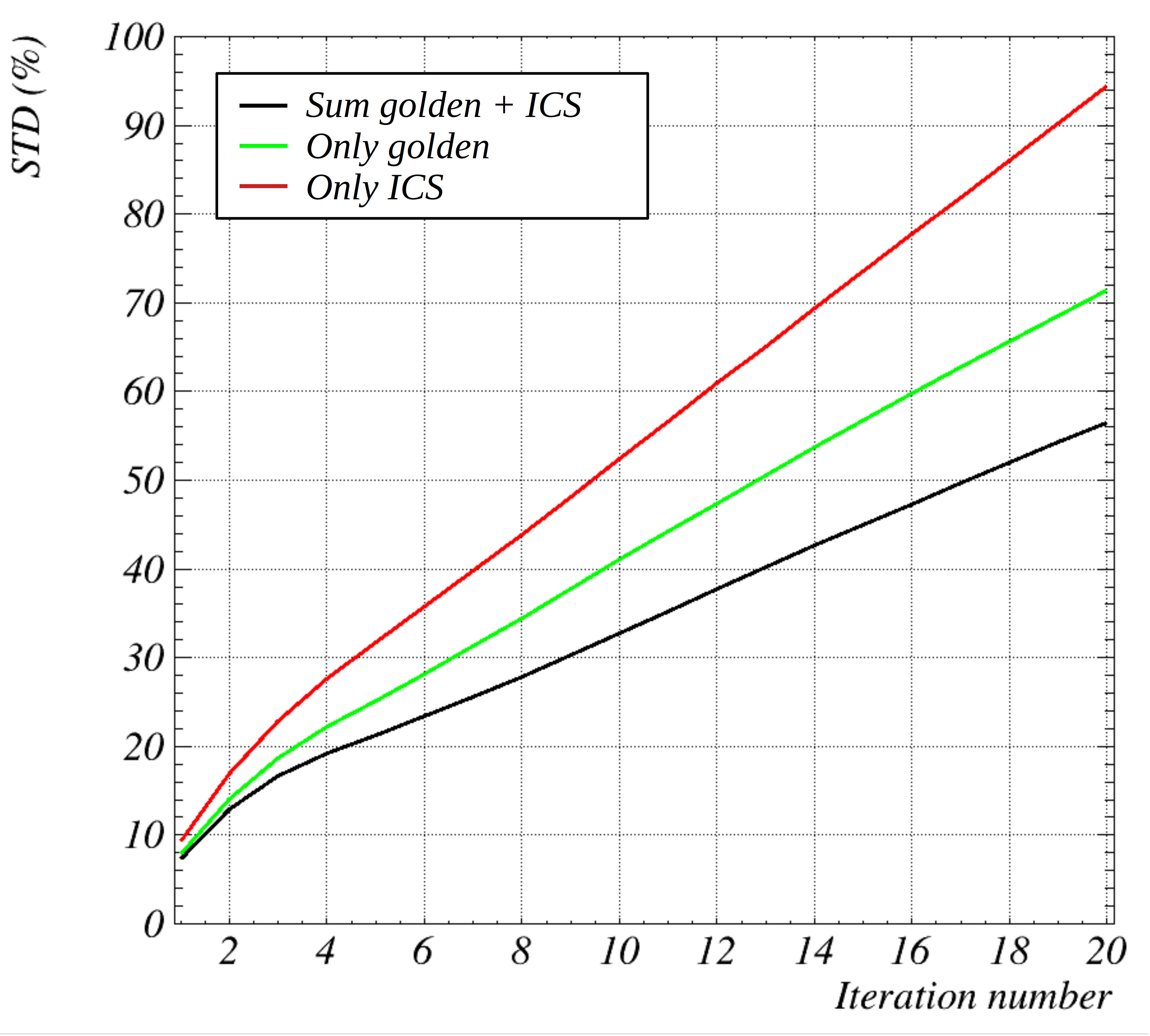}
\label{fig:expMetrics:1}}
\hfil
\subfloat[]{\includegraphics[height=1.45in]{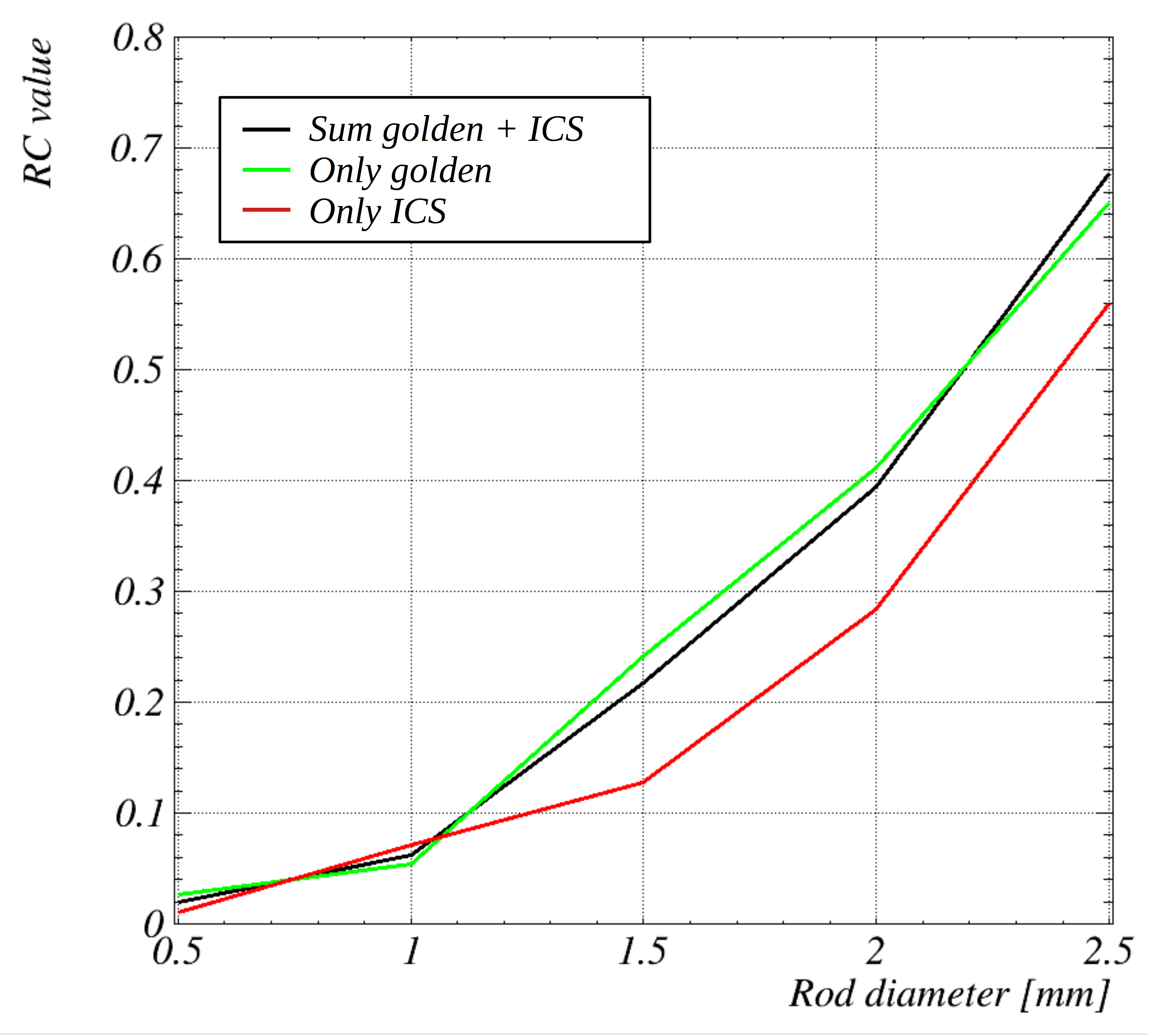}
\label{fig:expMetrics:2}}
\caption{Image quality metrics obtained with experimental data. (a) Evolution of the percentage of standard deviation in the uniformity \ac{roi} in a range of \ac{lmmlem} iterations (b) Recovery coefficients for the five \acp{roi} in the rods region at iteration 5.}
\label{fig:expMetrics}
\end{figure}

As expected, the overall image quality of the reconstructed experimental data is inferior to that of the \ac{mc} simulated data, which can be attributed to data-degradation factors not present in the simulation, such as non-uniform SiPM response, limited energy resolution, etc. Nevertheless, similar observations as the ones made with \ac{mc} simulations can be made; the joint usage of \ac{ics} and golden events improves the uniformity of the images when compared to only golden event images. Figure \ref{fig:expMetrics} shows the corresponding \ac{iq} metrics computed in the uniform and rods regions; table \ref{tab:rcexp} shows the recovery coefficients and their corresponding $\%~\textrm{STD}$.

\setlength{\tabcolsep}{0.01mm}
\begin{table}[h]
\centering
\caption{Recovery coefficients obtained with experimental data for sum golden + ICS (S), only golden (G) and only ICS (I) events}\label{tab:rcexp}
\begin{tabular}{ccccccccccc} 
\toprule%
& \multicolumn{2}{c}{0.5 mm}   & \multicolumn{2}{c}{1.0 mm} & \multicolumn{2}{c}{1.5 mm}  & \multicolumn{2}{c}{2.0 mm}  & \multicolumn{2}{c}{2.5 mm}  \\
\midrule
 & RC & $\%~\textrm{STD}$ & RC & $\%~\textrm{STD}$   & RC & $\%~\textrm{STD}$   & RC & $\%~\textrm{STD}$  & RC & $\%~\textrm{STD}$\\
\midrule
(S) & $0.02$  & $84.25$    & $0.06$  & $98.01$ & $0.22$  & $86.57$ & $0.39$  & $39.39$ & $0.68$  & $35.84$\\
(G) & $0.02$  & $114.49$    & $0.05$  & $119.00$ & $0.24$  & $109.60$ & $0.41$  & $49.60$ & $0.65$  & $40.01$\\
(I) & $0.01$  & $108.56$    & $0.07$  & $136.25$ & $0.13$  & $89.24$ & $0.29$  & $51.67$ & $0.56$  & $56.61$\\
\bottomrule
\end{tabular}
\end{table}

\section{Discussion}\label{sec:discussion}

This work proposes a physics-based model of \ac{ics} events formation which leads to analytical expressions for the \ac{ics} sensitivity image and the system matrix needed for iterative reconstruction. The approach's main feature is that \ac{ics} events are not \textit{selected} according to an algorithm that looks for the correct Compton scattering site or the \textit{correct} \ac{lor}. \ac{ics} event-selection approaches are common in the literature, and their accuracy typically depends on the assumed PET geometry and readout, as well as the nature of the event-selection method. It follows that, in a number of \ac{ics} events, the Compton scattering site is incorrectly determined and the event contributes to degrade the image, introducing noise and worsening the spatial resolution; the identification accuracy of these algorithms determines the degree of degradation.
By contrast, the approach proposed here models all \ac{ics} events as V-\acp{lor}, which can be regarded as a more appropriate way to describe the information about the spatial emission distribution encoded in these events. The necessary trade-off resides in the fact that V-\acp{lor} are less informative than \acp{lor}. When compared with conventional energy-based selection methods, the proposed V-\ac{lor} approach yields generally better-quality images (see Appendix \ref{sec:conventionalICS}), but, clearly, this would not be the case when considering an \ac{ics} event-selection approach with accuracy equal or close to $100~\%$. It must be noted, however, that the addition of \ac{ics} events as regular \acp{lor} via an ICS event-selection approach should entail, in theory, a corresponding change in the sensitivity image which would then depend on the accuracy value of the approach; otherwise, the quantitative accuracy of the obtained images could be compromised. Another point to consider is that the proposed V-\acp{lor} approach does not require previous training based on large datasets or significant computational resources, as most machine-learning based approaches do.  

Evidently, the construction of V-\acp{lor} requires the ability of the PET scanner to discriminate hits occurring in different PET detectors. As such, the proposed approach is particularly well-suited for \ac{pet} concepts with one-to-one crystal-to-photosensor coupling, as they can provide information about the multiple interactions within the same module \cite{OTA2017, SEEGER2022, CUCARELLA2025}. In the case of pixelated detector modules with shared readout or multiplexing architectures, the proposed model can be directly applied to inter-module \ac{ics} events. Furthermore, the blurring effects of \ac{ics} within the same module or within a monolithic crystal could be incorporated into the model also proposed here to handle conventional (golden) coincidence events.

The concept of V-\ac{lor} was already introduced in a previous work \cite{GILLAM2014}, whereby simplified detection models were used both for golden and ICS events. This present study exploits and expands this concept by providing a strong physical foundation. As a consequence, the proposed analytical models, when used in a statistical iterative algorithm such as \ac{lmmlem} with \ac{ics} data obtained under ideal conditions, lead to accurate quantification of the total reconstructed activity. This feature stands out from most existing techniques proposed to calculate system matrix and sensitivity image elements. Moreover, the work shows how such V-\ac{lor}-modeled ICS events can be used together with \ac{lor}-modeled golden events in a joint algorithm, increasing the sensitivity of the PET scanner while maintaining the quantitative accuracy of the total reconstructed activity in the images (see figure \ref{fig:cylinder:2}). To this end, analytical expressions for the golden sensitivity image and the system matrix are also proposed (see Appendix \ref{sec:golden}), essentially based on the ICS model without the scattered photon probability terms. The aforementioned quantitative accuracy of the total reconstructed activity is ensured under ideal detection conditions, i.e., assuming that the sets of data are composed of those (and only those) events that the models are based upon - and in particular assuming that these events do not suffer attenuation and/or object scatter, as these  effects are not modeled. Obviously, experimental data do not fulfill these ideal conditions; indeed, images obtained with experimental data exhibit lower quality than their \ac{mc}-simulation counterparts and are not quantitatively exact. The rationale for this result can be attributed to several factors. First, the experimental measurements were acquired using a system that is still under development (e.g., a correction for non-uniform SiPM channel efficiencies has not yet been developed). In addition, a possible misalignment between the phantom and the scanner axes may have introduced some blurring when projecting all axial slices into a single image or inaccuracies when applying 3D ROIs; some defects in the 3D printing procedure might have also affected the empty cavities. Furthermore, the limited energy resolution of the system, together with potential light crosstalk effects, may have further degraded the performance of the approach as well as reducing the number of events detected. Discrepancies between the LYSO density and composition assumed in the simulation (Table \ref{tab:materials}) and the experimental values might introduce further differences in the respective images.

Other experimental data-degradation effects which affect the performance of the reconstruction are attenuation, random coincidences (both for golden and ICS), and object-scattered events, none of which are corrected for in this work. A particular example, which could be accounted for by extending the models correspondingly, is represented by those events detected as golden, but actually being \ac{ics} events in which the Compton scattering interaction remains undetected because of the energy windows and the limited energy resolution. More generally, conventional estimation techniques for random and object scatter might fail when broadening the energy window to capture ICS events, as shown for randoms in \cite{TORRES2008}. The implementation of strategies to compensate for such effects are out of the scope of this work, but will be required to allow for accurate quantification of experimental PET data.

The validation of the models, carried out by comparing the predictions of the ICS sensitivity image model with \ac{mc} simulations of one-voxel sources, suggests that the former slightly overestimate the probability of ICS detection extracted from \ac{mc} simulations. This observation is supported by the results with the \ac{mc}-simulated cylinder source (figure \ref{fig:cylinder:2}), in which the sum of all voxel values in the images yields values slightly smaller than the ground truth simulated activity. The reason for the discrepancy (of approximately $3-5~\%$ in the ICS sensitivity image) can be traced back to the attenuation coefficient values used to estimate the sensitivity image and the system matrix, here extracted from the NIST database using linear interpolation; these values are not exactly the same employed by the \ac{mc} simulation, as shown in Appendix \ref{sec:golden}. In any case, the proposed models fulfill their purpose, which is not to provide values of the sensitivity image and the system matrix identical to those of a particular \ac{mc} simulation platform, but rather to provide a reasonable compromise between accuracy and flexibility to efficiently adapt to any PET geometry.

Results clearly indicate that the joint usage of \ac{ics} and golden events improves the uniformity of homogeneous regions compared to the individual usage of golden events. This difference is observed both with \ac{mc} simulations of the \ac{iq} phantom (figure \ref{fig:simMetrics:1}) and with experimental data (figure \ref{fig:expMetrics:1}), and becomes more pronounced at higher iterations of the iterative algorithm. In contrast, the recovery coefficients slightly decrease when including \ac{ics} events modeled as V-\acp{lor}; again, this result is observed both with \ac{mc} simulations of the \ac{iq} phantom (figure \ref{fig:simMetrics:2}) and with experimental data (figure \ref{fig:expMetrics:2}). Recovery coefficients are associated to the spatial resolution of the system and, although its characterization via the low-activity, downscaled \ac{iq} phantom is suboptimal (including the procedure followed to obtain the recovery coefficients, as mentioned in section \ref{sec:materials}), degradation in the image spatial resolution can be a consequence of combining two channels of information: the highly informative channel corresponding to the golden events (modeled as \acp{lor}), and a less informative channel (the \ac{ics} events modeled as V-\acp{lor}). Indeed, this effect is also observed in other usages of joint algorithms, e.g., with the different two- and three-interaction channels of a Compton camera \cite{ROSER2022}, and also plays a role in the \ac{sor} metric, see Appendix \ref{sec:sor}. However, the slight loss of resolution is not visually appreciable in the reconstructed images, while there is an observable reduction in background noise. The noise reduction could be even more significant for PET configurations suffering from very high ICS fractions, such as those with dual or triple detector layers.

It is important to highlight that the reconstruction process was not optimized regarding computational time. A reduction of these times can be achieved by computing the system matrix once and storing it (e.g. in RAM), instead of recomputing it at each iteration. In this spirit, preliminary investigations with a  1x NVIDIA Quadro RTX 6000 GPU suggest a $20$-fold reduction of the computation time when compared with the employed CPU.

Finally, it must be noted that the formulation of the proposed model leaves the door open for a number of potential improvements. First, it is not restricted to voxels, and could be extended to other discretization schemes such as blobs \cite{LOUGOVSKI2015}. Second, acollinearity effects can be modeled by replacing the Dirac delta $\delta(\Omega_0 + \Omega_0')$ in (\ref{eq:omegas}) by a suitable term accounting for this effect, e.g. a Gaussian term with FWHM~$0.50^{\circ}$ \cite{COLOMBINO1965} or slightly higher \cite{SHIBUYA2007}. Such change increases the complexity of the model and its computation, but could have an impact on the spatial resolution for PET scanners with larger FOV than the one employed here \cite{MOSES2011}. Third, different attenuation values of  detector and non-detector materials, including object attenuation, can be readily included in the model via suitable exponential survival terms. Lastly, the current formulation of the model does not make use of the detected  deposited energy in the two singles where the \ac{ics}-photon is involved; instead, synthetic deposited energies are estimated via the Compton equation and used in the estimation of the attenuation coefficients of the models. This redundant information between the Compton-equation derived deposited energies and the actually measured ones can be exploited by incorporating an energy-uncertainty term to the model \cite{SAUVE1999, DU2001}, which could improve the system matrix rows intricacy and hence the spatial image resolution, or through the weights $w_1$ and $w_2$ in (\ref{eq:sm3}). The implementation and evaluation of potential improvements to the \ac{ics} model, together with its comparison with event-selection methods available in the literature, is foreseen in a future study.

\section{Conclusion}\label{sec:conclusion}

In this work, a physics-based model of \ac{ics} in \ac{pet} is proposed and validated. The model is shown to yield reasonably accurate analytical expressions of the sensitivity image and the system matrix, where ICS events are modeled as V-\ac{lor}s without the need to identify the Compton scattering site. The work shows that such V-\ac{lor}-modeled \ac{ics} events, when included to conventional events in a joint algorithm, allow improving the uniformity in an \ac{iq} phantom, at the cost of slightly decreasing the recovery coefficients; such results are obtained both with \ac{mc}-simulations and with experimental data acquired with MERMAID, the latter for a phantom with activity well below that of conventional small-animal scenarios. The proposed approach, which can be easily adapted to any PET geometry, could be particularly beneficial in low-count PET scenarios, where the detection sensitivity is a limiting factor.

\appendices
\section{Nomenclature}\label{sec:nomenclature}

Table \ref{tab:nomenclature} summarizes the main symbols employed in this work and their definitions.

\begin{table}[ht]
\centering
\caption{Nomenclature}\label{tab:nomenclature}
\begin{tabular}{cl} 
\toprule%
Symbol  & Description   \\
\midrule
$\vec{r}_0$ & Annihilation photon emission position    \\
$\Omega_0$, $\Omega_0'$ & Annihilation photon emission directions   \\
$\vec{r}_1$ & Photoabsorption position undergone by the non-ICS photon   \\
$\vec{r}_2$, $\vec{r}_3$ & Interaction positions undergone by the ICS photon \\
$D_i$ & PET detector where photon interaction given by $\vec{r}_i$ occurs\\
$\ell_1$  & Length in the $\vec{r}_1 - \vec{r}_0$ direction  \\
$\Delta \ell_1$  & Amount of  $\ell_1$ in detector material    \\
$\ell_2$  & Length in the $\vec{r}_1 - \vec{r}_0$ direction  \\
$\Delta \ell_2$  & Amount of  $\ell_2$ in detector material    \\
$\ell_3$  & Length in the $\vec{r}_3 - \vec{r}_2$ direction  \\
$\Delta \ell_3$  & Amount of  $\ell_3$ in detector material    \\
$\ell_3', \ell_1' $  & Length in the $\vec{r}_3 - \vec{r}_1$ direction  \\
$\xi$ & Length in the $\vec{r}_2 - \vec{r}_1$ direction \\
$\Delta \ell_3' + \Delta \ell_1'$  & Amount of  $\ell_3'$, $\ell_1'$ in detector material    \\
$\mu_i$ & Total attenuation in detector after $\vec{r}_i$ \\
$\mu_i^e$ & Photoelectric attenuation in detector after interaction in $\vec{r}_i$ \\
$d\sigma_0^C/d\Omega_i$ & Differential Klein-Nishina cross section \\
$n_e$ & Effective number density of electrons in the detector \\
\bottomrule
\end{tabular}
\end{table}

\section{Analytical model for conventional coincidences (golden events)}\label{sec:golden}

The differential probability for an emission to be detected through two energy depositions within the photopeak (or, in other words, to become a golden event) can be expressed as

\begin{IEEEeqnarray}{lCl}
dP^{G}
&=& 
\frac{d^3r_0}{V}  \Theta_v(\vec{r}_0)  
\frac{d\Omega_0}{4\pi} \frac{d\Omega_0'}{4\pi} 
4\pi \delta(\Omega_0 + \Omega_0')
 e^{-\mu_{0} \Delta \ell_1}
   \mu_{0}^{e}d\ell_1 
\nonumber \\
&\cdot& 
e^{-\mu_{0} \Delta \ell_2}  \mu_{0}^{e}d\ell_2,
\end{IEEEeqnarray}
where the absence of the leading factor $2$ should be noted. Next, the change of variables 
$\left\{ 
\vec{r}_0, \Omega_0, \ell_1, \ell_2 
\right\}
\longrightarrow 
\left\{ 
\vec{r}_1, \vec{r}_2, \xi  
\right\}$ allows obtaining the following expression for the sensitivity image element:

\begin{IEEEeqnarray}{lCl}
s^{G}_v &=& \frac{1}{4\pi V}
\int_{\vec{r}_1 \in PET} d^3r_1
\int_{\vec{r}_2 \in PET} d^3r_2
\int_{\xi \in \mathbb{R}^+} d\xi \Theta_v(\xi) 
\nonumber \\ 
&\cdot  & 
\frac{e^{-\mu_0 (\Delta\ell_1 + \Delta\ell_2) } (\mu_0^e)^2}{|\vec{r}_1 - \vec{r}_2 |^2},
\label{eq:sensGolden}
\end{IEEEeqnarray}
and the following probability density:

\begin{IEEEeqnarray}{lCl}
\frac{dP^G}{d^3r_1 d^3r_2} &=& \frac{1}{4\pi V}
\int_{\xi \in \mathbb{R}^+} d\xi \Theta_v(\xi) 
\frac{e^{-\mu_0 (\Delta\ell_1 + \Delta\ell_2) } (\mu_0^e)^2}{|\vec{r}_1 - \vec{r}_2 |^2}, \nonumber \\*
\label{eq:smGolden}
\end{IEEEeqnarray}
from which the system matrix elements $h_{iv}^{G}$ can be derived:

\begin{IEEEeqnarray}{lCl}
h^{G}_{iv} &=&  
\int_{\vec{r}_1, \vec{r}_2 \in \Delta \eta_i} d^3r_1 d^3r_2  
\frac{dP^{G}}{d^3r_1 d^3r_2 } 
  \nonumber \\*
\label{eq:smGolden2}
\end{IEEEeqnarray}

\begin{figure}[!t]
\centering
\includegraphics[width=2.7in]{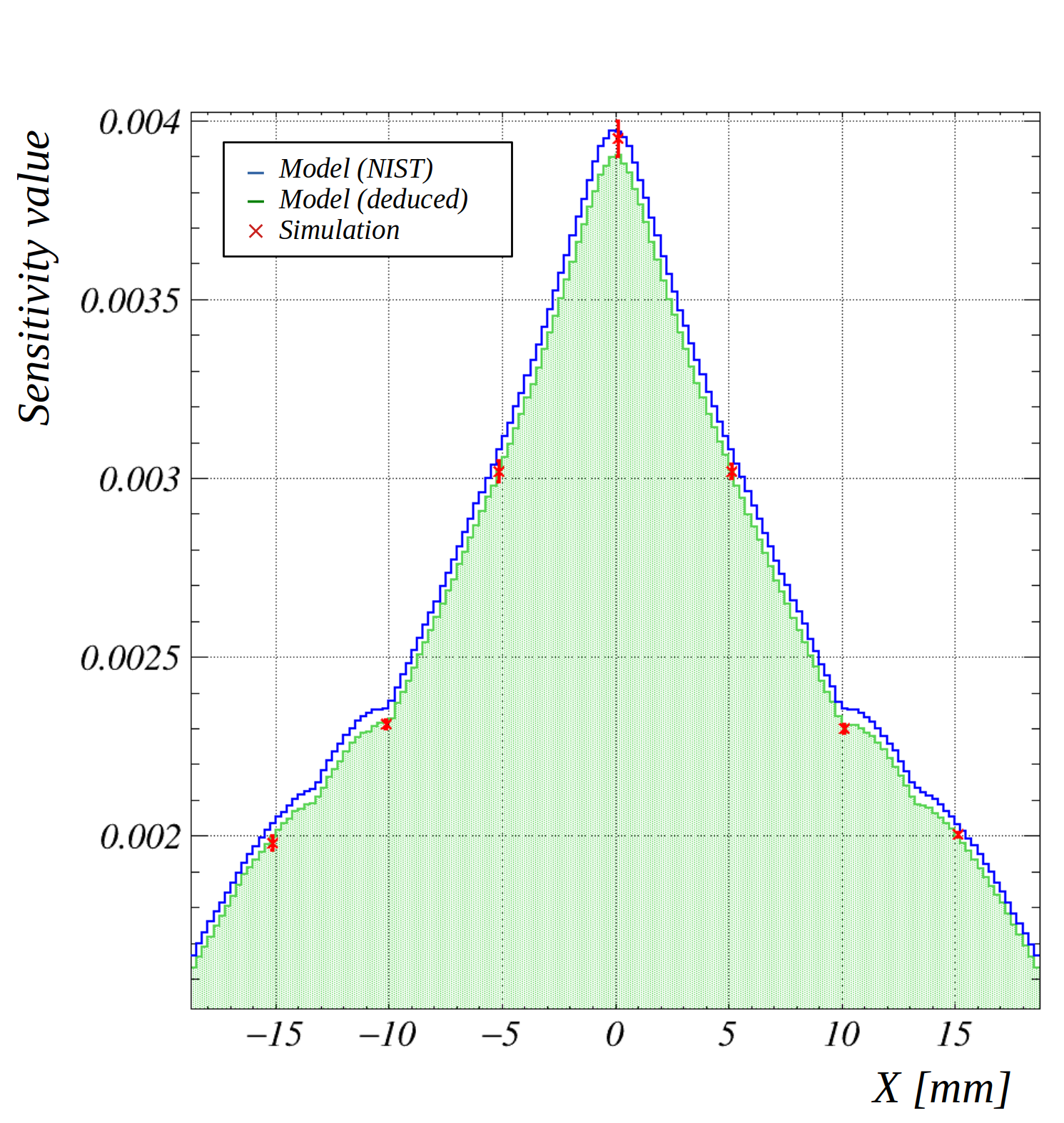}
\caption{Validation of the golden analytical sensitivity image model with the attenuation coefficient extracted from the NIST database (blue line) and deduced from simulations (green line), against Monte Carlo simulations (red crosses). Each simulated $X$-position represents the mean of five independent simulations; the displayed error bars correspond to the standard error of the mean.}
\label{fig:validationGolden}
\end{figure}

The mathematical expression for the sensitivity image (\ref{eq:sensGolden}) was validated against Monte Carlo simulations in a manner equivalent to the \ac{ics} counterpart. Figure \ref{fig:validationGolden} shows the results of such validation. In addition, the figure shows the resulting model sensitivity image when the total and photoelectric attenuation coefficients at $511~\textrm{keV}$, $\mu_0$ and $\mu_0^e$, are estimated directly from a dedicated \ac{mc} simulation, instead of using the NIST database. The simulation to obtain an estimate of the attenuation coefficients consisted of a $511~\textrm{keV}$ collimated photon source impinging on a single LYSO elongated detector; in this way, the distribution of the first photon interaction (if any) within the detector follows the distribution $e^{-\mu_0 \Delta x}\cdot \mu_0^e$. The fit yielded the values  $\mu_0^e = 0.02646 \pm 0.00010~\textrm{mm}^{-1}$, $\mu_0 = 0.0857 \pm 0.0003~\textrm{mm}^{-1}$, resulting in the green-line distribution of figure \ref{fig:validationGolden}. These numbers are to be compared with the NIST-interpolated values of $\mu_0^e = 0.0266467 \pm 0.000007~\textrm{mm}^{-1}$, $\mu_0 = 0.08527 \pm 0.00007~\textrm{mm}^{-1}$ resulting in the blue-line distribution shown also in figure \ref{fig:validationGolden}.

Note that these results apply to the attenuation coefficients for photons with $511~\textrm{keV}$ energy, which are employed in the analytical models for golden events; the small discrepancy found between the NIST and simulation-derived attenuation coefficient might not necessarily be the same at different photon energies.

% you can choose not to have a title for an appendix
% if you want by leaving the argument blank
%\section{}
%Appendix two text goes here.

\section{\ac{iq}-phantom insert region results}\label{sec:sor}

\begin{figure}[!t]
\centering
\subfloat[]{\includegraphics[height=1.08in]{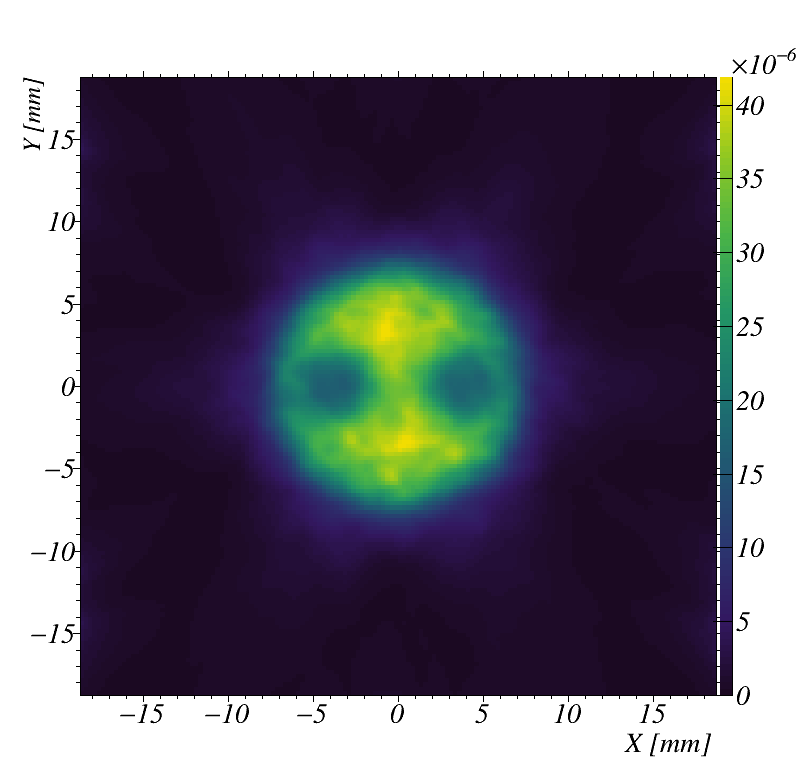}
\label{fig:simusor:1}}
\hfil
\subfloat[]{\includegraphics[height=1.08in]{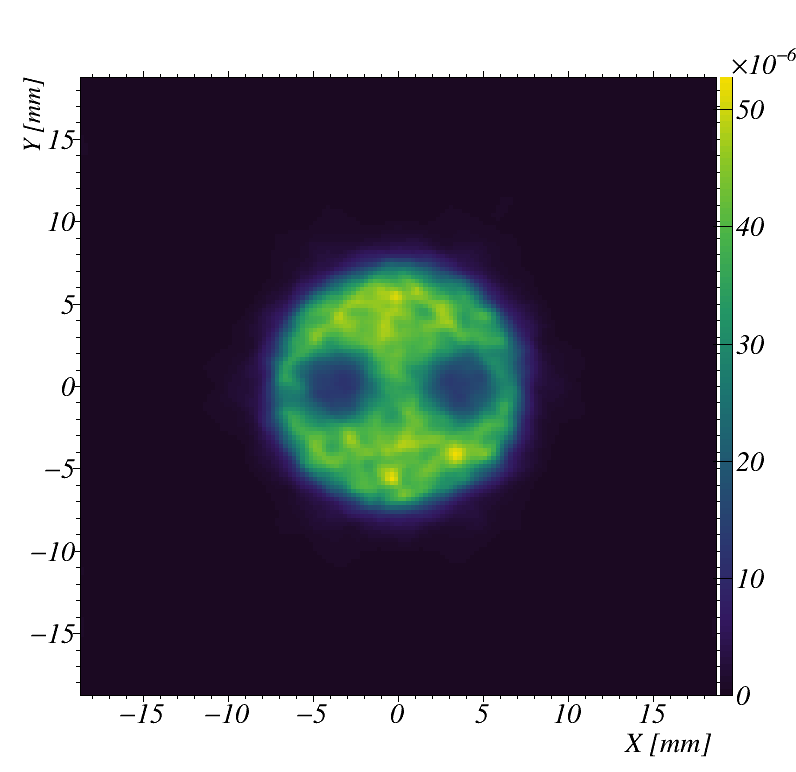}
\label{fig:simusor:2}}
\hfil
\subfloat[]{\includegraphics[height=1.08in]{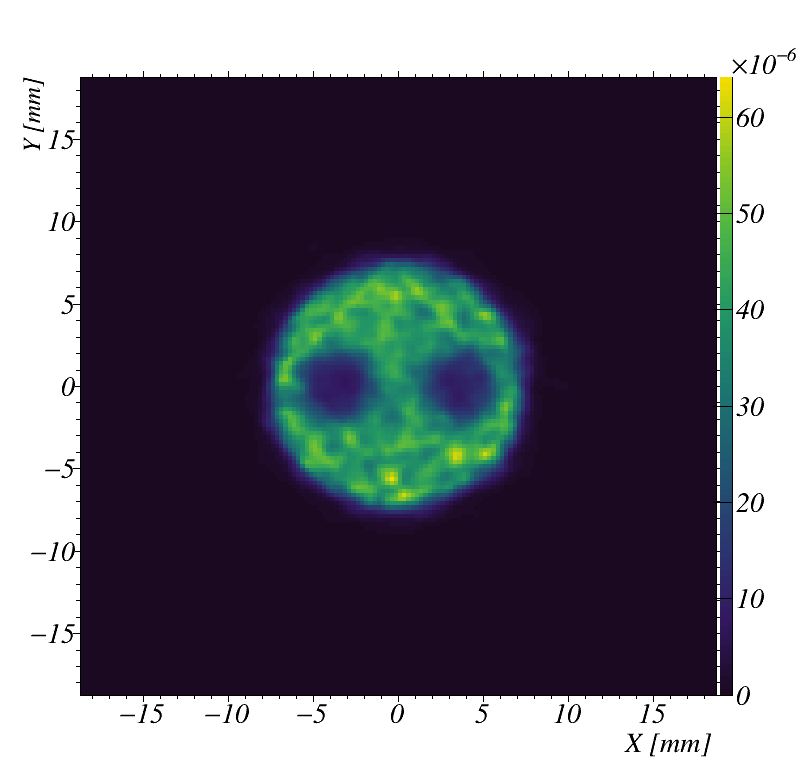}
\label{fig:simusor:3}}
\\ %forces next subfloat to the next line
\subfloat[]{\includegraphics[height=1.08in]{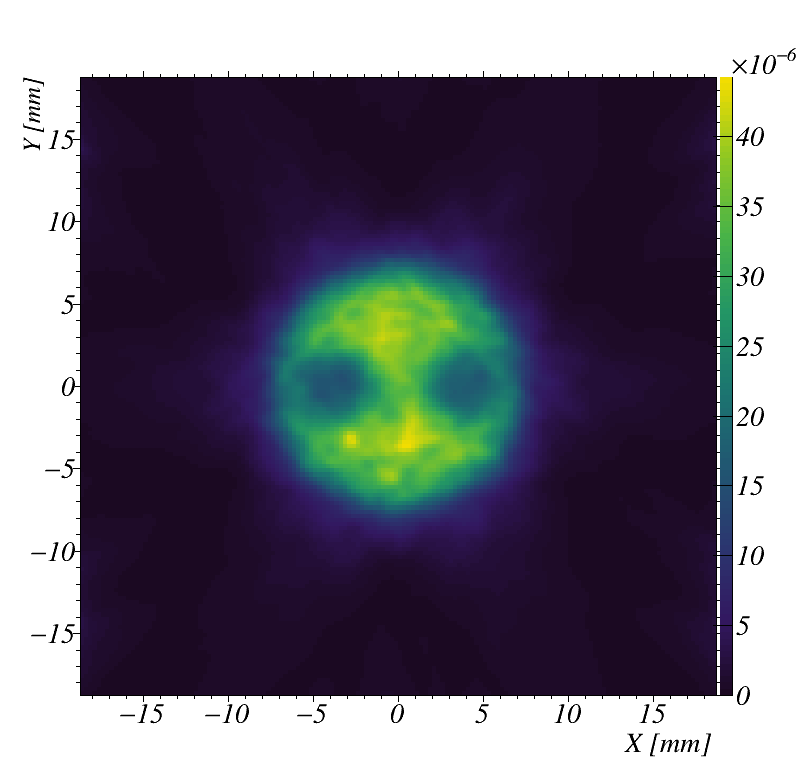}
\label{fig:simusor:4}}
\hfil
\subfloat[]{\includegraphics[height=1.08in]{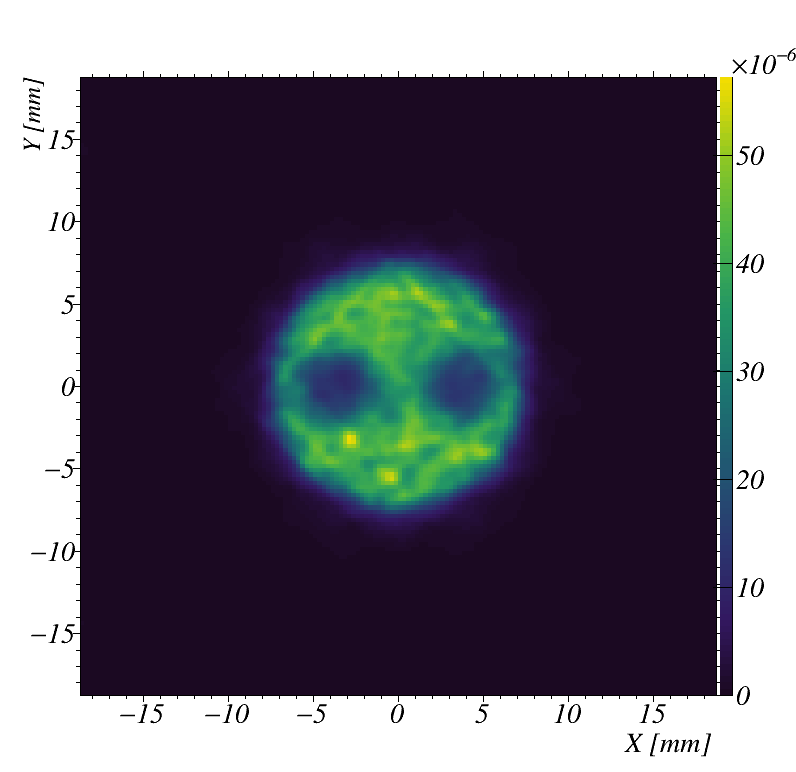}
\label{fig:simusor:5}}
\hfil
\subfloat[]{\includegraphics[height=1.08in]{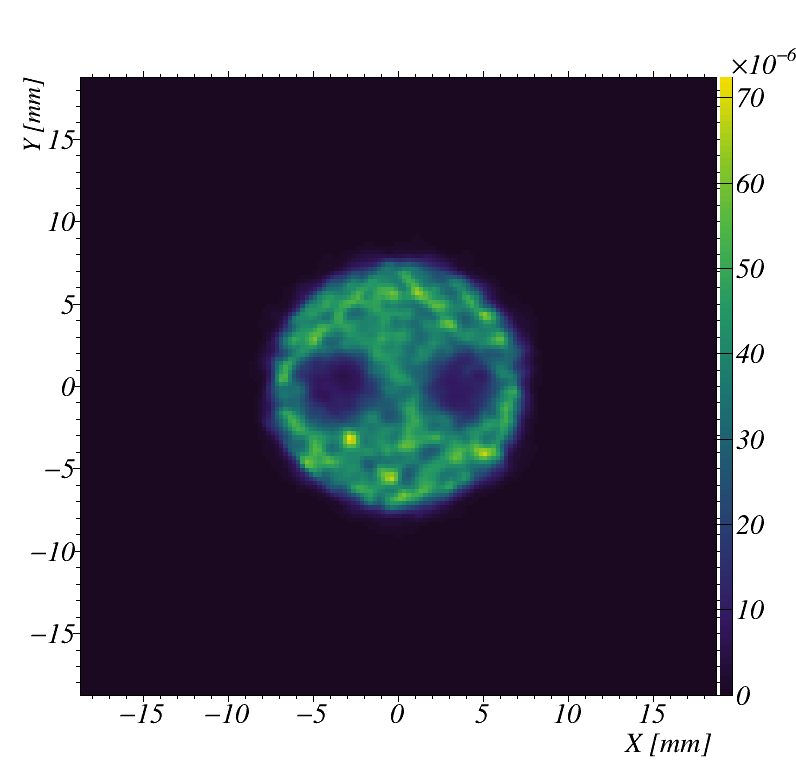}
\label{fig:simusor:6}}
\\ %forces next subfloat to the next line
\subfloat[]{\includegraphics[height=1.08in]{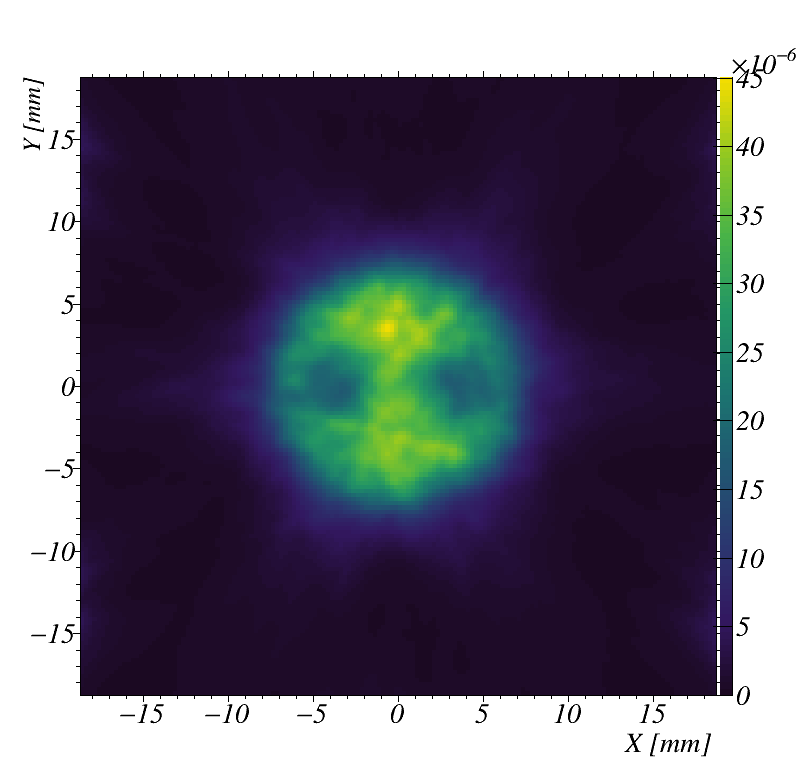}
\label{fig:simusor:7}}
\hfil
\subfloat[]{\includegraphics[height=1.08in]{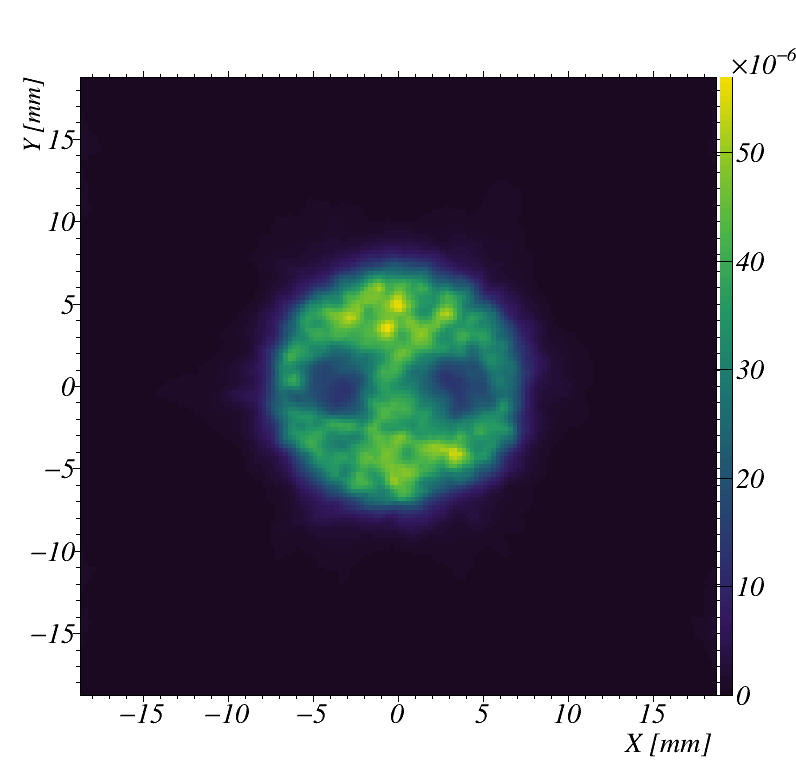}
\label{fig:simusor:8}}
\hfil
\subfloat[]{\includegraphics[height=1.08in]{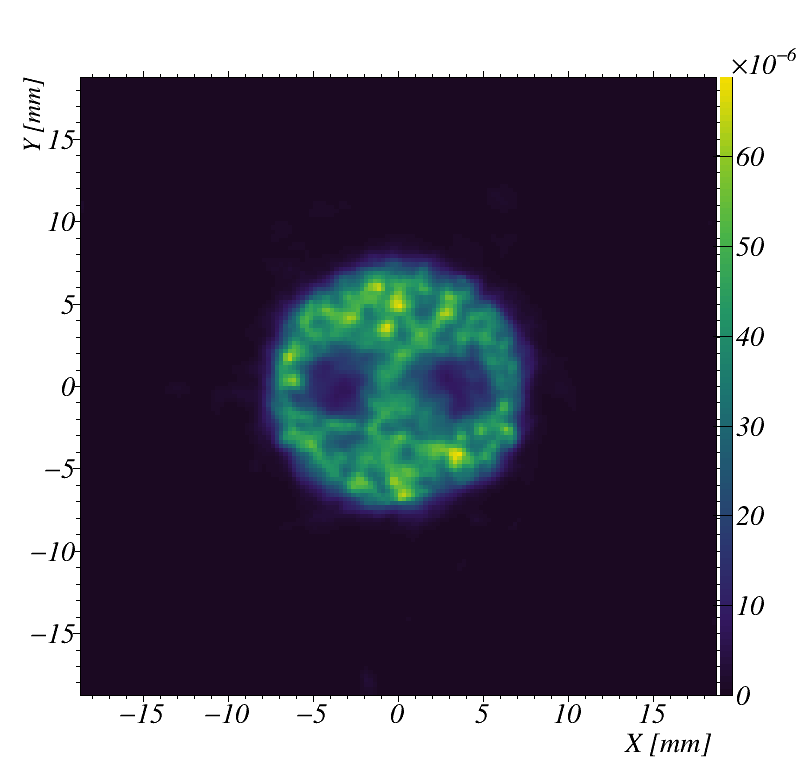}
\label{fig:simusor:9}}
\caption{Transversal ($XY$) projections over the NEMA inserts region at iteration $5$ (first column), $10$ (second column) and $20$ (third column) for golden and \ac{ics} (first row), only golden (second row) and only \ac{ics} (third row) events, obtained with \ac{mc}-simulated data. The scale represents the activity in MBq recovered in the sum of voxels involved in each projection (e.g. a Z-row of voxels for the transversal projection).}
\label{fig:simusor}
\end{figure}

\begin{figure}[!t]
\centering
\subfloat[]{\includegraphics[height=1.08in]{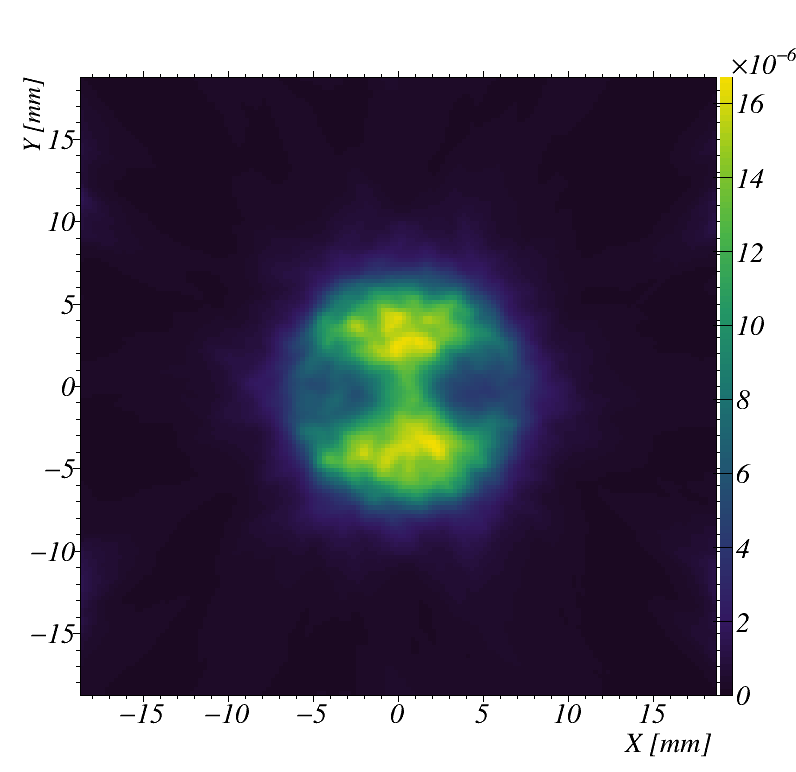}
\label{fig:expsor:1}}
\hfil
\subfloat[]{\includegraphics[height=1.08in]{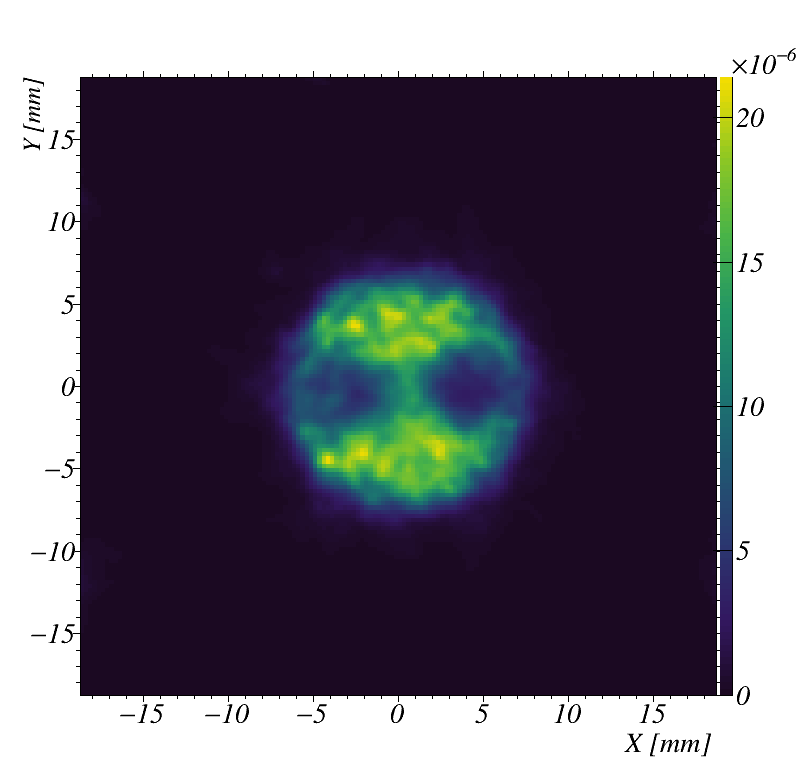}
\label{fig:expsor:2}}
\hfil
\subfloat[]{\includegraphics[height=1.08in]{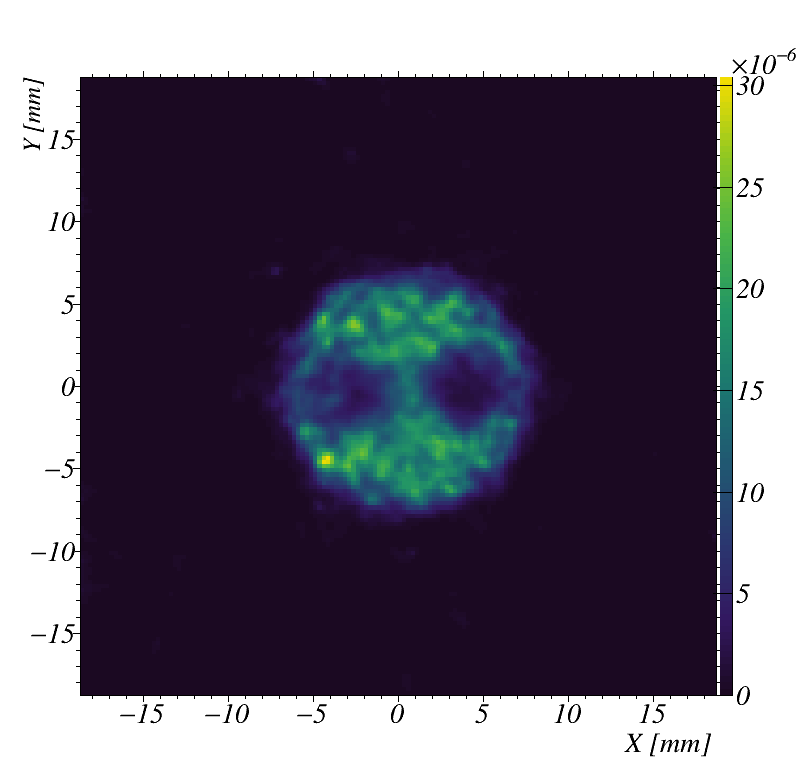}
\label{fig:expsor:3}}
\\ %forces next subfloat to the next line
\subfloat[]{\includegraphics[height=1.08in]{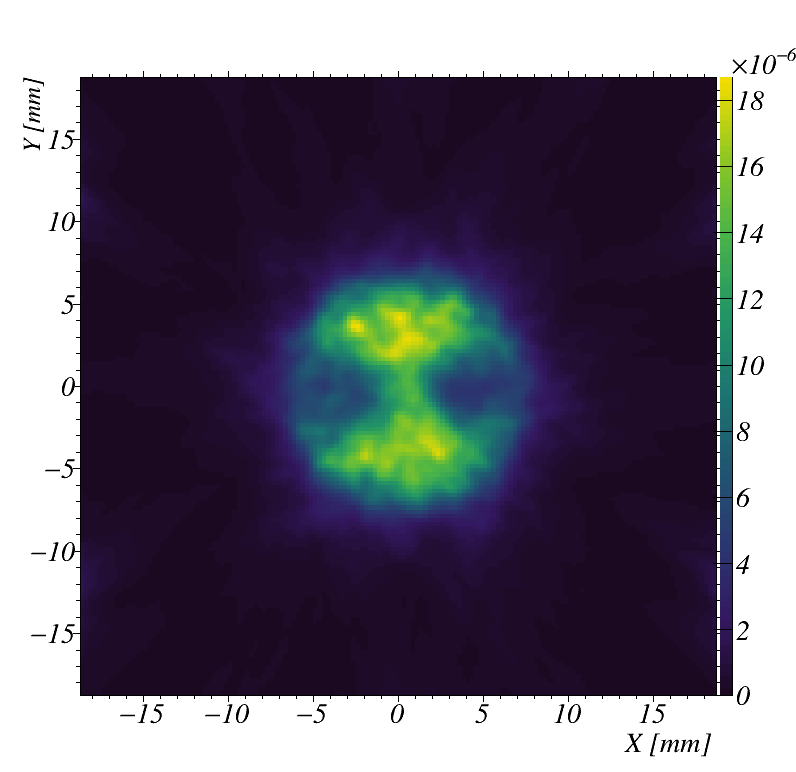}
\label{fig:expsor:4}}
\hfil
\subfloat[]{\includegraphics[height=1.08in]{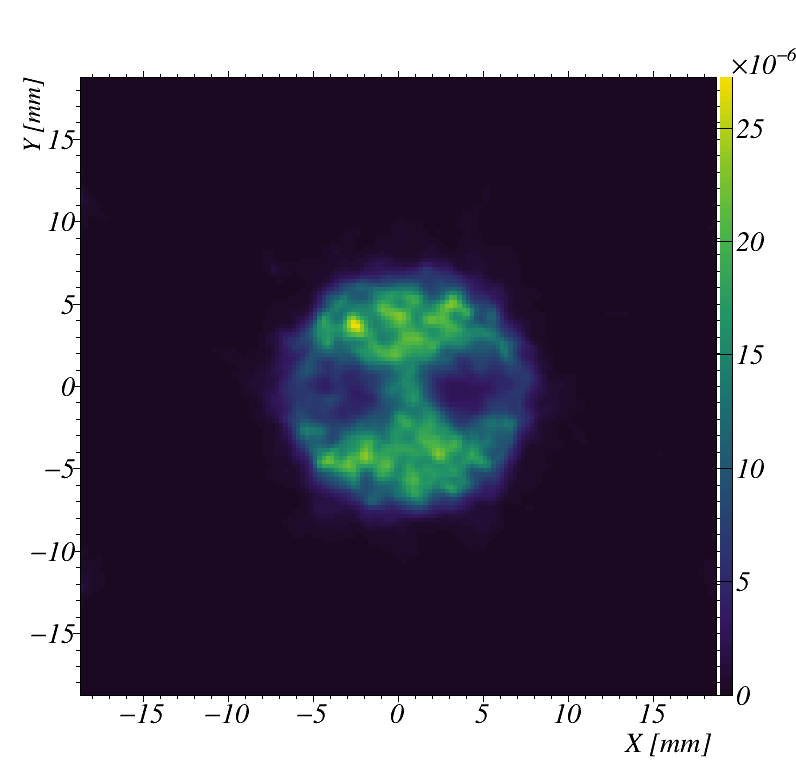}
\label{fig:expsor:5}}
\hfil
\subfloat[]{\includegraphics[height=1.08in]{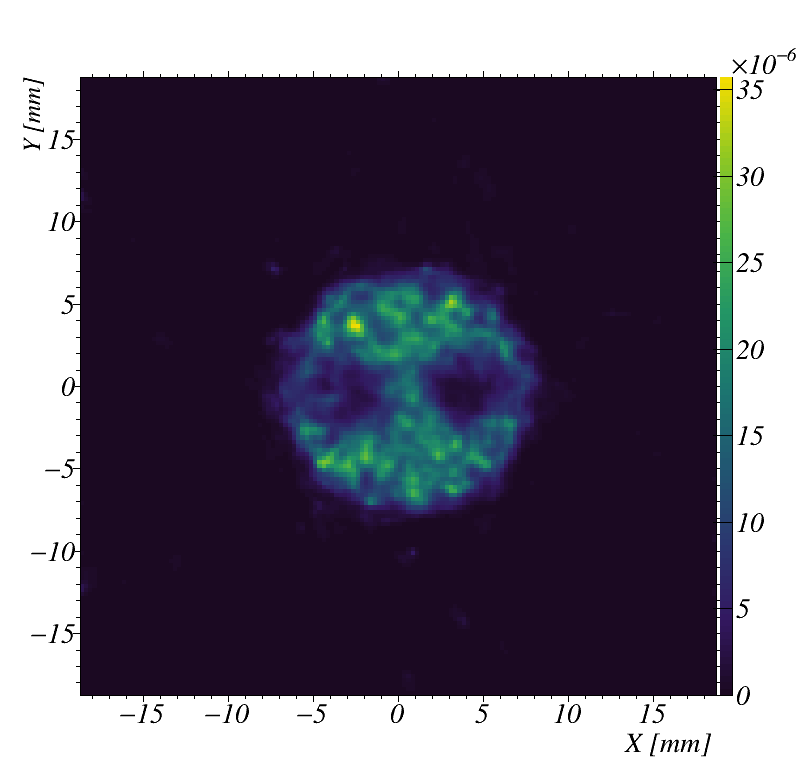}
\label{fig:expsor:6}}
\\ %forces next subfloat to the next line
\subfloat[]{\includegraphics[height=1.08in]{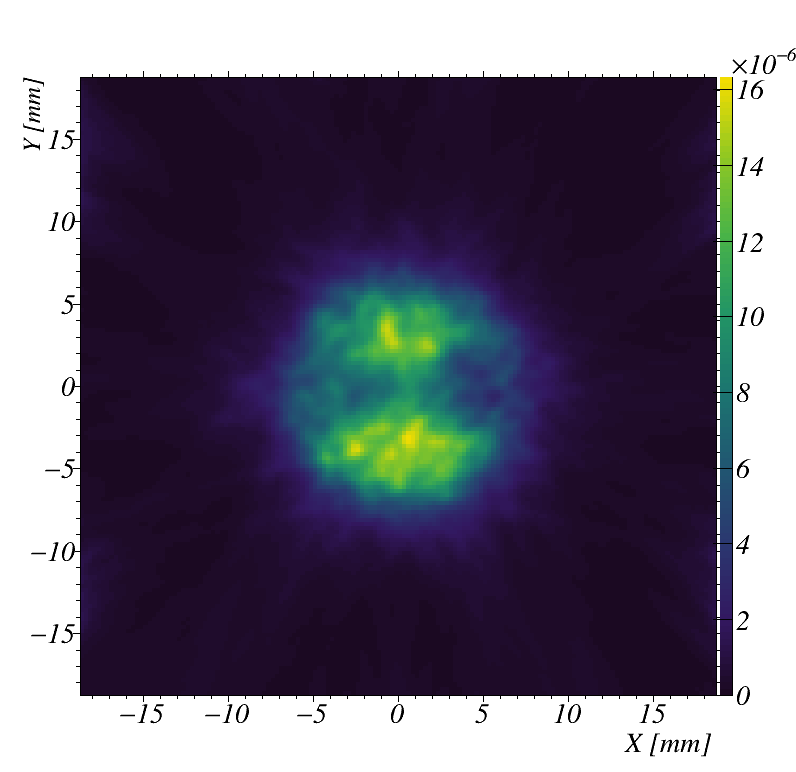}
\label{fig:expsor:7}}
\hfil
\subfloat[]{\includegraphics[height=1.08in]{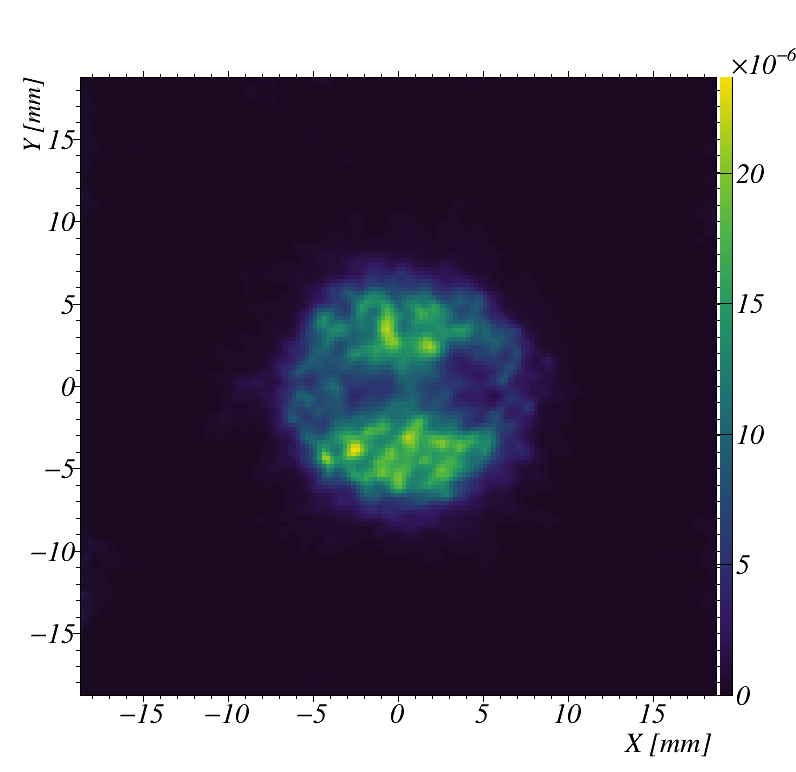}
\label{fig:expsor:8}}
\hfil
\subfloat[]{\includegraphics[height=1.08in]{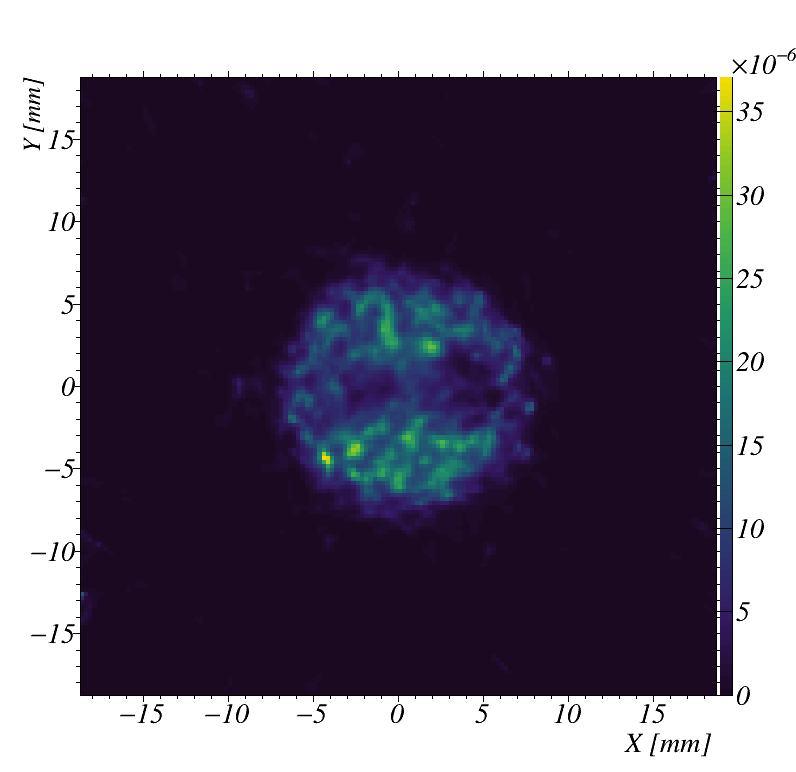}
\label{fig:expsor:9}}
\caption{Transversal ($XY$) projections over the NEMA inserts region at iteration $5$ (first column), $10$ (second column) and $20$ (third column) for golden and \ac{ics} (first row), only golden (second row) and only \ac{ics} (third row) events, obtained with MERMAID. The scale represents the activity in MBq recovered in the sum of voxels involved in each projection (e.g. a Z-row of voxels for the transversal projection).}
\label{fig:expsor}
\end{figure}

\begin{figure}[!t]
\centering
\subfloat[]{\includegraphics[height=1.45in]{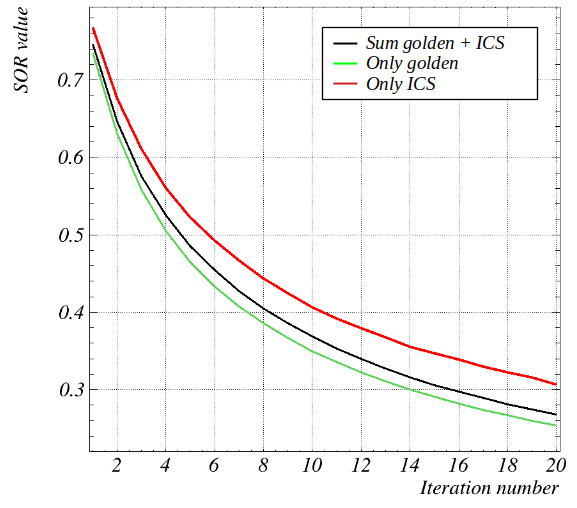}
\label{fig:sor:1}}
\hfil
\subfloat[]{\includegraphics[height=1.45in]{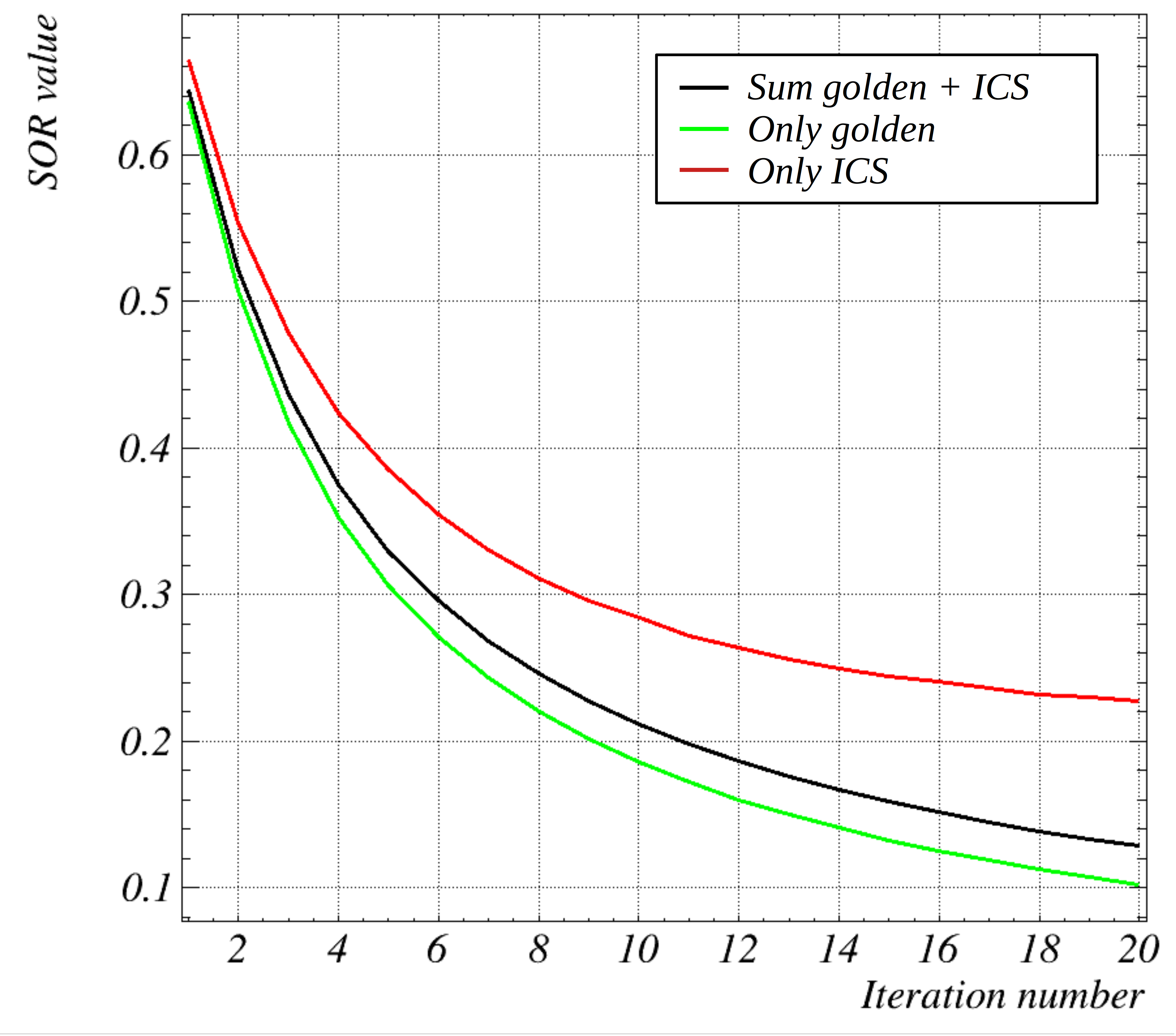}
\label{fig:sor:2}}
\caption{SOR metrics obtained in one of the \ac{iq}-phantom inserts with (a) \ac{mc}-simulated data; (b) experimental data.}
\label{fig:sor}
\end{figure}

Figure \ref{fig:simusor} shows the $XY$-projections over the inserts region of the \ac{iq} phantom at three different iterations ($5$, $10$ and $20$) and for different sets of \ac{mc}-simulated data: only \ac{ics} events (bottom row), only golden events (central row) and both \ac{ics} and golden reconstructed through the joint algorithm (\ref{eq:joint}). Figure \ref{fig:expsor} shows the corresponding information for experimental data. Finally, figure \ref{fig:sor} shows the evolution of the \ac{sor} computed following the NEMA protocol adapted to the downscaled size of the phantom, for one of the two air inserts with both simulated and experimental data.

As with the rods and uniform regions of the \ac{iq} phantom, the images obtained with experimental data exhibit higher degradation when compared with the simulated data ones. Similar to the results with the RCs, the SOR values are slighlty better when using only golden events as compared to the situation where \ac{ics} events are included. Again, this outcome can be regarded as a logical consequence of combining the highly informative channel of golden events  (modeled as \acp{lor}), and a less informative channel (the \ac{ics} events modeled as V-\acp{lor}); in particular, the section of the V-\ac{lor} not containing the emission point might cross the inserts, thus placing activity there. This difference is particularly noticeable quantitatively in the experimental case; as such, the influence of further degradation effects (optical sharing, randoms, object scatter) over the different channels of information (golden and \ac{ics} events) remains to be investigated.

\section{Comparison with energy-based ICS recovery methods}\label{sec:conventionalICS}

Among the different conventional ICS recovery strategies, energy-based criteria such as the maximum energy and second maximum energy \cite{COMANOR1996} are common and relatively easy to implement. Given that this work focus on three-hit ICS events, whereby a single Compton scattering occurs, there are only two possible sites to which this interaction can be assigned after a detection (i.e. $\vec{r}_2$ or $\vec{r}_3$ in Figure \ref{fig:ICSevent}). In these circumstances, the maximum energy criterion identifies the Compton scattering site with the one where the most energy is deposited, whereas the second maximum energy criterion assumes the opposite.

\begin{figure}[!t]
\centering
\subfloat[]{\includegraphics[height=1.05in]{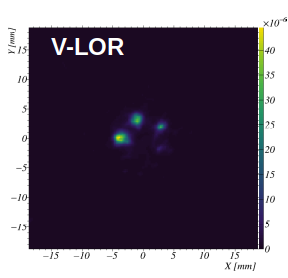}
\label{fig:conventionalICSsim:1}}
\hfil
\subfloat[]{\includegraphics[height=1.05in]{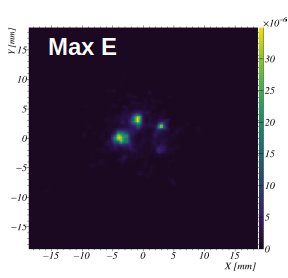}
\label{fig:conventionalICSsim:2}}
\hfil
\subfloat[]{\includegraphics[height=1.05in]{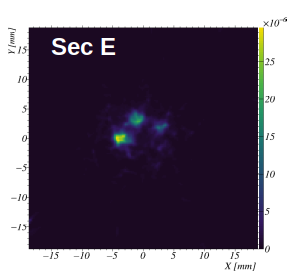}
\label{fig:conventionalICSsim:3}}
\\ %forces next subfloat to the next line
\subfloat[]{\includegraphics[height=1.05in]{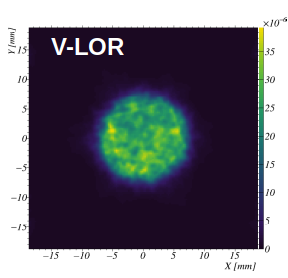}
\label{fig:conventionalICSsim:4}}
\hfil
\subfloat[]{\includegraphics[height=1.05in]{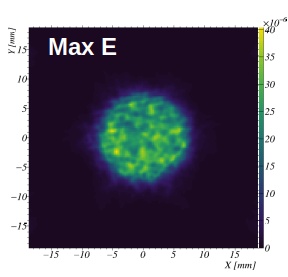}
\label{fig:conventionalICSsim:5}}
\hfil
\subfloat[]{\includegraphics[height=1.05in]{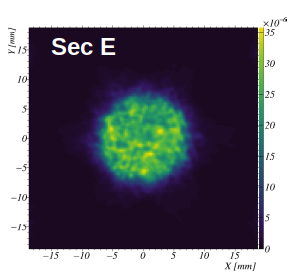}
\label{fig:conventionalICSsim:6}}
\\ %forces next subfloat to the next line
\subfloat[]{\includegraphics[height=1.05in]{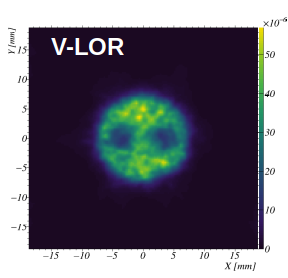}
\label{fig:conventionalICSsim:7}}
\hfil
\subfloat[]{\includegraphics[height=1.05in]{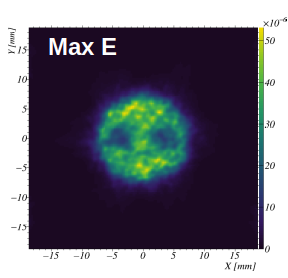}
\label{fig:conventionalICSsim:8}}
\hfil
\subfloat[]{\includegraphics[height=1.05in]{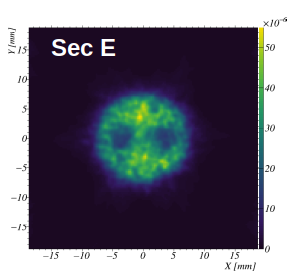}
\label{fig:conventionalICSsim:9}}
\caption{Transversal ($XY$) projections over the NEMA rods (first row), uniform (second row) and inserts (third row) regions at iteration $10$ and only-\ac{ics} simulated data.}
\label{fig:conventionalICSsim}
\end{figure}

\begin{figure}[!t]
\centering
\subfloat[]{\includegraphics[height=1.05in]{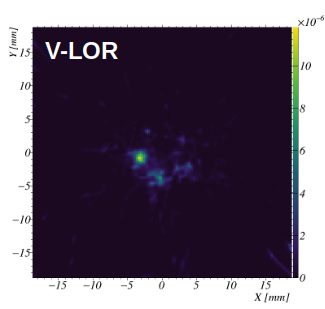}
\label{fig:conventionalICSexp:1}}
\hfil
\subfloat[]{\includegraphics[height=1.05in]{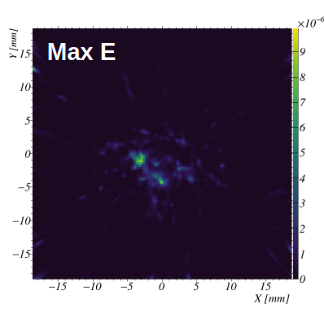}
\label{fig:conventionalICSexp:2}}
\hfil
\subfloat[]{\includegraphics[height=1.05in]{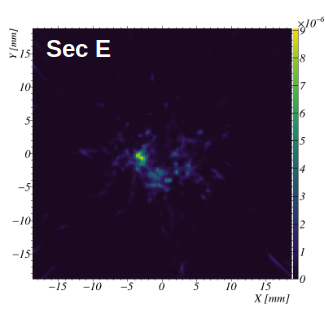}
\label{fig:conventionalICSexp:3}}
\\ %forces next subfloat to the next line
\subfloat[]{\includegraphics[height=1.05in]{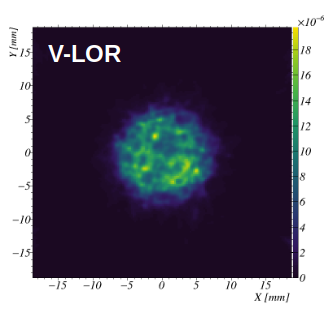}
\label{fig:conventionalICSexp:4}}
\hfil
\subfloat[]{\includegraphics[height=1.05in]{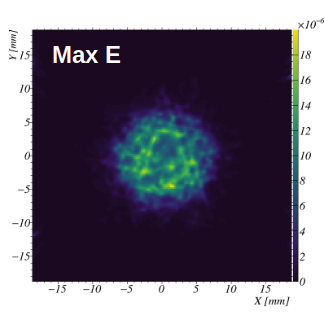}
\label{fig:conventionalICSexp:5}}
\hfil
\subfloat[]{\includegraphics[height=1.05in]{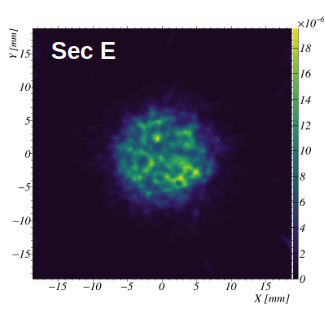}
\label{fig:conventionalICSexp:6}}
\\ %forces next subfloat to the next line
\subfloat[]{\includegraphics[height=1.05in]{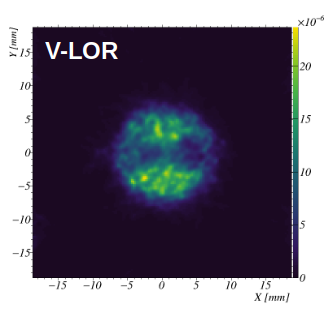}
\label{fig:conventionalICSexp:7}}
\hfil
\subfloat[]{\includegraphics[height=1.05in]{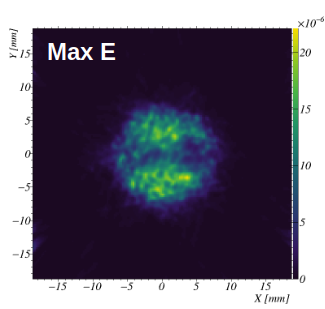}
\label{fig:conventionalICSexp:8}}
\hfil
\subfloat[]{\includegraphics[height=1.05in]{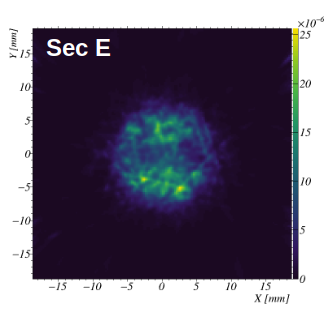}
\label{fig:conventionalICSexp:9}}
\caption{Transversal ($XY$) projections over the NEMA rods (first row), uniform (second row) and inserts (third row) regions at iteration $10$ and only-\ac{ics} MERMAID experimental data.}
\label{fig:conventionalICSexp}
\end{figure}

\begin{figure}[!t]
\centering
\subfloat[]{\includegraphics[height=0.99in]{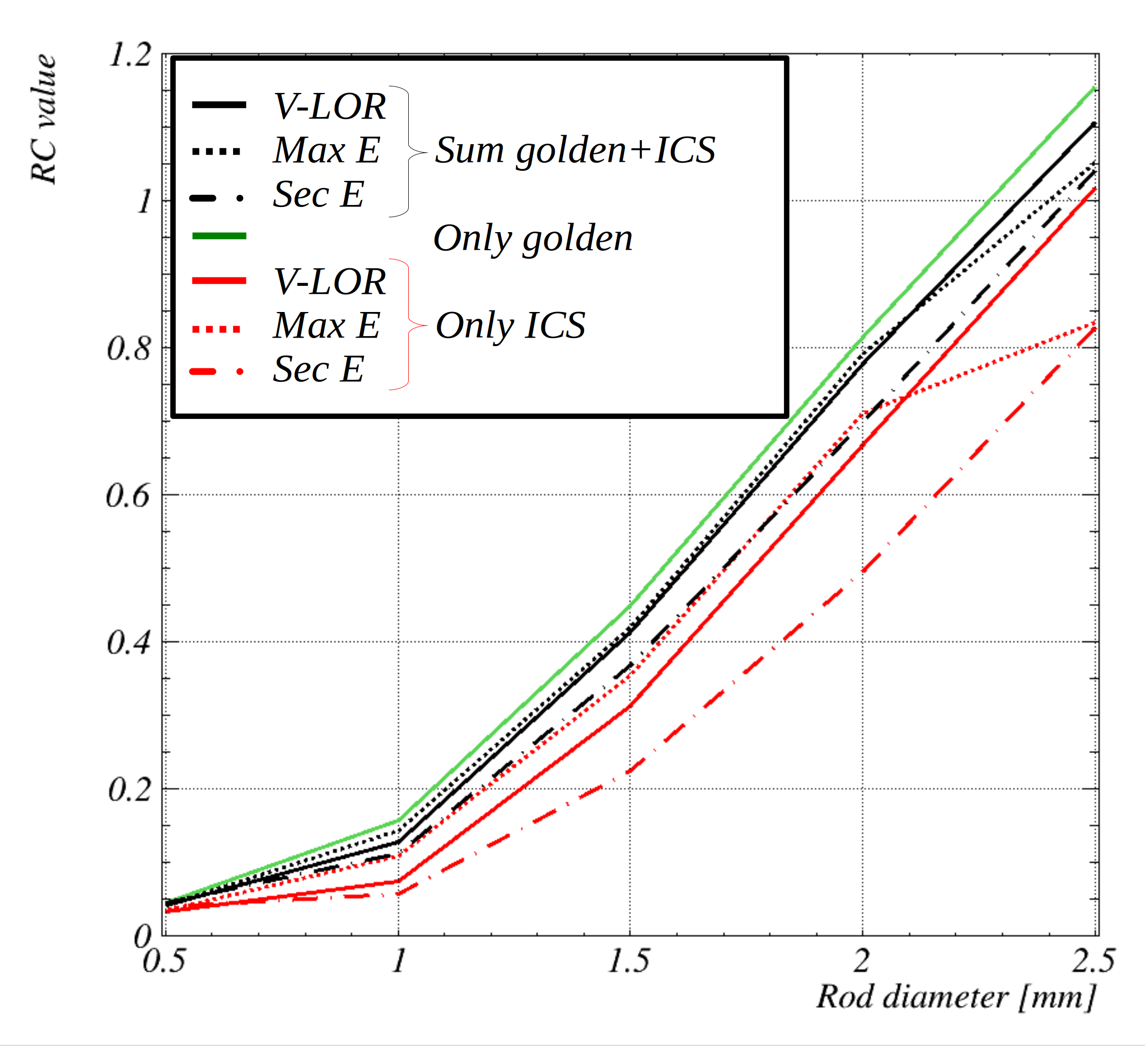}
\label{fig:conventionalICSsim:10}}
\hfil
\subfloat[]{\includegraphics[height=0.99in]{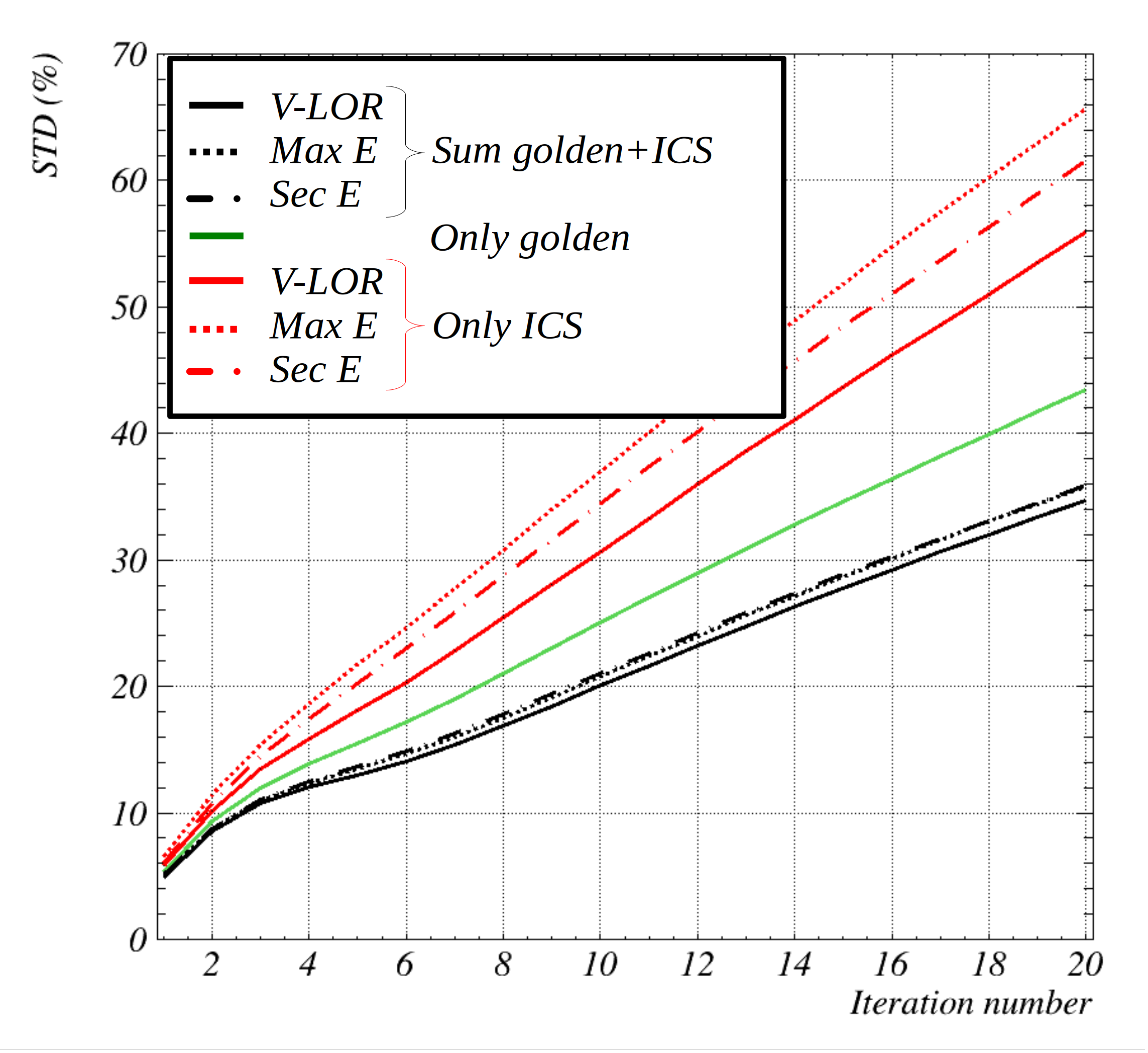}
\label{fig:conventionalICSsim:11}}
\hfil
\subfloat[]{\includegraphics[height=0.99in]{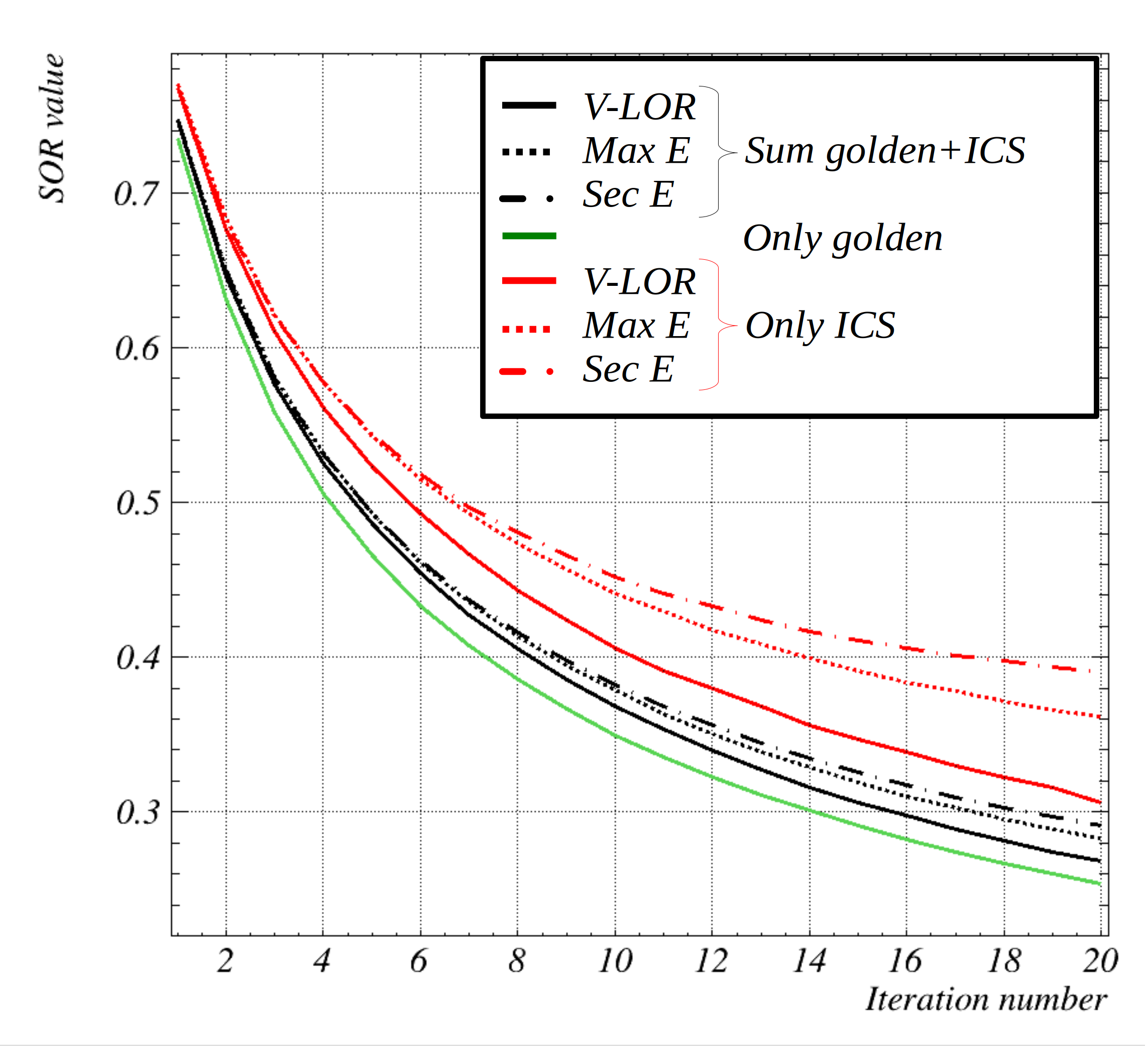}
\label{fig:conventionalICSsim:12}}
\\ %forces next subfloat to the next line
\subfloat[]{\includegraphics[height=0.99in]{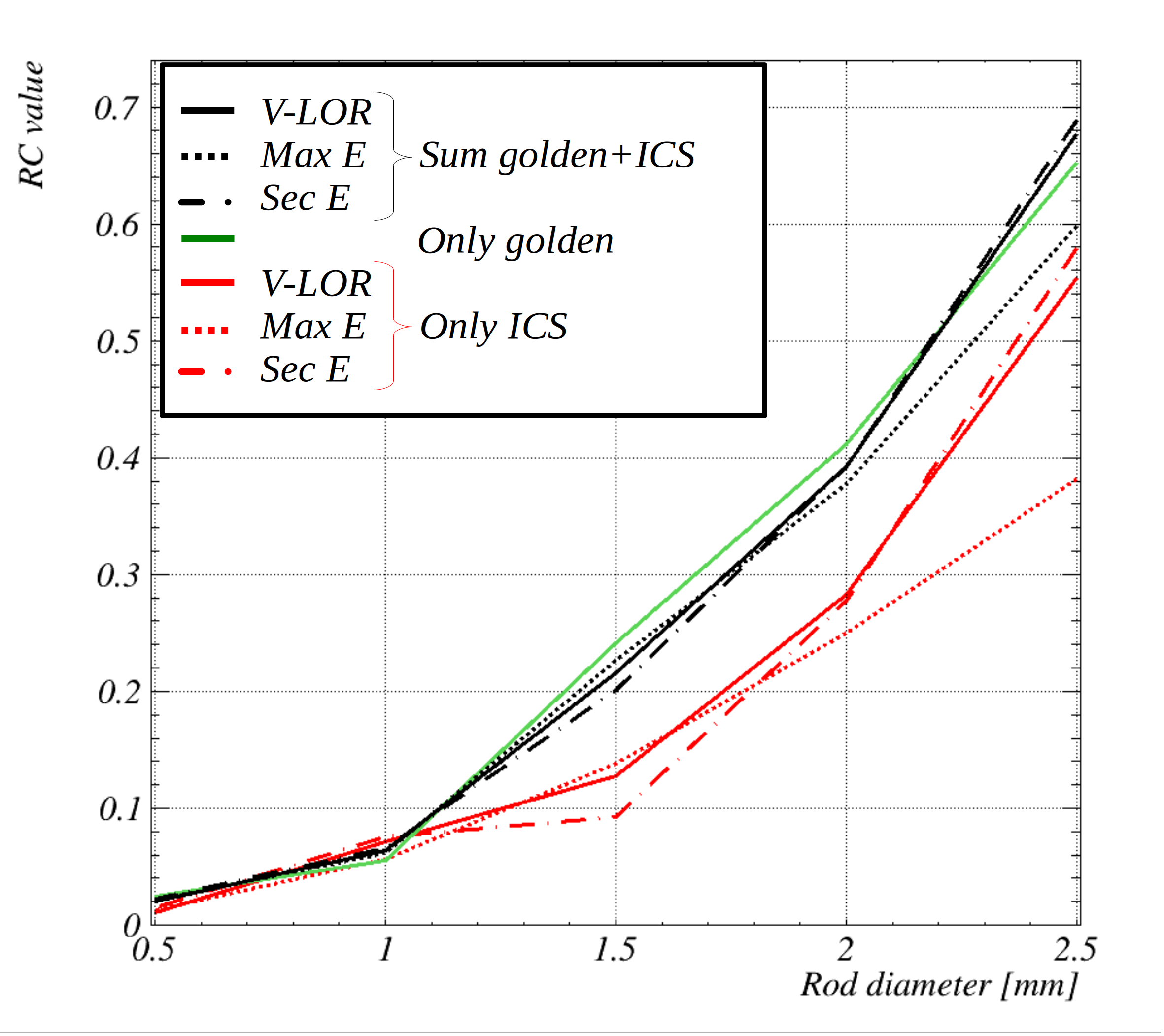}
\label{fig:conventionalICSexp:10}}
\hfil
\subfloat[]{\includegraphics[height=0.99in]{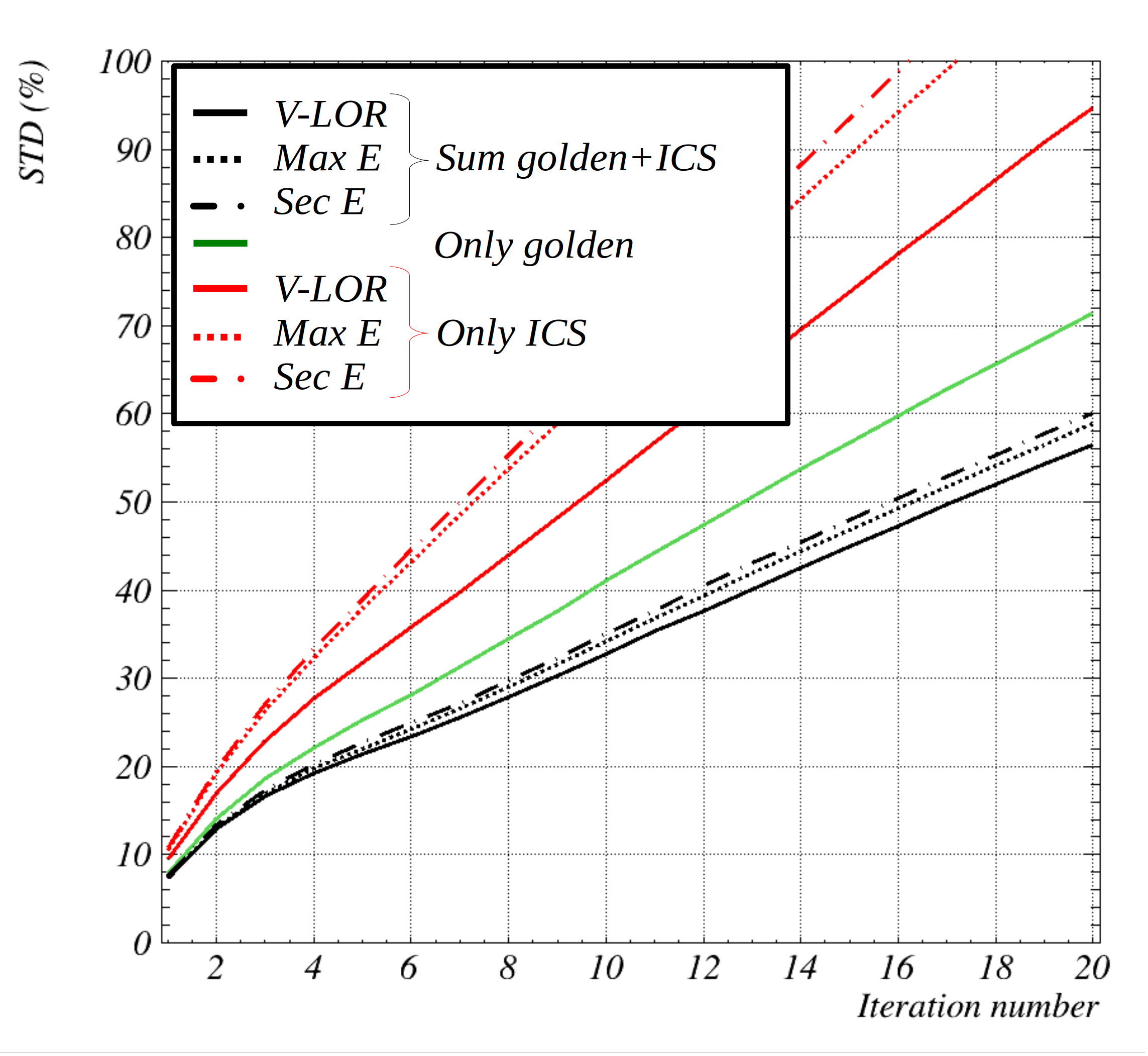}
\label{fig:conventionalICSexp:11}}
\hfil
\subfloat[]{\includegraphics[height=0.99in]{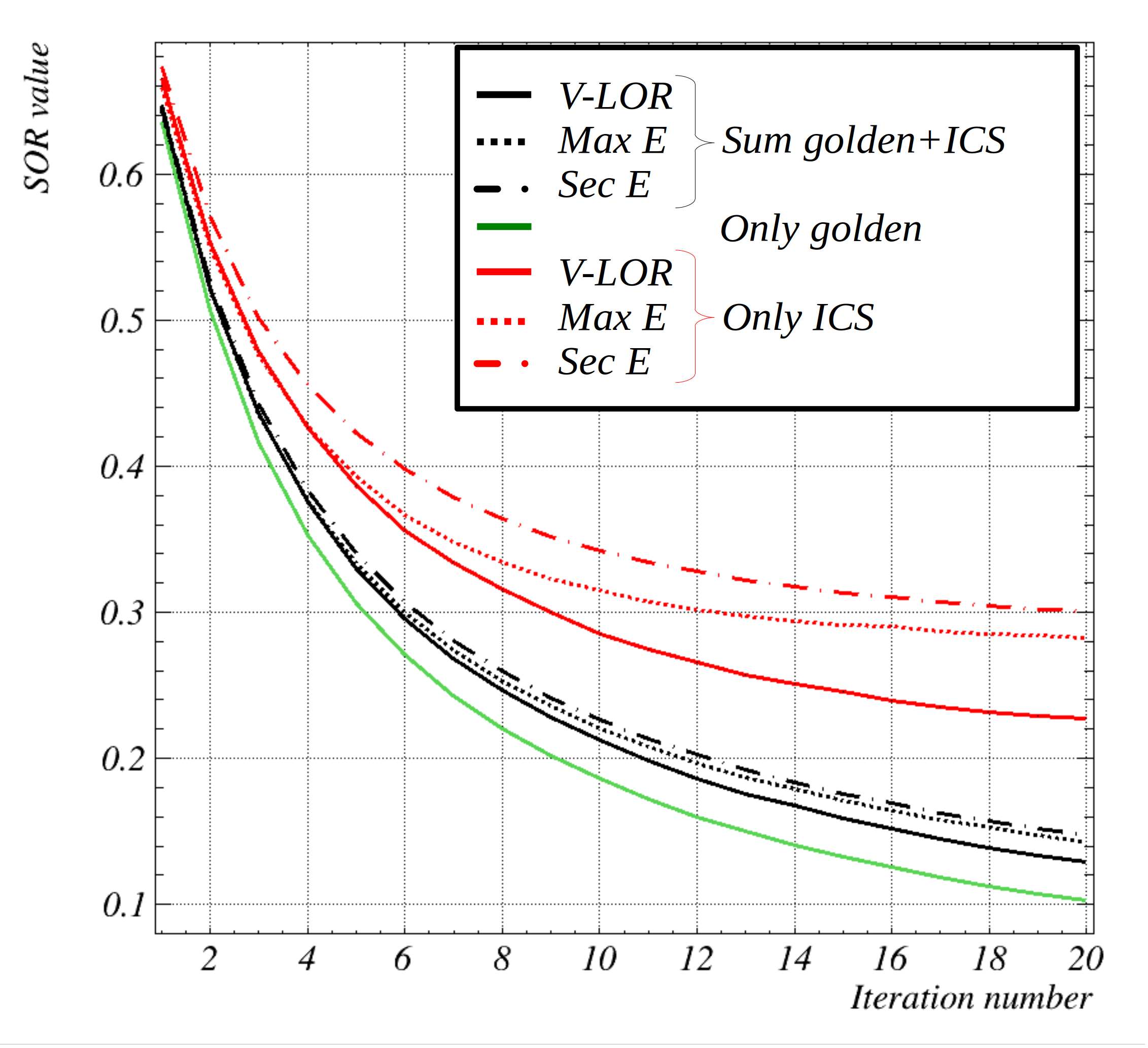}
\label{fig:conventionalICSexp:12}}
\caption{Recovery coefficients (iteration 5), percentage of uniformity-standard deviation, and SOR metrics, for simulated data (top row) and for MERMAID experimental data (bottom row).}
\label{fig:conventionalICSmetrics}
\end{figure}

Figures \ref{fig:conventionalICSsim}-\ref{fig:conventionalICSmetrics} show the comparison of these two energy-based ICS selection criteria with the proposed analytical model. In particular, figure \ref{fig:conventionalICSsim} shows XY-projections for ICS-only simulated data in the three regions of the \ac{iq} phantom (figures \ref{fig:conventionalICSsim:1}-\ref{fig:conventionalICSsim:9}), while figure \ref{fig:conventionalICSexp} shows the same information with MERMAID experimental data. Finally, figure  \ref{fig:conventionalICSmetrics} shows the \ac{iq} phantom associated metrics for ICS-only, golden-only, and sum golden + ICS data, in both the simulated and experimental scenarios. It must be noted that, when using the energy-based ICS selection criteria, the reconstruction with sum golden + ICS data via the joint algorithm (\ref{eq:joint}) becomes essentially conventional LM-MLEM, using LORs both for golden and ICS-selected events; in these cases, only the golden sensitivity image $\mathbf{s}^{\text{G}}$ was employed.

Results suggest that the proposed methodology, based on an analytical model leading to V-LORs, keeps similar recovery coefficients values compared to the two energy-based ICS selection methods (both for the only-ICS data and for the sum), while improving clearly the uniformity and the SOR. This improvement is also visible qualitatively in the $XY$-projections.

%

% use section* for acknowledgment
\section*{Acknowledgment}

This work was supported by the Alexander von Humboldt foundation, by the  European Union’s Horizon 2024 research and innovation programme under the Marie Sk\l{}odowska-Curie grant agreement no. 101207067, and by the German Research Foundation (DFG) under grant agreement no. 496099829.

%The authors would like to thank...

% Can use something like this to put references on a page
% by themselves when using endfloat and the captionsoff option.
\ifCLASSOPTIONcaptionsoff
  \newpage
\fi

% trigger a \newpage just before the given reference
% number - used to balance the columns on the last page
% adjust value as needed - may need to be readjusted if
% the document is modified later
%\IEEEtriggeratref{8}
% The "triggered" command can be changed if desired:
%\IEEEtriggercmd{\enlargethispage{-5in}}

% references section

% can use a bibliography generated by BibTeX as a .bbl file
% BibTeX documentation can be easily obtained at:
% http://mirror.ctan.org/biblio/bibtex/contrib/doc/
% The IEEEtran BibTeX style support page is at:
% http://www.michaelshell.org/tex/ieeetran/bibtex/
%\bibliographystyle{IEEEtran}
% argument is your BibTeX string definitions and bibliography database(s)
%\bibliography{IEEEabrv,../bib/paper}
%
% <OR> manually copy in the resultant .bbl file
% set second argument of \begin to the number of references
% (used to reserve space for the reference number labels box)
%\begin{thebibliography}{1}
%
%\bibitem{IEEEhowto:kopka}
%H.~Kopka and P.~W. Daly, \emph{A Guide to \LaTeX}, 3rd~ed.\hskip 1em plus
%  0.5em minus 0.4em\relax Harlow, England: Addison-Wesley, 1999.
%
%\end{thebibliography}

\bibliographystyle{IEEEtran}
\bibliography{bare_jrnl}

% biography section
% 
% If you have an EPS/PDF photo (graphicx package needed) extra braces are
% needed around the contents of the optional argument to biography to prevent
% the LaTeX parser from getting confused when it sees the complicated
% \includegraphics command within an optional argument. (You could create
% your own custom macro containing the \includegraphics command to make things
% simpler here.)
%\begin{IEEEbiography}[{\includegraphics[width=1in,height=1.25in,clip,keepaspectratio]{mshell}}]{Michael Shell}
% or if you just want to reserve a space for a photo:

%\begin{IEEEbiography}{Michael Shell}
%Biography text here.
%\end{IEEEbiography}

% if you will not have a photo at all:
%\begin{IEEEbiographynophoto}{John Doe}
%Biography text here.
%\end{IEEEbiographynophoto}

% insert where needed to balance the two columns on the last page with
% biographies
%\newpage

%\begin{IEEEbiographynophoto}{Jane Doe}
%Biography text here.
%\end{IEEEbiographynophoto}

% You can push biographies down or up by placing
% a \vfill before or after them. The appropriate
% use of \vfill depends on what kind of text is
% on the last page and whether or not the columns
% are being equalized.

%\vfill

% Can be used to pull up biographies so that the bottom of the last one
% is flush with the other column.
%\enlargethispage{-5in}

% that's all folks
\end{document}